\documentclass[11pt]{article}

\usepackage{a4}

\usepackage[breaklinks,colorlinks,citecolor=blue]{hyperref}
\usepackage{url}
\usepackage{etoolbox}
\newcommand{\pasp}{Proceedings of the Astronomical Society of the Pacific}
\newcommand{\mnras}{Monthly Notices of the Royal Astronomical Society}
\newcommand{\apj}{Astrophysical Journal}
\newcommand{\apjs}{Astrophysical Journal Supplement} 
\newcommand{\aj}{Astronomical Journal}
\newcommand{\apjl}{Astrophysical Journal Letters}
\newcommand{\nat}{Nature}
\newcommand{\aap}{Astronomy \& Astrophysics}
\newcommand{\prd}{Physical Review D}
\newcommand{\actaa}{Acta Astronomica}
\usepackage{graphicx}
\usepackage{tikz}
\usetikzlibrary{decorations.pathreplacing}
\usepackage{pgfplots}

\usepackage[utf8]{inputenc}
\usepackage{amsmath}
\usepackage{amsfonts}
\usepackage{amssymb}

\newtoggle{arxiv}
\toggletrue{arxiv}
\iftoggle{arxiv}{
\usepackage{natbib}
\bibpunct{(}{)}{,}{a}{}{;}

\newcommand{\citenp}[1]{\citealt{#1}}

}{

\usepackage[
    backend=bibtex, 
    style=authoryear, 
    natbib=true,
    sortlocale=en_US,
    url=true, 
    doi=true,
    eprint=true
]{biblatex}
\addbibresource{art-bibfile.bib}

\newcommand{\citenp}[1]{\cite{#1}}

}

\usepackage{animate}

\usepackage[nottoc,numbib]{tocbibind}

\newcommand{\be}{\begin{equation}}
\newcommand{\ee}{\end{equation}}
\newcommand{\bea}{\begin{eqnarray}}
\newcommand{\eea}{\end{eqnarray}}
\newcommand{\Dd}{\mathrm{d}}
\usepackage{footnote}
\makesavenoteenv{tabular}

\usepackage{listings}
\definecolor{dkgreen}{rgb}{0,0.6,0}
\definecolor{gray}{rgb}{0.5,0.5,0.5}
\definecolor{mauve}{rgb}{0.58,0,0.82}

\usepackage{caption}
\usepackage{makeidx}
\makeindex 

\newcommand{\Index}[1]{\index{#1}{#1}}
\numberwithin{equation}{section}

\begin{document}

\title{The expanding universe: an introduction \\[0.5em]
\large Lecture at the WE Heraeus Summer School\\ ``Astronomy from Four Perspectives'', Heidelberg}
\author{Markus Pössel, Haus der Astronomie}
\date{28 August 2017} 

\maketitle

\tableofcontents

\newpage

\section{Introduction}

Modern cosmology is based on general relativity. Teaching general relativity is a challenge --- if you go all the way, you will need mathematical concepts so advanced that they are not even included in the usual mathematics courses for students of physics. Typical graduate-level lectures on general relativity thus need to include sections introducing the necessary mathematical formalism, notably concepts from differential geometry.

For introducing general relativity to undergraduates, or even in a high school setting, simplifications are necessary, leading to the central question: in general relativity, how far can you go without the full formalism? In this lecture, we will ask this question in the context of cosmology: Which aspects of the expanding universe, of the modern cosmological models, can you understand without using the formalism of general relativity? 

On the simplest level, this brings us to the various models commonly used to explain cosmic expansion -- the expanding \index{rubber balloon}{rubber balloon}, the linear \index{rubber band}{rubber band} as a one-dimensional universe, and the \index{raisin cake}{raisin cake} \citep{Eddington1930,Lotze1995,Price2001,UAYF,Strauss2016}. Used judiciously, these models can convey a basic understanding of what it means for a universe to be expanding. The main focus of this lecture is on quantitative results, though: How many of the calculations of standard cosmology can we reproduce without employing the formalism of general relativity?

\begin{figure}[htpb]
\includegraphics[width=0.9\textwidth]{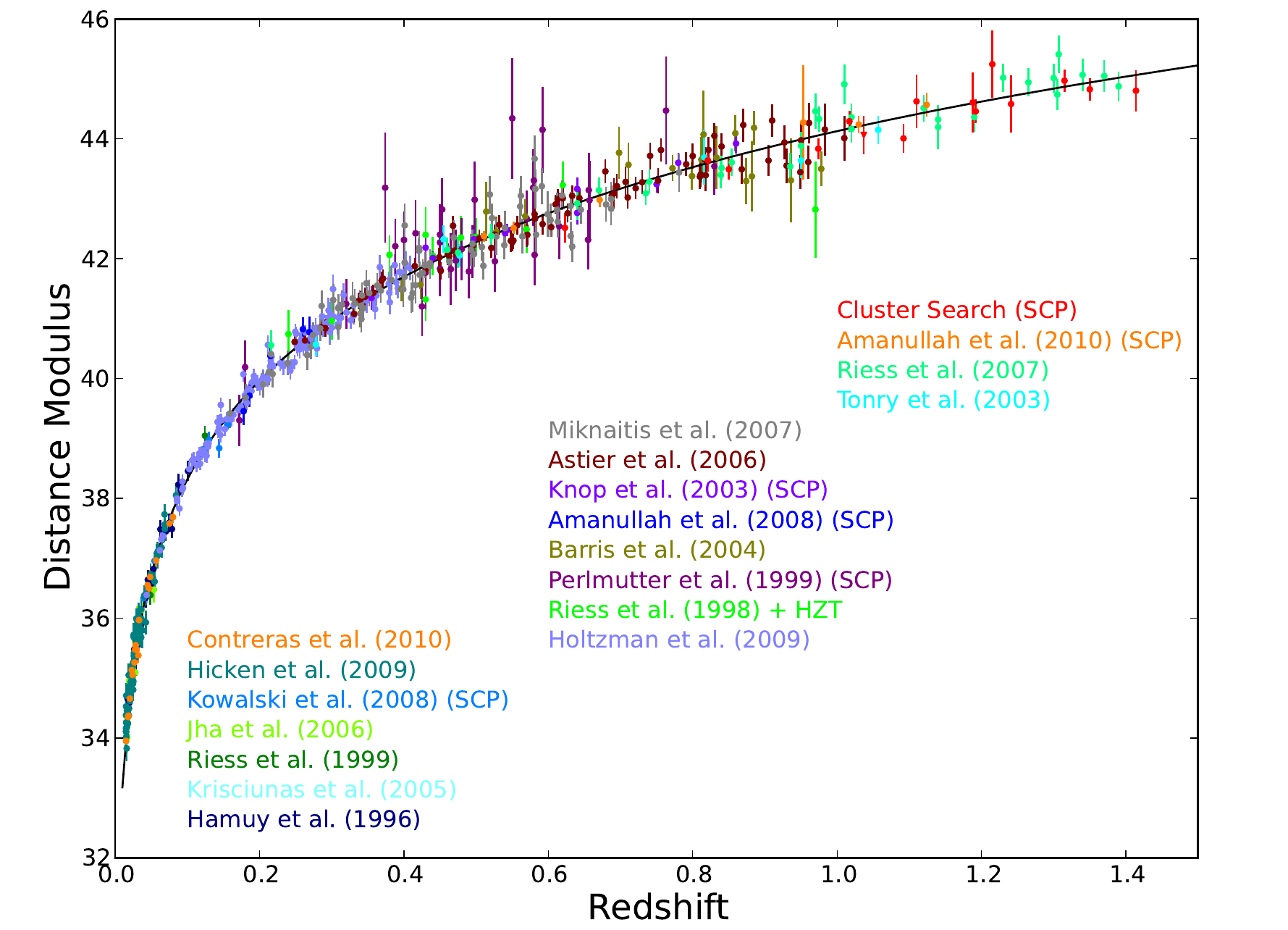}
\caption{Distance moduli plotted against redshift for various standard candles. Figure 4 in \citenp{SuzukiEtAl2012}, available at \href{http://www-supernova.lbl.gov/}{http://www-supernova.lbl.gov/}
}
\label{InitialDiagram}
\end{figure}

As it turns out, in the context of cosmology, the basic tenets of general relativity can take you quite a long way. Our goal in these lecture notes will be to understand the basic predictions of the {\em \Index{Friedmann-Lema\^{\i}tre-Robertson-Walker models}}, expanding homogeneous universes that form the backbone of the Big Bang models of modern cosmology, along these lines. More specifically, our goal will be to understand one of the most important links between cosmological models and astronomical observations: the generalized Hubble diagram, linking the distances of certain standard candles (that is, objects of known brightness) and their redshifts. Figure \ref{InitialDiagram} shows an example (namely figure 4 in \citenp{SuzukiEtAl2012}).

How come that all these sources lie along this particular curve? How do we derive the curve's shape, and how is it linked to fundamental properties of the universe?

The following notes have grown out of a lecture with the title ``Introducing the expanding universe without using the concept of a spacetime metric,'' which I held on 28 August 2017 at Haus der Astronomie, at our WE Heraeus Summer School ``Astronomy from four perspectives,'' a summer school for teachers, students training to be teachers, astronomers and astronomy students from Heidelberg, Padova, Jena, and Florence. This year's theme was ``The Dark Universe,'' an exploration of dark matter and dark energy.  An edited version of the lecture can be found on YouTube at
\begin{center}
\href{http://youtu.be/gA-0C-88WbE}{http://youtu.be/gA-0C-88WbE}
\end{center}
The aim of the lecture was to give a basic overview of modern cosmological models, in order to prepare our participants for more specialized lectures and tutorials. The lecture's goal of making cosmology understandable without introducing the underlying formalism was motivated both by the diverse backgrounds of the listeners and by the summer school's underlying goal of making modern astronomical research accessible to high school students.

On the part of the reader, I assume familiarity with basic classical mechanics, including Newton's law of gravity, and the basics of special relativity; in order to make the text more accessible, the content we need from special relativity is summarised in appendix \ref{SpecialRelativity}.

\section{Length scales and the realm of cosmology}

Seen naively, \Index{cosmology} is the most ambitious science. We aim to understand the universe as a whole! The universe, as everyday experience shows, is rather complex, with many different interesting scales, comprising everything from insects via humans, city-scale structures, moons, planets, stars, and galaxies -- and that list doesn't even list all the interesting stuff at the submicroscopic level!

Of course, cosmology is really much simpler than that (although still very ambitious!). We do what all astronomers and physicists do: We concentrate on a specific subset of phenomena, and formulate simplified models to describe what is happening in that subset --- do describe the physical objects involved, and their interactions.

\begin{figure}[htpb]
\includegraphics[width=\textwidth]{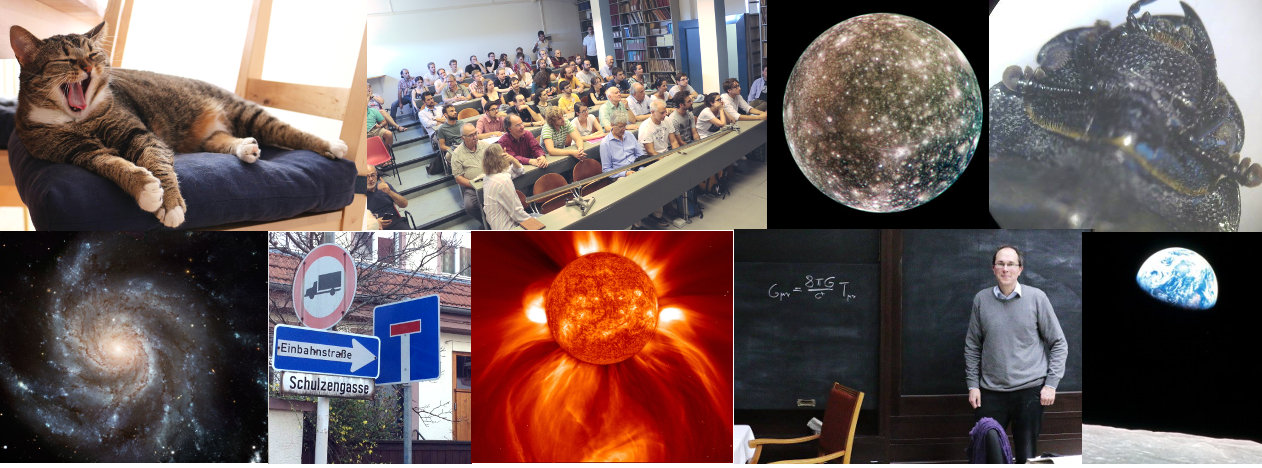}
\caption{Some examples of what would need to be included, were we to try to understand the universe at all scales.
\footnotesize
Image credit:  C. Liefke, M. Pössel, R. Pössel, NASA, NASA/ESA/CFHT/NOAO, ESA/NASA/SOHO}
\end{figure}

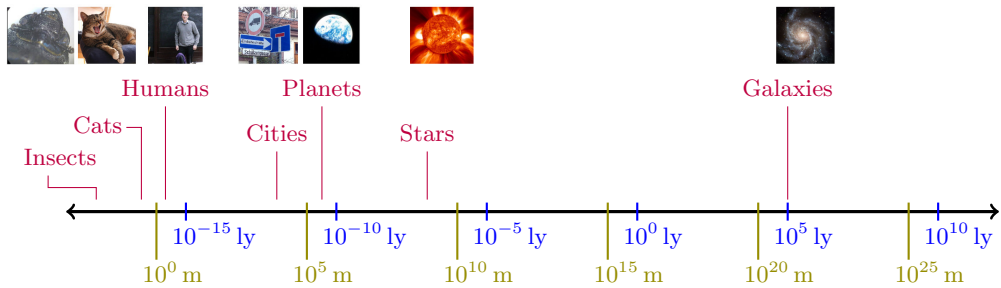
\begin{figure}[htpb]
\hspace*{-0.5em}\includegraphics[width=1.084\textwidth]{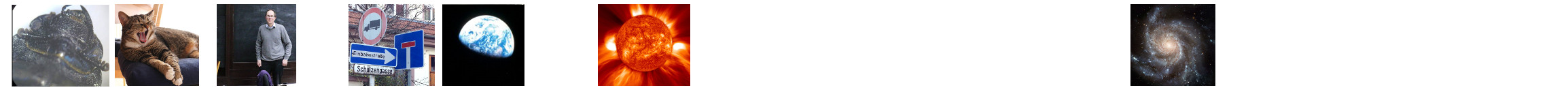}\\
\begin{tikzpicture}[scale=0.80]

\draw[<->,very thick] (-1.5,0) -- (14.,0);
\foreach \i in {0,5,...,25}
{
\draw[thick,olive] (0.5*\i,0.15) -- (0.5*\i,-0.8);
}

\foreach \i in {-15,-10,...,10}
{

\draw[thick,blue] ({0.5*(\i+15.98)},0.15) -- ({0.5*(\i+15.98)},-.15);
}

\foreach \s [count=\x from 0] in {{$10^0\,$m},{$10^5\,$m},{$10^{10}\,$m},{$10^{15}\,$m},{$10^{20}\,$m},{$10^{25}\,$m}}
{
\node at (2.5*\x-0.4,-1.0) [anchor=west,olive]  {\footnotesize{\s}};
}

\foreach \s [count=\x from 0] in {{$10^{-15}\,$ly},{$10^{-10}\,$ly},{$10^{-5}\,$ly},{$10^{0}\,$ly},{$10^{5}\,$ly},{$10^{10}\,$ly}}
{
\node at ({0.5*(5*\x+0.98)-0.4},-0.4) [anchor=west,blue]  {\footnotesize{\s}};
}

\def\x{-2}
\def\y{0.6}
\draw[purple] (0.5*\x,0.2) -- (0.5*\x,0.4) -- (-1.8,0.4) -- (-1.8,\y);
\node at(-1.6,\y) [anchor=south,purple] {\footnotesize Insects};

\def\x{-0.5}
\def\y{0.6}
\draw[purple] (0.5*\x,0.2) -- (0.5*\x,1.4) -- (-0.5,1.4);
\node at(-0.4,1.45) [anchor=east,purple] {\footnotesize Cats};

\def\x{0.3}
\def\y{1.75}
\draw[purple] (0.5*\x,0.2) -- (0.5*\x,\y);
\node at(0.5*\x,\y) [anchor=south,purple] {\footnotesize Humans};

\def\x{4}
\def\y{1.0}
\draw[purple] (0.5*\x,0.2) -- (0.5*\x,\y);
\node at(0.5*\x,\y) [anchor=south,purple] {\footnotesize Cities};

\def\x{5.5}
\def\y{1.75}
\draw[purple] (0.5*\x,0.2) -- (0.5*\x,\y);
\node at(0.5*\x,\y) [anchor=south,purple] {\footnotesize Planets};

\def\x{9}
\def\y{1.0}
\draw[purple] (0.5*\x,0.2) -- (0.5*\x,\y);
\node at(0.5*\x,\y) [anchor=south,purple] {\footnotesize Stars};

\def\x{20.98}
\def\y{1.75}
\draw[purple] (0.5*\x,0.2) -- (0.5*\x,\y);
\node at(0.5*\x,\y) [anchor=south,purple] {\footnotesize Galaxies};

\end{tikzpicture}
\caption{Different length scales \label{CosmicScales}}
\end{figure}

In the case of cosmology, the defining feature is \index{scale@scales!cosmic}{scale}. Figure \ref{CosmicScales} shows the various length scales, from the smallest objects we can still see with the naked eye all the way up to galaxies and beyond.

As cosmologists, when we formulate our simplest large-scale models, we leave the bugs to the entomologists, cats to the Internet community, humans to the life scientists, psychologists and sociologists, and even within astronomy, we are not all that interested in planets, stars, and the structure of galaxies.

In large-scale models, \index{galaxies!as point particles}{galaxies} are something akin to the point particles of classical mechanics: structureless objects whose only interesting properties are position, motion, and (total)  mass. 

Our coarse, large-scale view also determines the dominant interaction we shall model. It's gravity. As you learn in Astronomy 101, this is not because gravity is very strong. On the contrary, if you look at the elementary constituents of ordinary matter, namely at protons, atomic nuclei, and electrons, the other fundamental interactions are much stronger than gravity --- the electrostatic attraction between an electron and a proton, for instance, is a whopping $10^{39}$ stronger than their mutual gravitational attraction, and the discrepancy is even larger for nuclear forces and the particles on which they act.

But over long scales, \index{gravity!as dominant force}{gravity} wins out: The nuclear interactions have strictly limited range. Electromagnetism has positive and negative charges, and precisely because its forces are comparatively strong, charged particles combine into electrically neutral objects. There do not appear to be any large scale imbalances of electric charge --- say, a surplus of negative charges in the Andromeda galaxy and a corresponding deficit in our own galaxy. 
On the other hand, gravitational charges, that is, masses, will always add up. That is how, on the largest scales, gravity comes to dominate.

In order to describe gravity, we turn to the best current theory of gravity that we have: Albert Einstein's general theory of relativity.

\section{General relativity}

\index{general relativity}{General relativity}, first published by Einstein in late 1915, relates gravity to distortions of space and time. A famous, concise prose summary is due to John Wheeler, and states that spacetime tells matter how to move, while matter tells spacetime how to curve (\citenp{WheelerSciAm}). 

To formulate these statements more precisely, and to give a more precise meaning to terms like distortion and \Index{curvature}, simplified expositions often introduce pared-down geometric models, reducing four-dimensional spacetime to a two-dimensional surface. But these simple visualisations can only take us so far. At some point, in particular if we our goal is to make specific calculations, we will need to acquaint ourselves with the proper formalism.

But as it will turn out, for most of cosmology, you do not need to know what it means for spacetime to be curved. Instead, our calculations will make use of much more basic principles that are part of general relativity. The first is the {equivalence principle}, namely that in free fall, the most immediate effects of gravity are absent. The second is the Newtonian limit: under certain conditions, general relativity reduces to the \index{Newtonian limit}{Newtonian description of gravity}. The third is a statement about sources of gravity -- a generalisation of Newtonian gravity, where mass is the only physical quantity that produces gravity. 
These three pieces of information will turn out to be all that is needed to derive the standard model of an expanding universe.

\subsection{Equivalence principle}
\label{EquivalencePrincipleSection}

\begin{figure}[htbp]
\begin{center}
\includegraphics[width=0.27\textwidth]{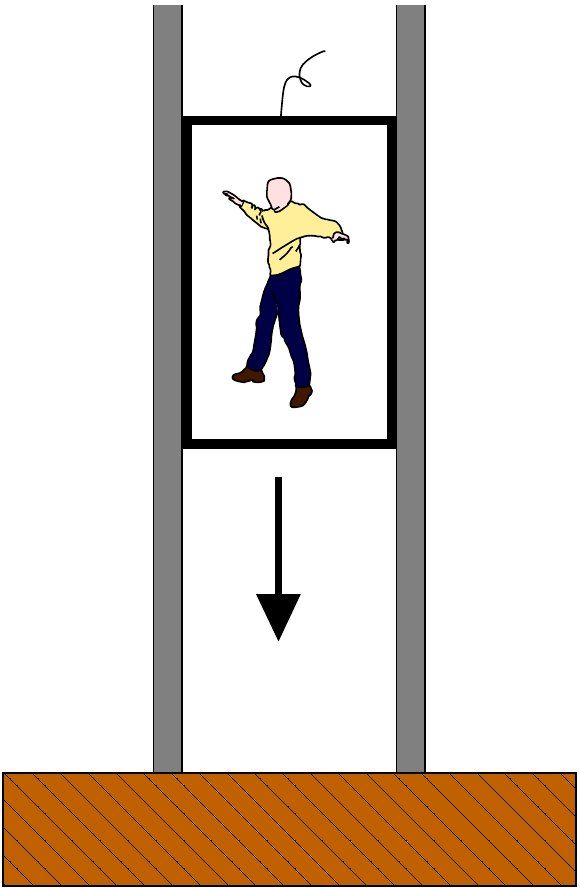} \hspace*{3em}
\includegraphics[width=0.54\textwidth]{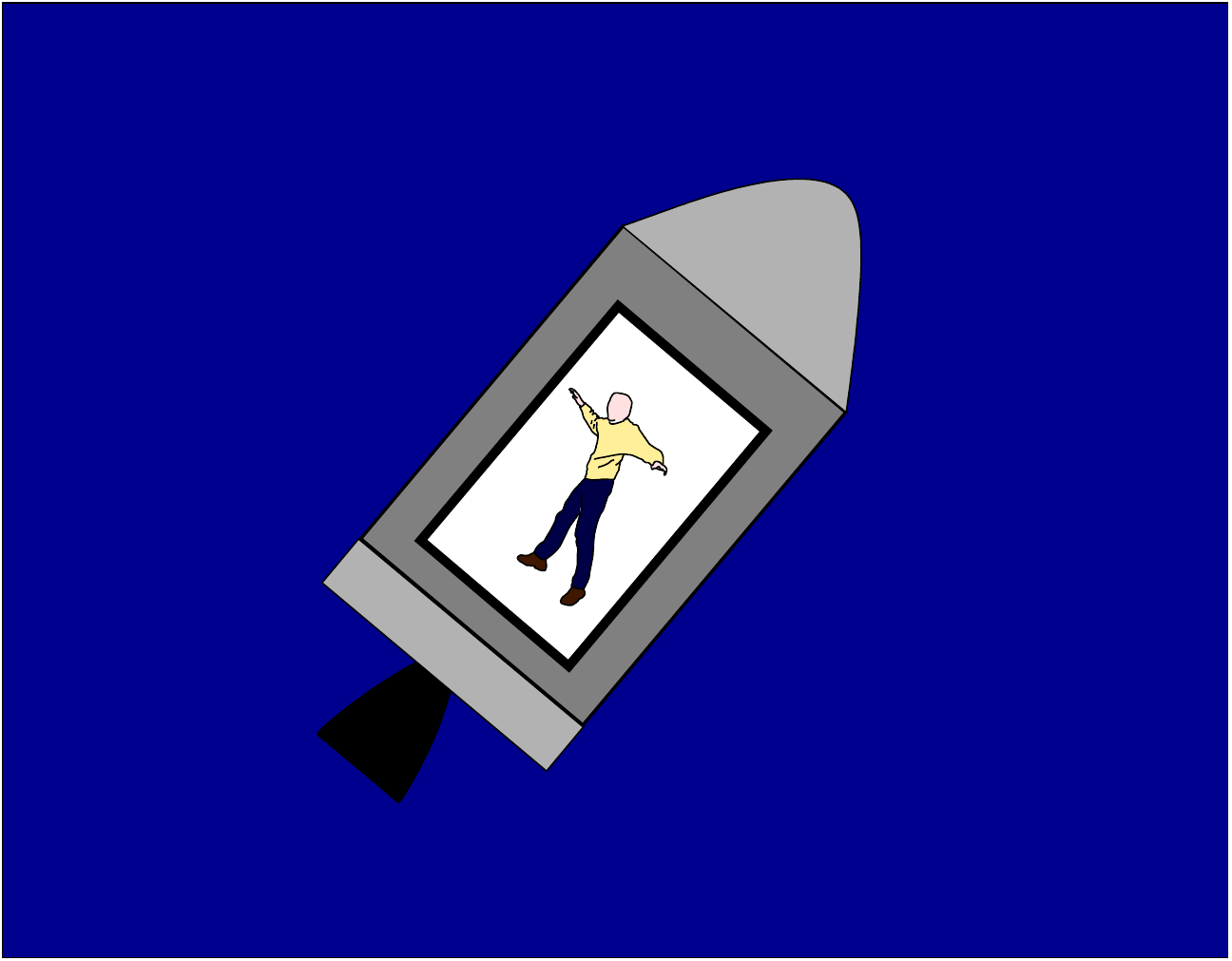}
\caption{Different observers in free fall: An observer in free fall in a gravitational field (left), and one who is far from any sources of gravity}
\label{FreeFall}
\end{center}
\end{figure}

When Einstein began to think about how to incorporate gravity into his special theory of relativity, he hit upon a simple thought experiment. In his own words, twice removed:\footnote{This is from a speech Einstein gave at Kyoto University in December 1922, which was in German, translated live into Japanese, documented in the same language, and an English translation published in \citenp{Ono1982}. }
\begin{quotation}
The breakthrough came suddenly one day. I was sitting on a chair in my patent office in Bern. Suddenly a thought struck me: If a man falls freely, he would not feel his weight. I was taken aback. This simple thought experiment made a deep impression on me. This led me to the theory of gravity.
\end{quotation}
In modern parlance, the outcome of this is the {\em (Einstein) \Index{equivalence principle}}. Consider two observers in \Index{free fall}, one in an elevator cabin, the other adrift in the cabin of a space-ship, far from any sources of gravity; these two cases are shown in figure \ref{FreeFall}.

A key question is: Can these two observers tell the difference? When they perform physics experiments in their little cabins, can they tell whether or not there are sources of gravity nearby?

To a large extent, the answer is no. After all, in free fall, the most common indicators of a gravitational field are absent. 

In everyday life, if I release a ball, it will fall to the ground. If I am in an elevator cabin in free fall, and gently release a ball, it will continue to float in front of me. Water will float, forming a wobbling giant droplet. If I position myself on a balance, that balance will show my weight to be zero. Behind all this is the fact that, in a Newtonian gravitational field, objects that are in the same place accelerate at the same rate.\footnote{In the Newtonian picture, this is because the same object mass occurs both in the formula linking force and acceleration, and in the formula specifying the gravitational force between a point mass $m$ and a much larger mass $M$. In the field of $M$, the point mass will be accelerated in the radial direction as
$$
a = \frac{1}{m}F = -\frac{1}{m} \frac{GMm}{r^2} = -\frac{GM}{r^2},
$$
which is independent of the object mass $m$. 
}

In fact, this rather good correspondence between a gravity-free situation and a free-fall situation is routinely used in physics. A widely known example is the \Index{International Space Station (ISS)}. At the cruising height of the ISS, at an altitude of about 400 km above sea level, the gravitational acceleration caused by the Earth is about 89\% as  strong as on Earth's surface. The reason the astronauts, and all unattached objects around them, are floating is not because gravity is weak, but because the ISS is in a free-fall orbit around Earth. \index{drop tower}{Drop towers}, where experiments are dropped inside a vacuum tube, can create similar \Index{microgravity} conditions, albeit for a much shorter time of a few seconds, and in a much smaller volume.

Our first rough version of the principle tries to summarize these observations as follows:
\begin{quotation}
\noindent {\em Einstein \Index{equivalence principle}, draft version:} Physics experiments performed by an observer in free fall will have the same outcome as experiments performed by an observer who is infinitely far from all sources of gravity. In particular, the rules governing space and time are those of special relativity.
\end{quotation}
\subsection{Tidal forces and the limits of the equivalence principle}
\begin{figure}[tbhp]
\begin{center}
\includegraphics[width=0.15\textwidth]{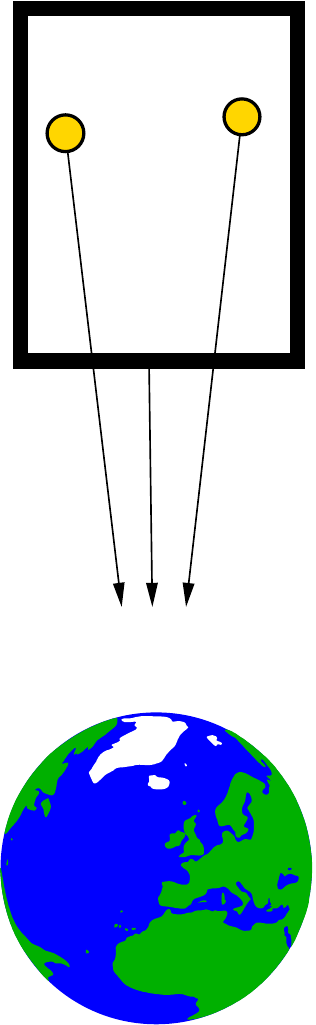} \hspace*{0.4\textwidth}
\includegraphics[width=0.15\textwidth]{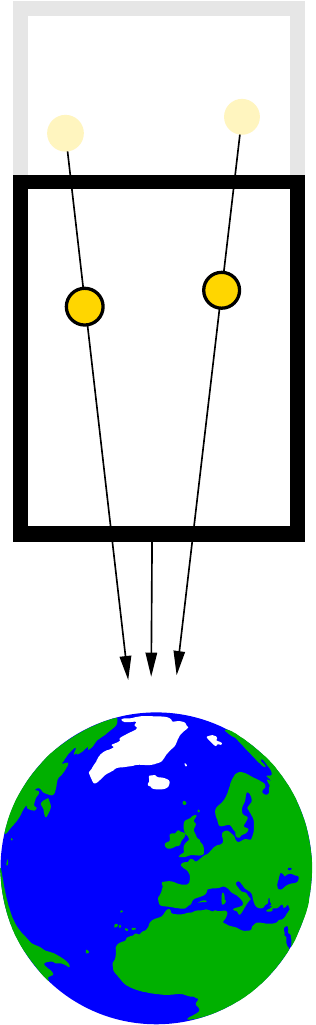}
\caption{A giant cabin containing two spheres, falling towards Earth, shown here at some time $t_1$ (left) and at a later time $t_2$ (right)}
\label{GiantCabin}
\end{center}
\end{figure}

If we look more closely, we will soon realise that there are fundamental problems with this version. Consider a truly gigantic elevator cabin falling towards Earth, with two giant spheres inside.\footnote{Worrying about the gigantic mass of such a cabin? That aspect of our thought experiment is admittedly inconsistent; so far, we have treated all falling particles as \Index{test particles}, whose own gravitational influence has no significant consequences. We will continue to do so. Just imagine that our giant cabin and giant spheres are made of truly fluffy, low-mass material.}
What happens next is shown in figure \ref{GiantCabin}.
Clearly, it's becoming important that the two spheres are not both falling downwards on parallel trajectories. Instead, they both fall towards the center of the Earth. This falling motion brings them closer together over time, as the figure shows. This is an effect an observer inside the cabin can detect. He or she need only let these two spheres float, making sure that, initially, they are at rest relative to each other, and wait until the two spheres have started to accelerate towards each other. An observer drifting along in a space-ship, far removed from all sources of gravity, will not see this effect.

What the observer in free fall in a gravitational field sees, and the gravity-free observer doesn't, are effects known as {\em tidal} \index{tidal effects}{effects}, which are due to the fact that gravitational fields typically vary from location to location and/or over time. As the term indicates, these varying gravitational fields are responsible for Earth's tides. The main reason our home planet's oceans have tides is because the Moon's gravity acting on the water directly below is slightly stronger than the Moon's gravity at the Earth's center-of-mass, which in turn is stronger than the Moon's gravity acting on water on the opposite side of the Earth.

From classical calculations in Newtonian gravity, it is clear that tidal forces have an important property.
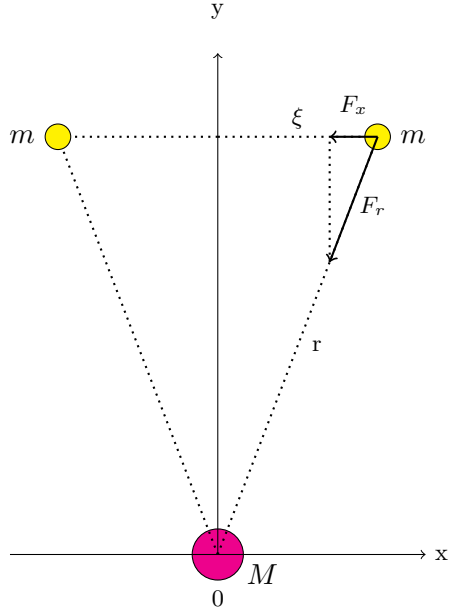
\begin{figure}[htbp]
\begin{center}
\begin{tikzpicture}[scale=0.85]
\pgfmathsetmacro{\initialRadius}{7}
\pgfmathsetmacro{\initialShift}{2.5}
\pgfmathsetmacro{\initialY}{sqrt(\initialRadius*\initialRadius - \initialShift*\initialShift)}
\draw[fill=magenta] (0,0) circle (0.4);
\draw[->] (0,0) -- (0, 1.2*\initialY);
\draw[->] (-1.3*\initialShift, 0) -- (1.3*\initialShift, 0);

\draw[thick, dotted] (0,0) -- (\initialShift,\initialY);
\draw[thick,dotted] (0,\initialY) -- (\initialShift,\initialY);
\draw[thick,dotted] (0,0) -- (-\initialShift,\initialY);
\draw[thick,dotted] (0,\initialY) -- (-\initialShift,\initialY);
\draw[fill=yellow] (\initialShift,\initialY) circle (0.2);
\draw[fill=yellow] (-\initialShift,\initialY) circle (0.2);
\node at (\initialShift+0.2,\initialY) [anchor=west] {$m$};
\node at (-\initialShift-0.2,\initialY) [anchor=east] {$m$};
\node at (0.7,-0.3) {$M$};

\pgfmathsetmacro{\kraftRatio}{0.3}
\pgfmathsetmacro{\einsMinusKraft}{1-\kraftRatio}
\pgfmathsetmacro{\einsMinushalbKraft}{1-0.5*\kraftRatio}

\draw[thick,->] (\initialShift,\initialY) -- (\initialShift*\einsMinusKraft,\initialY*\einsMinusKraft);
\draw[thick,->] (\initialShift,\initialY) -- (\einsMinusKraft*\initialShift,\initialY);
\draw[thick,dotted] (\initialShift*\einsMinusKraft,\initialY*\einsMinusKraft) -- (\einsMinusKraft*\initialShift,\initialY);
\node at (0,-0.4) [anchor=north] {\footnotesize 0};
\node at (1.4*\initialShift,0) {\footnotesize x};
\node at (0, 1.3*\initialY) {\footnotesize y};
\node at (0.5*\initialShift + 0.3, 0.5*\initialY) {\footnotesize r};
\node at (0.5*\initialShift,\initialY+0.3) {\footnotesize $\xi$};

\node at  (\initialShift*\einsMinushalbKraft + 0.3,\initialY*\einsMinushalbKraft-0.1) {\footnotesize $F_r$};
\node at  (\initialShift*\einsMinushalbKraft,\initialY+0.5) {\footnotesize $F_x$};
\end{tikzpicture}
\caption{Tidal forces: two test particles attracted to a mass $M$}
\label{TidalCalc}
\end{center}
\end{figure}
Take, for instance, the situation of two spheres as shown in figure \ref{TidalCalc}, of two test particles (in yellow) with masses $m$, which are attracted to a mass $M$ (magenta) at the spatial origin. The strength of the acceleration of each yellow sphere is given by Newton's formula,
\be
F_r = \frac{GMm}{r^2}
\ee
and directed toward the center of the mass $M$. Let $F_x$ be that part of the force which accelerates the right-hand yellow sphere towards its left counterpart, decreasing the distance between the spheres. That decrease in distance is what a free-falling observer could measure, deducing the presence of an (inhomogeneous) gravitational field. By elementary geometry, the ratio between $F_x$ and $F_r$ is the same as that of $\xi$ (the distance between the yellow sphere and the y axis) and r.\footnote{
The force triangle with hypotenuse $F_r$ and one leg $F_x$ is similar to the distance triangle whose hypotenuse is the line segment $r$, with one leg $\xi$: $F_x$ and $\xi$ are parallel, and so are $F_r$ and $r$, so the respective angles between them are the same. Both triangles have one right angle. Having two congruent angles is sufficient for the triangles to be similar.
}
Thus, we must have \be
F_x = \frac{\xi}{r}\cdot F_r = \frac{GMm}{r^3}\cdot \xi.
\ee
Two properties of this result are typical for tidal forces: they fall of faster than the ordinary gravitational force when it comes to the distance $r$ from the gravitational source, namely $1/r^3$ instead of $1/r^2$. And they are proportional to the separation $2\xi$ of the two test masses whose relative distance they change. 

Conversely, this means that tidal effects get smaller if we restrict our attention to smaller regions of space. In a small region, only small separations $2\xi$ are possible. Still, even a small acceleration will lead to considerable speeds, and observable effects, if we allow too much time to pass. We need to restrict observation time, as well. All in all, we need to restrict our attention to a small {\em spacetime} region.

\subsection{The equivalence principle, reformulated}

Even in a small, but finite spacetime region, there will in general be non-zero tidal effects. But in practice, our ability to detect small effects will be limited. All in all, here is a new version of the equivalence principle, which takes into account the limitations imposed by tidal forces:
\begin{quotation}
\noindent {\em Einstein \index{equivalence principle|textbf}{equivalence principle}:} Consider two observers whose measuring devices and instruments have a given limit of sensitivity. Then we can always find a maximum size $S$ (defining spatial extent as well as a maximum observation time), so that the following holds: Physics experiments performed by the first observer in free fall in a restricted spacetime region of size $S$ will have the same outcome as experiments performed by an observer in a restricted spacetime region of size $S$ who is infinitely far from all sources of gravity. In particular, the rules governing space and time are those of special relativity.
\end{quotation}
In the infinitesimal limit, where we make the experimental region infinitely small, tidal forces vanish altogether. In this limit, the effects of tidal forces are not even detectable with ideal measuring devices and instruments. This is less unrealistic than it sounds: Differential calculus teaches us about systematic ways of describing the infinitesimally small.

In this modified version, the equivalence principle is quite useful. It provides guidance when it comes to finding general-relativistic versions of existing laws of physics: If you know how these laws are defined in the context of special relativity, you know how these laws will be for a free-falling observer -- at least in an infinitesimal region.

This provides us with a powerful tool for deriving predictions of general relativity (or, for that matter, other theories as long as they incorporate the equivalence principle. In particular, the gravitational \index{redshift!gravitational}{redshift} of light in a gravitational field can be derived directly from the equivalence principle (\citenp{Schild1960,Schutz1985,Schroeter2002}).

\subsection{Tidal deformations and attraction}

So far, tidal forces have only been considered in their role as a limiting influence. Now, let us turn to what these forces actually do. Also, to be more precise, we should talk about tidal accelerations – after all, in a gravitational field, the acceleration is independent of an object's mass, and it makes sense to talk about gravitational acceleration, and the way it changes from location to location. 

One effect, we have already seen: Two test masses, transversally separated as they fall towards a point mass, will approach each other. There is another effect: the acceleration caused by the gravitational attraction of a point mass decreases with distance as $1/r^2$. Thus, the distance between two test masses that are separated in the direction of their fall will increase, since the lower test mass feels a greater acceleration than the upper mass.

\begin{figure}[htbp]
\begin{center}
\begin{tikzpicture}[scale=0.65]
\def\r{10}
\def\u{2.5}
\def\cr{0.4}
\def\ct{0.2}
\begin{scope}[xshift=-0.4\textwidth]
\fill[white] (0,1.63) circle (0.001);
\draw[thick] (0,0) circle (1.5);
\fill[magenta] (0,-\r) circle (\cr);
\draw (0,-\r) circle (\cr);
\draw[thick,loosely dotted] (0,-\r) -- (0,\u);
\draw[thick,loosely dotted] (0,-\r) -- ({1.5/\r*(\r+\u)},\u);
\draw[thick,loosely dotted] (0,-\r) -- ({-1.5/\r*(\r+\u)},\u);

\node at (0.1,-\r) [anchor=north west] {$M$};
\draw[thick] (0,1.5) circle (\ct);
\fill[yellow] (0,1.5) circle (\ct);
\node at (0,1.5) [anchor=south west] {\footnotesize $1$};
\draw[thick] (0,-1.5) circle (\ct);
\fill[yellow] (0,-1.5) circle (\ct);
\node at (0,-1.5) [anchor=north west] {\footnotesize $3$};
\draw[thick] (1.5,0) circle (\ct);
\fill[yellow] (1.5,0) circle (\ct);
\node at (1.6,0) [anchor=west] {\footnotesize $4$};
\draw[thick] (-1.5,0) circle (\ct);
\fill[yellow] (-1.5,0) circle (\ct);
\node at (-1.6,0) [anchor=east] {\footnotesize $2$};

\end{scope}
\begin{scope}[xshift=0.4\textwidth]

\draw[ultra thick, yellow!50!orange] (1.2,-2) -- (1.5,0);
\draw[ultra thick, yellow!50!orange] (-1.2,-2) -- (-1.5,0);
\draw[ultra thick, yellow!50!orange] (0,-0.3) -- (0,1.5);
\draw[ultra thick, yellow!50!orange] (0,-3.7) -- (0,-1.5);

\fill[magenta] (0,-\r) circle (\cr);
\draw (0,-\r) circle (\cr);
\node at (0.1,-\r) [anchor=north west] {$M$};
\draw[thick,loosely dotted] (0,-\r) -- (0,\u);
\draw[thick,loosely dotted] (0,-\r) -- ({1.5/\r*(\r+\u)},\u);
\draw[thick,loosely dotted] (0,-\r) -- ({-1.5/\r*(\r+\u)},\u);
\begin{scope}[yshift=-2cm]
\draw[thick] (0,0) ellipse (1.2 and 1.7);
\draw[thick] (0,1.7) circle (\ct);
\fill[yellow] (0,1.7) circle (\ct);
\node at (0,1.7) [anchor=south west] {\footnotesize $1$};
\draw[thick] (0,-1.7) circle (\ct);
\fill[yellow] (0,-1.7) circle (\ct);
\node at (0,-1.7) [anchor=north west] {\footnotesize $3$};
\draw[thick] (1.2,0) circle (\ct);
\fill[yellow] (1.2,0) circle (\ct);
\node at (1.3,0) [anchor=west] {\footnotesize $4$};
\draw[thick] (-1.2,0) circle (\ct);
\fill[yellow] (-1.2,0) circle (\ct);
\node at (-1.3,0) [anchor=east] {\footnotesize $2$};
\end{scope}
\end{scope}
\end{tikzpicture}
\caption{Falling particles that were originally at rest and arranged along a circle (left). After some time has passed, the particles have fallen and the circle has deformed (right). In the image on the right, the yellow traces show the distance each particle has fallen compared to the image on the left}
\label{TidalDeformation}
\end{center}
\end{figure}
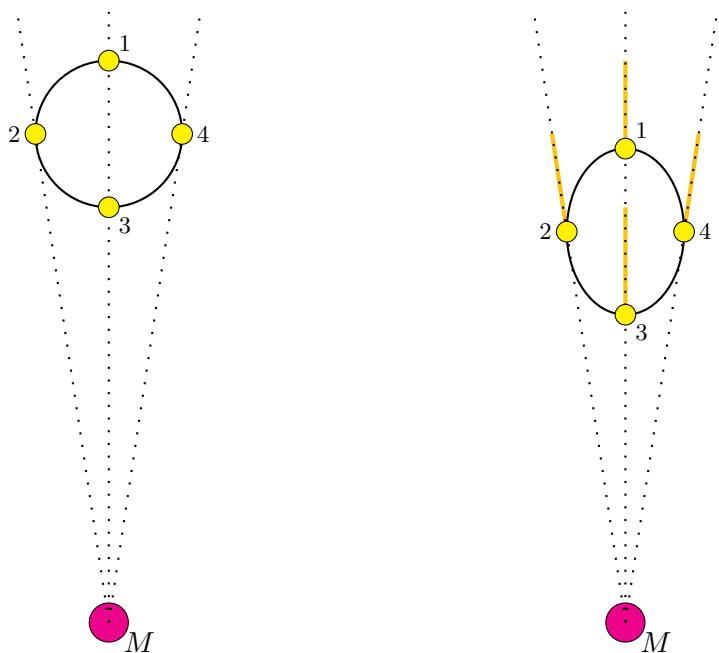

Figure \ref{TidalDeformation} shows the consequences. On the left, you can see four test particles (in yellow) arranged on a circle. Momentarily, these test particles are at rest. Below the test particles is an attracting mass $M$. 
The image on the right shows the situation a little while later. The test particles have fallen towards the attracting mass, but the overall shape of the particle swarm has changed: Test particles 2 and 4 have moved closer together, following as they do convergent paths towards the center of the attractive mass. Test particles 1 and 3 have moved apart, since 3 is closer to the attracting mass, and hence experiences greater gravitational acceleration than particle 1.

In Newtonian physics, one can show that at least within an infinitesimal time, this is the most general deformation that tidal accelerations cause: Start with a swarm of test particles arranged along the surface of a sphere, all of which are at rest relative to each other initially. In free fall, an external mass configuration that is outside the particle sphere will deform the sphere into an ellipsoid of the same volume. More concretely, the acceleration of the change in volume will initially be zero,\footnote{
Here and in the following, we denote time derivatives by a dot,
$$
\dot{f} \equiv \frac{\Dd f}{\Dd t}, \;\;\;\;\;\;\;\; \ddot{f} = \frac{\Dd^2 f}{\Dd t^2} \;\;\;\;\; \mbox{etc.}
$$
}
\be
\ddot{V}|_{t=0} = 0,
\ee
where $t=0$ is the time at which our particles were at rest relative to each other.

The only time when tidal forces will change the volume of a sphere of initially-at-rest test particles is if the attracting mass is {\em inside} the sphere. In this case, the tidal forces are due to the fact that the particles are getting pulled in different directions: each particle is getting pulled towards the center, where the attracting mass is. 

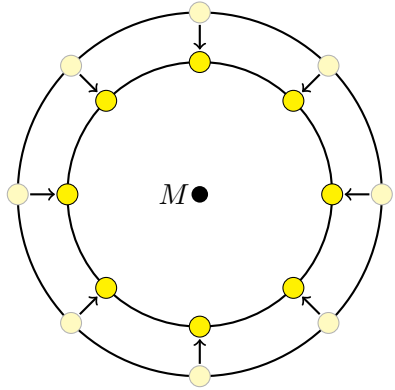
\begin{figure}[htbp]
\begin{center}
\begin{tikzpicture}[scale=1.1]

\draw[thick] (0,0) circle (2.2);
\draw[thick] (0,0) circle (1.6);
\foreach \x in {1,2,...,8}
  {
  \draw[thick,black!30] ({2.2*cos(45*\x)}, {2.2*sin(45*\x)}) circle (0.12);
  \fill[yellow!30!white] ({2.2*cos(45*\x)}, {2.2*sin(45*\x)}) circle (0.12);
  \draw[thick,->] ({2.05*cos(45*\x)}, {2.05*sin(45*\x)}) --({1.75*cos(45*\x)}, {1.75*sin(45*\x)});
  \draw[thick] ({1.6*cos(45*\x)}, {1.6*sin(45*\x)}) circle (0.12);
  \fill[yellow] ({1.6*cos(45*\x)}, {1.6*sin(45*\x)}) circle (0.12);
    }
\fill (0,0) circle (0.1);
\node [left] at (0,0) {$M$};
\end{tikzpicture}
\caption{Tidal forces acting on a spherical swarm of test particle due to a mass placed inside the sphere}
\label{AttractiveTidal}
\end{center}
\end{figure}

The result can be seen in figure \ref{AttractiveTidal}. For a point mass $M$ in the center of the sphere, the effect is readily calculated using the Newtonian gravitational force; we will do a very similar calculation in section \ref{DynamicsCalc}. The change in volume, expressed again as an acceleration (that is, a second derivative) is
\be
 \left.\frac{\ddot{V}}{V}\right|_{t=0} = -4\pi G\,\rho,
\ee
where $$\rho \equiv M/V$$
is the average density within the sphere.\footnote{This is an averaged integral version of sorts of the Poisson equation for the gravitational potential $\Phi$, namely $\triangle\Phi = 4\pi G\rho$. }

\subsection{Newtonian limit and Einstein's equation(s)}
\label{EinsteinsEquations}
In the preceding sections, we have performed a number of calculations using Newtonian gravity, and the concepts of motion and dynamics taken from classical mechanics. How much of this carries over to general relativity?

Quite a lot, as it turns out. Historically, when Einstein developed general relativity, all but one of the many astronomical and terrestrial observations involving gravity were explained very accurately using Newtonian mechanics and the Newtonian gravitational force. The one exception was the anomalous \Index{perihelion advance} of the planet Mercury, which Einstein took to be a fundamental effect, and used to shape his evolving theory of gravity.

But nonetheless, Newtonian gravity was highly successful. Any theory aiming to replace it had to be able to explain the successful predictions of Newtonian gravity. Under those conditions where the Newtonian predictions held --- all of which involved comparatively weak gravitational fields, and objects moving much more slowly than $c$, the speed of light in vacuum --- Einstein's theory needed to include Newtonian gravity as a limiting \index{Newtonian limit} case. 

Elementary derivations of this Newtonian limit are a standard topic in textbooks on general relativity. (``Post-Newtonian'' derivations that not only show the Newtonian limit, but add the various relativistic effects as a series development in $1/c^2$, \index{post-Newtonian (pN) gravity} are considerably more complicated; see e.g.\ \citenp{Poisson2014}.)

Concerning the Einstein equivalence principle, we have seen that in a local, free-falling reference frame, physics is special-relativistic --- at least up to tidal forces, which can be kept arbitrarily small by restricting the spacetime region under consideration, but are always present to some degree. But classical mechanics is itself a limit of special relativity, namely the limit where all particle motion is slow compared with the vacuum speed of light. This suggests that free-falling systems are likely to have another interesting limit, at least as long as this slow-motion condition is met: Whatever tidal forces remain can be described using Newtonian gravity.

This is indeed the case, and even better: even allowing for fast-moving particles, the full predictions of general relativity can be recovered with no more than a simple change in the source term for gravitational attraction! Before we talk about that simple change, for the record: The general description for how matter and gravity are linked are Einstein's equations, also called \Index{Einstein field equations (EFE)}, the centerpiece of his general theory of relativity; together with the definition of the terms involved (including their physical interpretation), they {\em are} general relativity. In their most general form, these equations can be written as 
\be
R_{\mu\nu} - \frac12 R\,g_{\mu\nu} = \frac{8\pi G}{c^4}\, T_{\mu\nu},
\label{EFE}
\ee
where the \Index{Ricci tensor} $R_{\mu\nu}$, \Index{Ricci scalar} $R$ and \Index{metric} tensor $g_{\mu\nu}$ represent spacetime geometry, while on the right-hand side, the \Index{energy-momentum tensor} $T_{\mu\nu}$ describes mass, energy, momentum and pressure of the matter contained within spacetime. Each of the indices $\mu$ and $\nu$ can take on values between $0$ and $3$, so equation (\ref{EFE}) is really a concise way of writing 16 equations at once --- or, in this case, 10 independent equations, since all terms involved remain the same if we switch $\mu$ and $\nu$. Together, these equations are the mathematical embodiment of Wheeler's summary of spacetime telling matter how to move, and matter telling spacetime how to curve.

As shown in a highly recommended article by \citenp{Baez2005}, from the full Einstein field equations linking spacetime curvature and the energy-momentum content of the spacetime, one can derive a much simpler form. In order to do so, one needs to go into free-fall and consider test particles arranged into a sphere. If these test particles are initially at rest relative to each other, one can show that the volume of the test particle sphere will change as 
\be
 \left.\frac{\ddot{V}}{V}\right|_{t=0} = -4\pi G\,(\rho + [p_x + p_y + p_z]/c^2),
 \label{EE}
\ee
\index{Einstein field equations (EFE)!simplified}
where the average density $\rho$ inside the test particle sphere includes contributions from all applicable kinds of energy (as per $E=mc^2$), and where we have added pressure terms $p_x, p_y, p_z$ for pressure into the x-, y-, and z-direction. Some additional information on the motivation of this in the context of special relativity is given in sections \ref{RelativisticMass} and \ref{MomentumPressure}.

In the isotropic case, no direction is special, and we have $p_x=p_y=p_z=p$, leading to the equation 
\be
 \left.\frac{\ddot{V}}{V}\right|_{t=0} = -4\pi G\,(\rho + 3p/c^2).
 \label{EEIso}
\ee
\index{pressure!gravitational effects}
As Bunn and Baez show, this form of Einstein's equation can be used to re-derive Newton's law of gravity for test particles surrounding a spherical mass.

If there is no mass, energy, pressure inside the test sphere, all that external gravitational sources can do is deform the sphere, keeping its volume unchanged as per
\be
 \left.\ddot{V}\right|_{t=0} = 0. 
 \label{EEvac}
\ee
That is the simplest form of the vacuum Einstein's equations. \index{Einstein field equations (EFE)!vacuum}

What should we make of the modified source terms in equations (\ref{EE}) and (\ref{EEIso})? Including energy as a source of gravity should come as no surprise, given $E=mc^2$. \index{mass-energy equivalence} 
But energy, even in special relativity, is not a scalar quantity. Instead, it is one aspect of relativistic four-momentum, which includes both energy and momentum. When we are looking at a gas, for instance, the particles involved will have energy as well as momentum. The latter becomes important whenever the particles bounce off the confining walls of a cylinder, creating pressure, or, in the absence of a confining wall, for keeping track of what part of the momentum is flowing out of a particular region, or into it.

Given that energy and momentum are part of the same relativistic quantity, and pressure directly related to momentum, we should not be too surprised that, in a relativistic description, all these quantities occur together as sources of gravity.

\subsection{What we need from general relativity}

Now we have all that we will need from general relativity! We know that:
\begin{itemize}
\item in a reference frame in free fall, the laws of physics are almost the same as in special relativity; the deviations from special relativity are due to tidal forces, grow smaller as we look at smaller and smaller spacetime regions, and vanish as such a region becomes infinitesimally small.
\item in such a reference frame in free fall, as long as all motions are much slower than the vacuum speed of light $c$, what happens is described by the Newtonian limit of general relativity, and can make use of Newtonian calculations. Having all motions much slower than $c$ also includes the condition $\rho\gg p/c^2$ for pressure terms, since pressure is the result of random particle motion.
\item even if we have fast-moving matter (notably matter with a large pressure component), the only correction to the usual Newtonian formula is that the source of gravity is not the mass density $\varrho_M$ alone. It's the mass density $\varrho_{M+E}$ including energy terms, and there is an extra pressure term,
\be
\varrho_M \;\;\; \to \;\;\; \varrho_{M+E} + 3p/c^2,
\ee
or the corresponding, slightly more complicated right-hand side of (\ref{EE}) for the non-isotropic case.
\end{itemize}
With this knowledge, let us consider cosmology.

\section{A homogeneous, isotropic, expanding universe}

\begin{figure}[htbp]
\begin{center}
\includegraphics[width=0.95\textwidth]{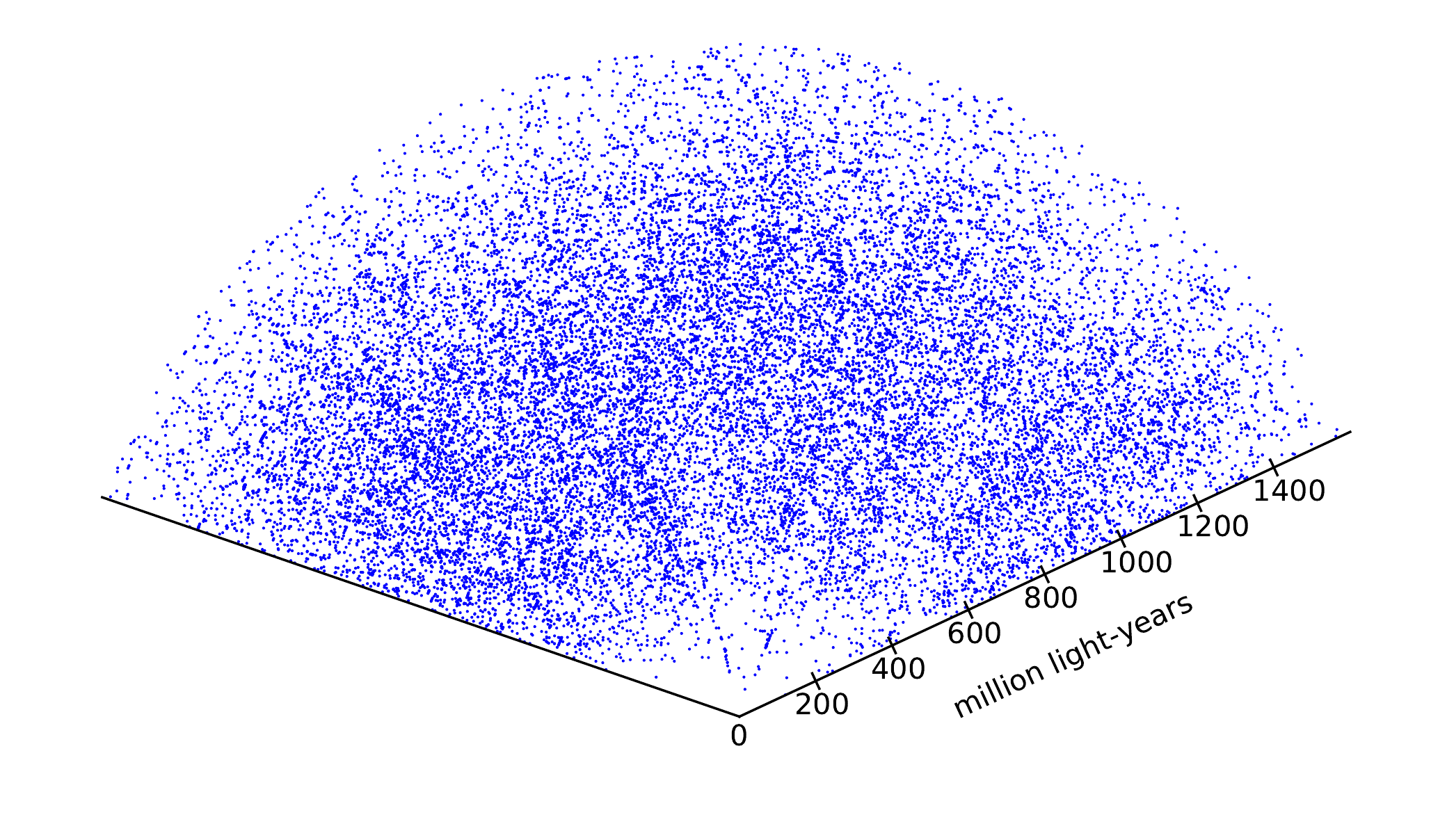}\\[-1.8em]
\includegraphics[width=0.95\textwidth]{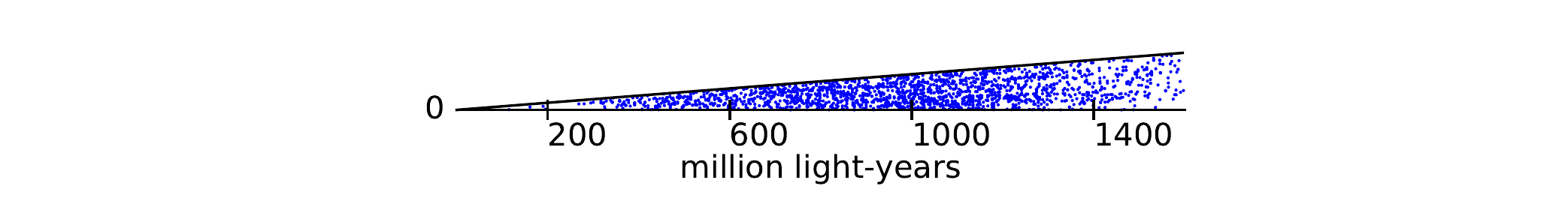}
\caption{Galaxy distribution in a wedge-like subregion of the universe: top view (top) and view from the side (bottom). Data from combining the NED distance catalogue and the \Index{2MASX catalogue}, 25231 data points in total. We as the observers are at the point $0$ at the apex of the wedge}
\label{2MASX}
\end{center}
\end{figure}

What is the simplest realistic large-scale model of the universe as a whole? As per our earlier considerations, we will be satisfied with our model representing only the very largest scales, treating galaxies as point particles with no relevant inner structure. So how are galaxies distributed, at large scales? Figure \ref{2MASX} shows two views of a wedge-like subregion of the universe, out to a distance of 1.4 billion light-years, giving a good overview of the \Index{large-scale structure}.\footnote{
The diagram shows galaxies in the 2MASX catalogue (``2Micron All-Sky Survey, Extended source catalogue,'' \citenp{JarrettEtAl2000}) (catalog VII/233 on Vizier) that have an entry in the NED-D tabulation of extragalactic, redshift-independent distances, Version 14.1.0 February 2017, \href{http://ned.ipac.caltech.edu/Library/Distances/}{http://ned.ipac.caltech.edu/Library/Distances/}, as compiled by Ian Steer, Barry F. Madore, and the NED Team.
}
Every blue dot is a galaxy; we as the observers are at the apex of the wedge. 

This is not perfectly, but fairly homogeneous, \index{homogeneity} at least on average. On average, the properties of this universe are the same, regardless of an observer's location --- or are they? At least from this wedge diagram, it's not so easy to be sure one way or the other. For instance, the somewhat emptier region near the apex is due to the fact that the wedge is particularly slim near its apex, thus contains fewer galaxies. The emptier regions at great distances, on the other hand, are a typical selection effect. At greater distances, it is more difficult to detect, and measure the distances to, less luminous galaxies; thus, only the brighter very distant galaxies will be included in our catalogue. In astronomy, this kind of selection effect due to brightness limitations is known as \Index{Malmquist bias}. Also, there are some radial strips where there are somewhat fewer galaxies, probably due to the fact that, in that particular direction, our view into the distance is obscured by dust in our own galaxy.

\begin{figure}[htbp]
\begin{center}
\includegraphics[width=0.75\textwidth]{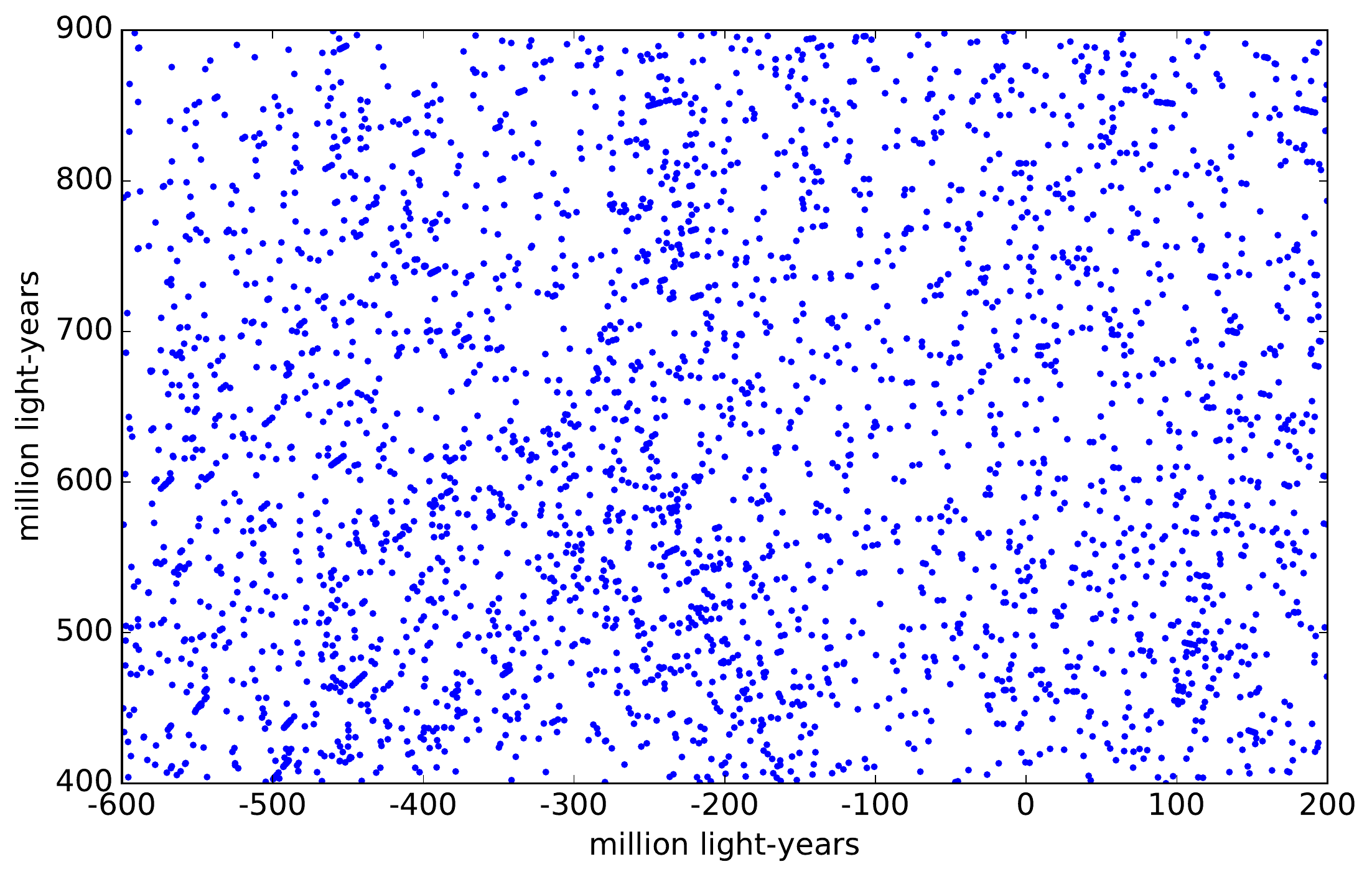}
\caption{Galaxy distribution in a box 800 times 500 times 30 million light-years in size; data and x-y coordinates are the same as in figure \ref{2MASX} }
\label{2MASXsquare}
\end{center}
\end{figure}

Figure \ref{2MASXsquare} shows galaxies from the same data set in a rectangular box of space, viewed from the top. 
There are some hints of structure, some slightly-denser-than-average areas, but overall, the galaxy distribution looks fairly homogeneous. There is no drastic clumping, say, with parts of the diagram devoid of galaxies altogether. Additional observations bear this out: Matter in our universe is distributed fairly uniformly -- at least on average, on sufficiently large scales. 

Truly three-dimensional statements in astronomy are always difficult, since distance measurements are anything but easy. But there is one direct consequence of a homogeneous universe that is more straightforward to test. If the universe is basically the same everywhere, then the universe should {\em look} the same, regardless where in the sky we point our telescopes. That is indeed the case: In whatever direction we look, we will, on average, see the same number of distant objects, with similar average properties. From our point of view, the universe is, on average, \index{isotropy} isotropic. (As an exercise, how are isotropy and homogeneity related? Convince yourself that a universe that is isotropic for at least two observers at different locations is necessarily isotropic, as well.)

Taking all these observations into account, we arrive at what Hermann Bondi, \index{Bondi, Hermann}, in or before 1952, called the {\em \Index{cosmological principle}}. In his own summary (``broadly speaking''; \citenp{Bondi1960}), ``[T]he universe presents the same aspect
from every point except for local irregularities.'' This extends the {\em \Index{Copernican principle}}, in that the Earth is nothing special --- not when it comes to our home planet's role in the solar system, nor when it comes to our position in the larger universe.

\begin{figure}[htbp]
\begin{center}
\includegraphics[width=0.4\textwidth]{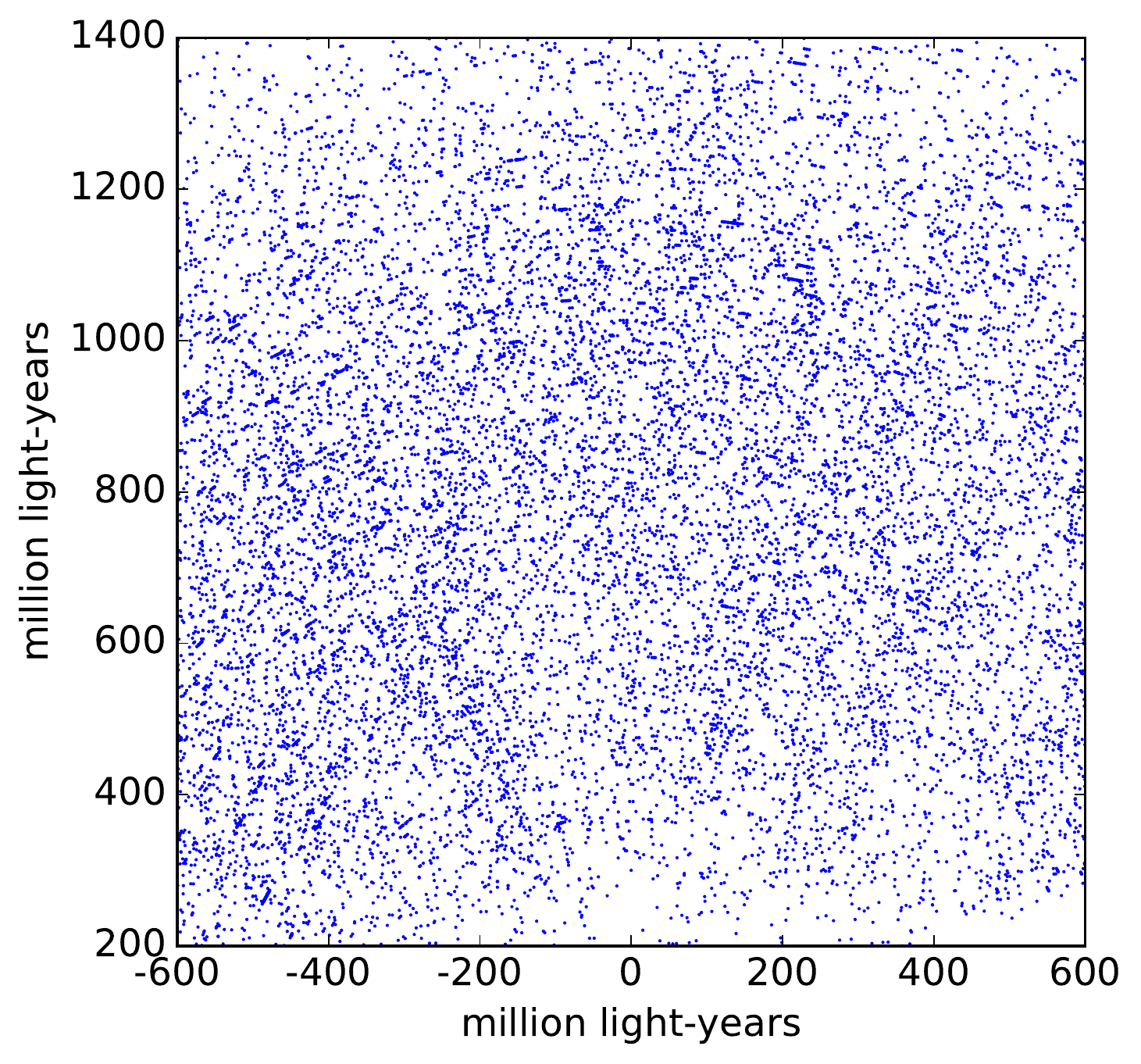}
\hspace*{4em}
\begin{tikzpicture}[scale=1.46]
\def\x{1.6}
\fill[blue!10!white] (-\x,-\x) -- (\x,-\x) -- (\x,\x) -- (-\x,\x) -- cycle;
\node at (0,0) {$\rho$};
\def\bb{1.99}
\draw [white] (-0.5,-\bb) -- (2,-\bb);
\end{tikzpicture}
\caption{Two complementary views of the large-scale universe: galaxies as point particles dotted throughout space (left) and an idealized, exactly homogeneous universe with a universal density $\rho$ (right)}
\label{CosmosModels}
\end{center}
\end{figure}
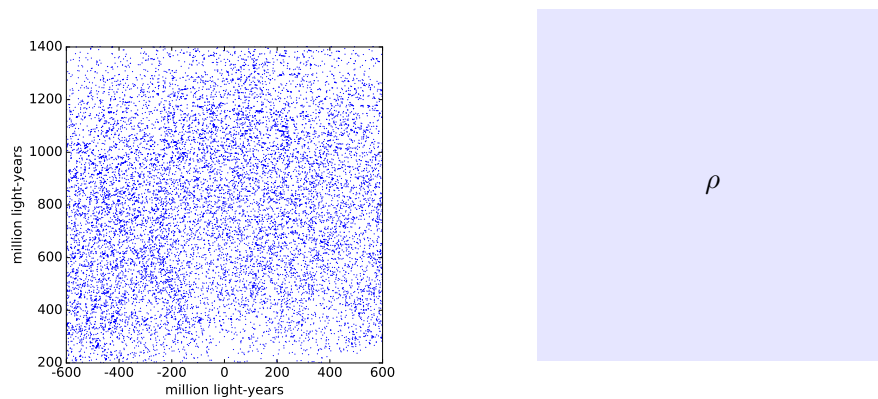

The simplest model for such a universe is a cosmos that is homogeneous not on average, but {\em exactly}, on all length scales. In the following, we will need to keep both pictures of the universe in mind, cf. figure \ref{CosmosModels}: \index{model universe!homogeneous and isotropic} the on-average homogeneous universe filled with point-particle galaxies, and the perfectly homogeneous universe. 
For some purposes, notably when it comes to the propagation of light, we will keep talking about separate galaxies. We will be asking, for instance, how long it takes light to travel from one galaxy to another, and how that light is redshifted. Let us call this the {\em \Index{galaxy dust}} picture of the universe. But when we talk about overall properties of the universe, such as the density values for matter or radiation, we will switch to the {\em continuum picture}. We will assume that the density is constant throughout the universe; after all, that is what (idealized) homogeneity means. 

If you take these pictures too literally, they contradict each other. But you shouldn't, really. Both are models, which are meant to map certain aspects of the large-scale universe. Physicists are allowed to use models --- simple, yet suitable models are what allows us to make powerful and predictive calculations in the first place! 

\subsection{What can change in a homogeneous universe?}

The homogeneity condition is a powerful constraint on how the universe can change over time. If we demand continued homogeneity, certain types of change are ruled out. After all, in that case, matter cannot move in any way that results in large-scale inhomogeneities. 

\begin{figure}[htbp]
\begin{center}
\includegraphics[width=0.43\textwidth]{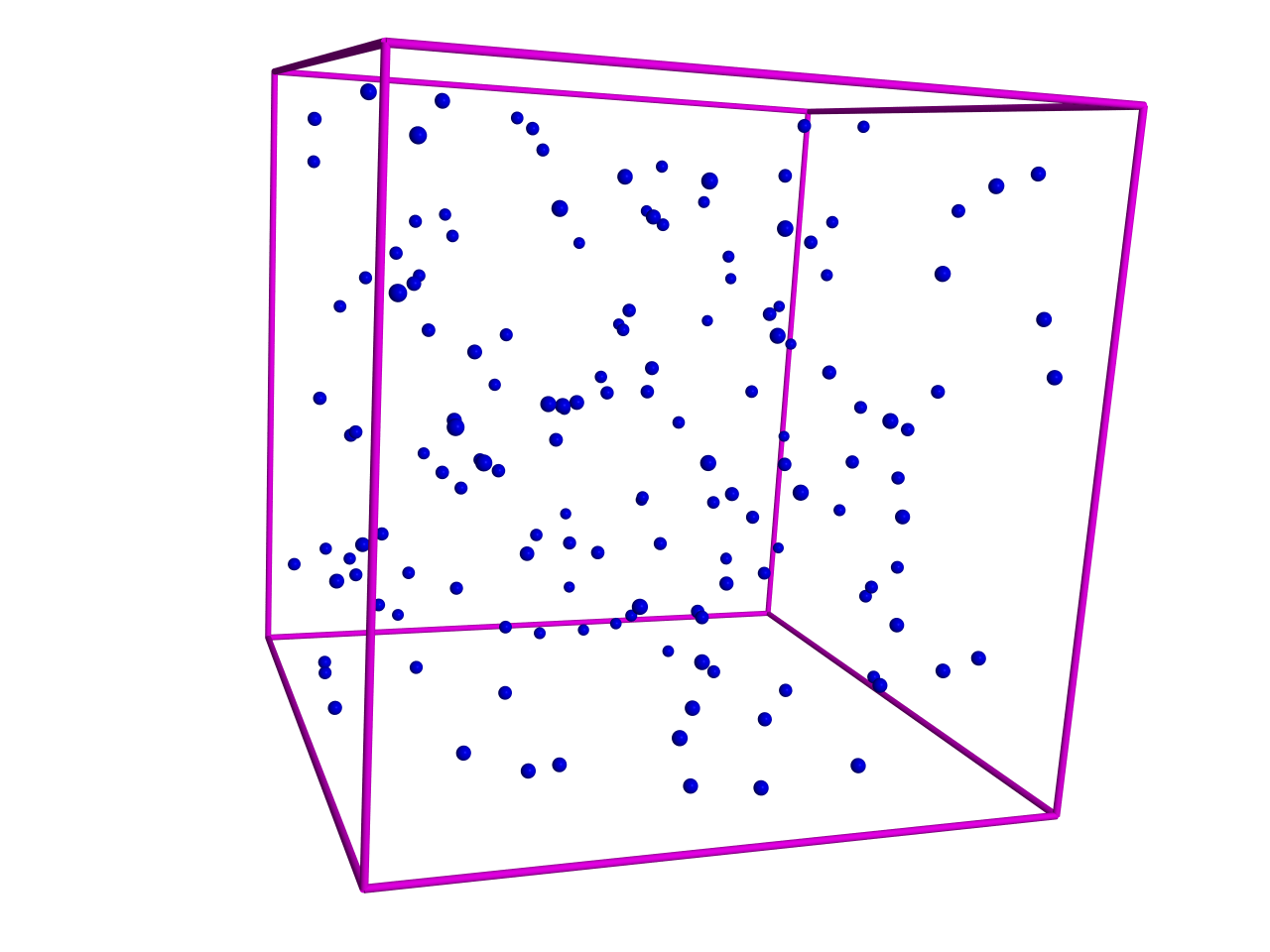}
\caption{In an unchanging (=static) universe, the pattern of galaxies does not change at all}
\label{Static}
\end{center}
\end{figure}

The simplest evolution for a homogeneous universe, guaranteed to keep homogeneity intact, would be no change at all. In the galaxy dust picture, that would correspond to an unchanging pattern, as shown in figure \ref{Static}.
We might be tempted to describe this situation by saying that ``all galaxies are staying where they are,'' but that is somewhat problematic. After all, points in space do not carry little markers that would allow us to state with certainty that a specific galaxy has remained at one particular point in space. A better way of describing a static universe involves consequences that are, in principle, observable, namely that {\em all pair-wise distances between galaxies remain constant}. This condition is sufficient to preserve the pattern that our point-like galaxies trace out in space.

Einstein, in a pioneering work that marked the first application of general relativity to the universe as a whole, and thus the birth of modern cosmology, presented a static model for the universe \index{Einstein's static universe} \index{static universe} (\citenp{Einstein1917}). On closer inspection, this static universe turned out to be unstable, though, prone to either collapse or expand, at any rate: it would depart from its static state in response to the smallest perturbations (\citenp{Eddington1930}).

What, then, is the next simplest kind of change that preserves homogeneity? It is when all pairwise distances between galaxies change in the same way, in proportion to one and the same time-dependent factor $a(t)$, commonly called the 
\begin{figure}[htbp]
\begin{center}
\includegraphics[width=0.43\textwidth]{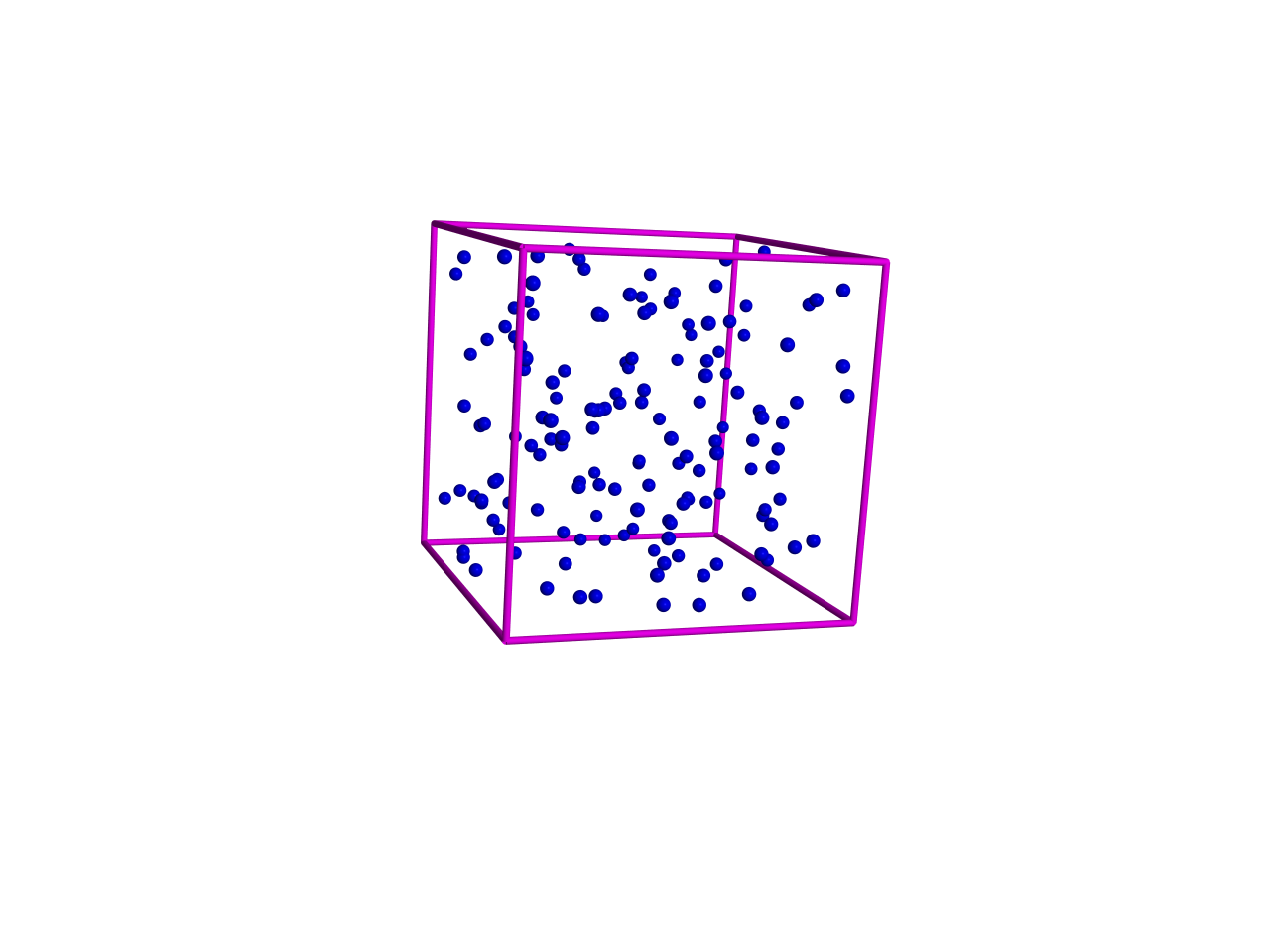}
\includegraphics[width=0.43\textwidth]{expanding-particles2.png}
\caption{Expansion with a universal scale factor: a portion of the universe, pictured at an earlier (left) and a later time (right)}
\label{ScaleFactorExpansion}
\end{center}
\end{figure}
{\em (universal) cosmic scale factor}, \index{universal scale factor} \index{scale factor} \index{cosmic scale factor} which depends on a suitably defined time parameter $t$ that is commonly called {\em cosmic time} (about which more below). That way, ratios between pair-wise galaxy distances do not change over time. The pattern of all these galaxies' locations in space does not change except for its overall scale. In an expanding universe as in figure \ref{ScaleFactorExpansion}, all distances become larger over time. But in terms of preserving homogeneity, a contracting universe where all distances shrink in proportion to the same scale factor works just as well. 

The systematic motions associated with scale-factor expansion, and its characteristic change of pair-wise distances, is called the {\em \Index{Hubble flow}}. Galaxies whose pair-wise distances change in exactly the way described by a changing overall scale factor are said to be ``in the Hubble flow'' or ``part of the Hubble flow.'' These systematic pair-wise distance changes are what is meant when cosmologists talk about \Index{cosmic expansion}.

The motions of real galaxies can deviate from the Hubble flow for different reasons. Many are part of galaxy clusters, orbiting each other; in this case, while the cluster's center of mass is in the Hubble flow, individual galaxies' motions will be slightly different. And even galaxies that are not in a cluster (so-called ``field galaxies'') will typically deviate from the Hubble flow at least a little bit. Collectively, these deviations from the Hubble flow are known as {\em \Index{peculiar motion}}. In the following sections, we will disregard peculiar motion, and assume that all galaxies are faithfully following the Hubble flow. 

A natural follow-up question is the following: In an expanding cosmos, what about bound systems? Do galaxies themselves grow larger, too? How about humans? Or atoms? This question is particularly natural whenever cosmic expansion is explained in terms like ``space itself is expanding'' --- language which I have avoided, precisely because I think that it leads to misconceptions, and does not evoke a faithful image of what is happening. I will address the question of bound systems, and what happens to them during cosmic expansion, later on in section \ref{BoundSystems}.

\subsection{Scale-factor expansion}
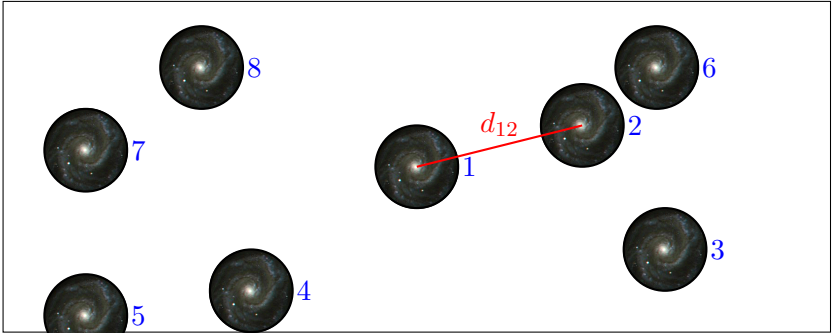
\begin{figure}[htbp]
\begin{center}
\begin{tikzpicture}[scale=1.1]
\def\scaleFactor{1}
\def\zeit{$t= t_0$}
\def\dist{$d_{12}$}
\draw (-5,-2) rectangle (5,2);

\begin{scope}
\clip (-5,-2) rectangle (5,2);

\pgfmathtruncatemacro{\galNumber}{1}

\foreach \x/\y in {0.0/0.0, 2.0/0.5, 3.0/-1.0, -2.0/-1.5, -4.0/-1.8, 2.9/1.2, -4.0/0.2, -2.6/1.2} {
	\coordinate (B) at ({\scaleFactor*\x},{\scaleFactor*\y});
	\node (gal) at (B) {\includegraphics[width=1.1cm]{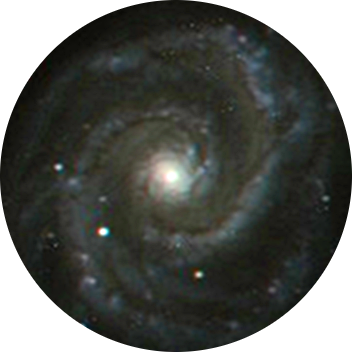}};
	\node[blue, right of = gal, xshift =-0.3cm] {\galNumber };
	\pgfmathsetmacro{\nextGalNumber}{\galNumber + 1.0}
	\pgfmathtruncatemacro{\galNumber}{\nextGalNumber}
	\global\let\galNumber=\galNumber
	\draw[thick]  (B) circle(0.5cm);
	}
\draw[red, thick] (0,0) -- ({2.0*\scaleFactor},{0.5*\scaleFactor});
\node[red, above] at ({0.5*2.0*\scaleFactor},{0.5*0.5*\scaleFactor}) {\dist};
\end{scope}
\end{tikzpicture}
\caption{Setup for tracing pairwise distances in an expanding universe}
\label{ScaleFactor1}
\end{center}
\end{figure}

\index{cosmic expansion|(}
Let us describe scale-factor expansion \index{scale-factor expansion!properties} in more detail. To that end, let us assign identifying numbers our galaxies, as in figure \ref{ScaleFactor1}, which shows a small region within an expanding universe.
Pairwise distances are readily specified by giving the indices of the two galaxies involved; notably, let $d_{ij}$ be the distance between the galaxies $i$ and $j$. As we shall see in the later sections, there are several different concepts of distance in cosmology; the distances $d$ we use here and in the following, and which grow in proportion to the scale factor, are known as \index{distance!proper} {\em \Index{proper distance}s}. 

\begin{figure}[htbp]
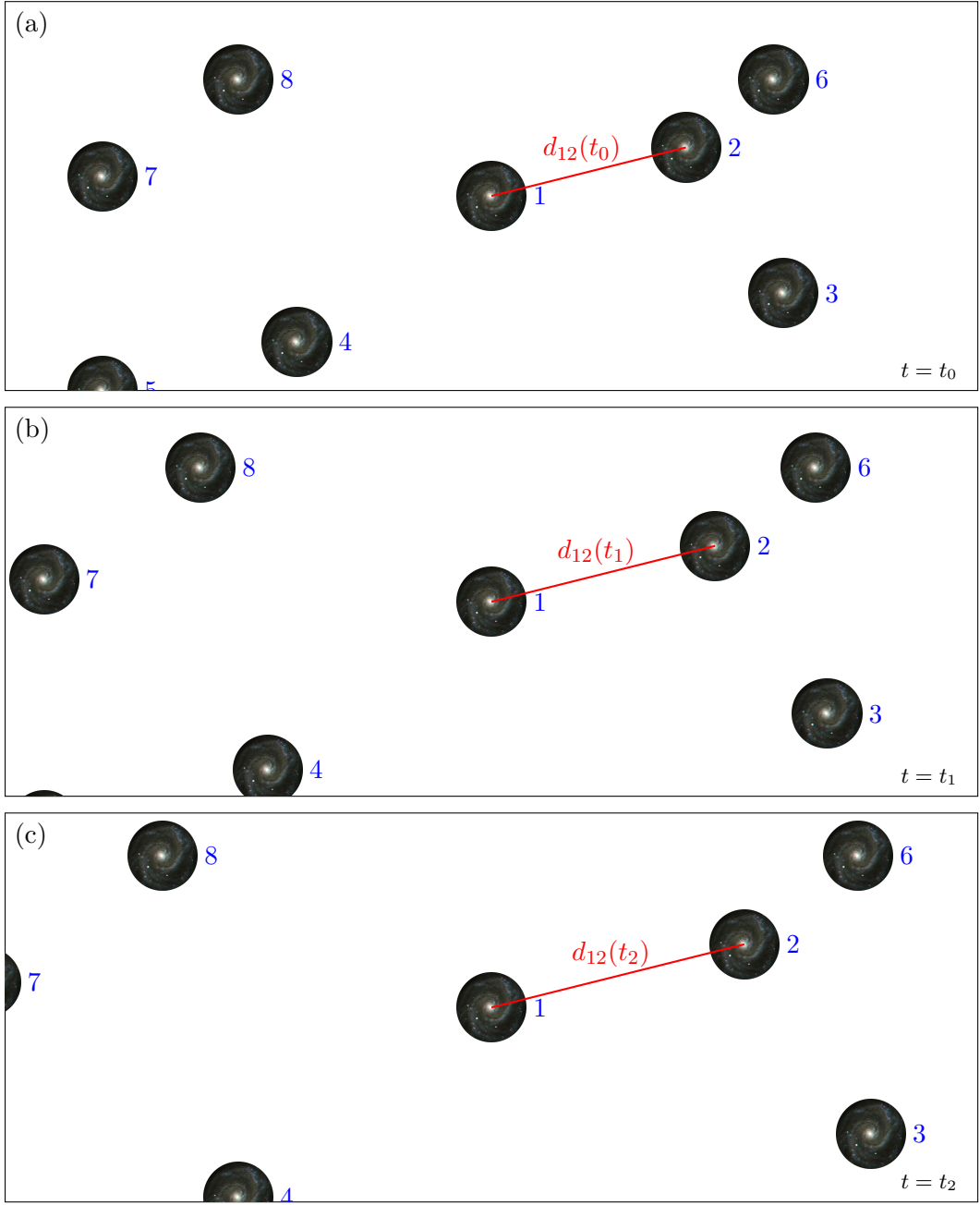

\def\tikzscale{1.4}
\begin{center}
\begin{tikzpicture}[scale=\tikzscale]
\def\scaleFactor{1}
\def\zeit{$t= t_0$}
\def\dist{$d_{12}(t_0)$}
\draw (-5,-2) rectangle (5,2);

\begin{scope}
\clip (-5,-2) rectangle (5,2);
\pgfmathtruncatemacro{\galNumber}{1}

\node at (-5,2) [anchor=north west] {(a)};
\foreach \x/\y in {0.0/0.0, 2.0/0.5, 3.0/-1.0, -2.0/-1.5, -4.0/-2.0, 2.9/1.2, -4.0/0.2, -2.6/1.2} {
	\coordinate (B) at ({\scaleFactor*\x},{\scaleFactor*\y});
	\node (gal) at (B) {\includegraphics[width=1cm]{galaxy-small.png}};
	\node[blue, right of = gal, xshift =-0.3cm] {\galNumber };
	\pgfmathsetmacro{\nextGalNumber}{\galNumber + 1.0}
	\pgfmathtruncatemacro{\galNumber}{\nextGalNumber}
	\global\let\galNumber=\galNumber
	}
\node at (4.5,-1.8) {\footnotesize\zeit};
\draw[red, thick] (0,0) -- ({2.0*\scaleFactor},{0.5*\scaleFactor});
\node[red, above,xshift=-0.1cm] at ({0.5*2.0*\scaleFactor},{0.5*0.5*\scaleFactor}) {\dist};
\end{scope}
\end{tikzpicture}

\vspace*{0.5em}

\begin{tikzpicture}[scale=\tikzscale]
\def\scaleFactor{1.15}
\def\zeit{$t= t_1$}
\def\dist{$d_{12}(t_1)$}
\draw (-5,-2) rectangle (5,2);

\begin{scope}
\clip (-5,-2) rectangle (5,2);
\pgfmathtruncatemacro{\galNumber}{1}

\node at (-5,2) [anchor=north west] {(b)};
\foreach \x/\y in {0.0/0.0, 2.0/0.5, 3.0/-1.0, -2.0/-1.5, -4.0/-2.0, 2.9/1.2, -4.0/0.2, -2.6/1.2} {
	\coordinate (B) at ({\scaleFactor*\x},{\scaleFactor*\y});
	\node (gal) at (B) {\includegraphics[width=1cm]{galaxy-small.png}};
	\node[blue, right of = gal, xshift =-0.3cm] {\galNumber };
	\pgfmathsetmacro{\nextGalNumber}{\galNumber + 1.0}
	\pgfmathtruncatemacro{\galNumber}{\nextGalNumber}
	\global\let\galNumber=\galNumber
	}
\node at (4.5,-1.8) {\footnotesize \zeit};
\draw[red, thick] (0,0) -- ({2.0*\scaleFactor},{0.5*\scaleFactor});
\node[red, above,xshift=-0.1cm] at ({0.5*2.0*\scaleFactor},{0.5*0.5*\scaleFactor}) {\dist};
\end{scope}
\end{tikzpicture}

\vspace*{0.5em}

\begin{tikzpicture}[scale=\tikzscale]
\def\scaleFactor{1.3}
\def\zeit{$t= t_2$}
\def\dist{$d_{12}(t_2)$}
\draw (-5,-2) rectangle (5,2);

\begin{scope}
\clip (-5,-2) rectangle (5,2);
\pgfmathtruncatemacro{\galNumber}{1}

\node at (-5,2) [anchor=north west] {(c)};
\foreach \x/\y in {0.0/0.0, 2.0/0.5, 3.0/-1.0, -2.0/-1.5, -4.0/-2.0, 2.9/1.2, -4.0/0.2, -2.6/1.2} {
	\coordinate (B) at ({\scaleFactor*\x},{\scaleFactor*\y});
	\node (gal) at (B) {\includegraphics[width=1cm]{galaxy-small.png}};
	\node[blue, right of = gal, xshift =-0.3cm] {\galNumber };
	\pgfmathsetmacro{\nextGalNumber}{\galNumber + 1.0}
	\pgfmathtruncatemacro{\galNumber}{\nextGalNumber}
	\global\let\galNumber=\galNumber
	}
\node at (4.5,-1.8) {\footnotesize \zeit};
\draw[red, thick] (0,0) -- ({2.0*\scaleFactor},{0.5*\scaleFactor});
\node[red, above,xshift=-0.1cm] at ({0.5*2.0*\scaleFactor},{0.5*0.5*\scaleFactor}) {\dist};
\end{scope}
\end{tikzpicture}
\caption{Snapshots of a small region of the universe undergoing scale-factor expansion}
\label{ScaleFactor2}
\end{center}
\end{figure}

In an expanding universe, all pairwise distances grow larger in proportion to the cosmic scale factor $a(t)$: as $a(t)$ changes, so do the distances. Distance ratios remain constant, as shown in figure \ref{ScaleFactor2}, where we have arbitrarily kept the position of galaxy 1 fixed.
We can give a concrete mathematical description by noting that, for scale factor expansion, the ratio of a particular such distance, evaluated at some time $t_0$, and the same pairwise distance, evaluated at some other time $t_1$, will be equal to the ratio of scale factor values at those two times,
\be
\frac{d_{ij}(t_0)}{d_{ij}(t_1)} \;\; = \;\; \frac{a(t_0)}{a(t_1)}.
\label{ScaleFactorExpansionEquation}
\ee
Conversely, this means that if we know all the pairwise distances at one specific time $t_0$, and know the scale factor $a(t)$, we can determine all pairwise distances at any other time $t$, namely as
\be
d_{ij}(t) = \frac{a(t)}{a(t_0)}\; d_{ij}(t_0).
\label{ChangingPairwise}
\ee
In astronomy, the $t_0$ chosen for reference is usually the present time.
\index{cosmic expansion|)}

Figure \ref{ScaleFactor2} shows some of the typical patterns of scale factor expansion. Notably, galaxy 2 does not appear to move away from galaxy 1 all that fast. Galaxy 7, in contrast, appears to be moving much faster. We will explore that systematic correlation in section \ref{HubbleRelationSection}. But before we do, it's time for a closer look at the parameter $t$.

\subsection{Cosmic time $t$ and proper distance $d$}
\label{CosmicTime}

So far, we have introduced the \Index{cosmic time} $t$ only via the cosmic scale factor $a(t)$, and we have implicitly assumed that $t$ is a useful time coordinate, to be used in statements like ``the position of galaxy $i$ at time $t_0$'' or about positions changing with time. Likewise, we have blithely talked about (proper) distances $d$ as if that were a well-defined concept.

If the only purpose of $t$ were to parametrize the universal scale factor $a$, then any other $t'=f(t)$, with $f$ a monotonically increasing function (i.e. a function with derivative $\Dd f/\Dd t \; >0$ everywhere) would serve as well. But $t$ can be defined to be much more than an arbitrary parameter.

In order to connect $t$ to the physical concept of time, let us zoom in on one of our galaxies. For each galaxy, we can define a notion of time by considering clocks that are moving along with the galaxy in question --- for each galaxy, we consider a clock that is at rest relative to that galaxy. Time as measured by an object's co-moving clock \index{co-moving clock!galaxy} is called that object's {\em \Index{proper time}}. In the vicinity of one particular galaxy, we can see how $a(t)$ changes, and we can determine how the proper time clock ticks. So why not combine the two, and use that galaxy's proper time to parametrize the scale factor $a(t)$?

If we do so for one galaxy, though, the same needs to hold for every other galaxy in the Hubble flow, as well. After all, in our simplified model, the universe is homogeneous. No location, no galaxy is special. If the parameter $t$ in $a(t)$ corresponds to proper time for one particular galaxy, it must correspond to proper time for any other galaxy in the Hubble flow. Otherwise, we could devise an experiment, namely comparing the evolution of $a(t)$ with the passage of proper time, that would have different outcomes depending on where, in which galaxy, we perform the experiment. In other words, the physical properties of our cosmos would vary with location, which would make the universe in question {\em in}homogeneous.

Note that, by defining cosmic time in this manner, we have also implicitly supplied another necessary part of defining a global time coordinate: a definition of \Index{simultaneity}. Switching to the continuum picture, we can track how the density of the universe $\rho(t)$ changes over time. In a homogeneous universe, by definition, $\rho(t)$ at some constant time $t$ will be the same at any location, wherever we measure the local density. You can turn this around, and use $\rho$ values to define which events in our universe are simultaneous.\footnote{In fact, the proper relativistic definition is that a homogeneous universe
\index{homogeneity!relativistic definition} is a universe in which it is {\em possible} to define  simultaneity in such a way that the local density $\rho$ will have the same value at all simultaneous events. Exercise: Convince yourself using reasoning from special relativity that in a homogeneous universe, an observer moving relative to the matter content of the cosmos will detect local inhomogeneities on a large scale. 
}

Note that, given our particular definition, the global time coordinate $t$ will have some unusual properties. Recall that in special relativity, we have the effect of {\em \Index{time dilation}}: when inertial observers in relative motion use their reference frames to describe each other's (proper time) clocks, each will deduce that the other's clocks are ticking more slowly than his own, giving rise to directly observable effects as in the case of the space-travelling twin (cf. section \ref{TimeDilation} for a lightning review of the pertinent concepts of special relativity). If one were to combine such clocks in relative motion, the result would be markedly different from any of the usual time coordinates defined by inertial observers. Statements that are true using the usual inertial time coordinates, notably those about the motion of light or material objects, do not necessarily hold for such an unusual combined time coordinate.\footnote{In fact, if you combine inertial time coordinates of systems in relative motion in this way, you will end up with a toy model that mimics most of the defining properties of modern cosmological models --- using mathematical tools no more complicated than the simplest form of the Lorentz transformations, and solving linear equations (\citenp{Poessel2017}).}

Similarly, in defining cosmic time to correspond to proper time for all galaxies in the Hubble flow, we have created such an unusual ``compound time coordinate''.  As long as we restrict our attention to a single galaxy and its direct neighbourhood, we can invoke the equivalence principle. Our cosmic time coordinate, which in that neighbourhood corresponds to that galaxy's proper time, can be used in calculations of speeds, accelerations and the like. But as soon as we zoom out and consider $t$ as a global coordinate, we need to be careful. 

Also, we should not forget that the choice of time coordinate affects the notion of distance. Whenever we talk about the distance of two galaxies at cosmic time $t$, we are implicitly defining a three-dimensional space, namely the subset of all points in spacetime that have this particular value of the cosmic time coordinate. Choose an unusual time coordinate, and that subset, in other words: space, will have unusual properties as well. At small scales, all should be well, in line with the equivalence principle. But as one goes to larger distances, these distances need to be interpreted carefully. They will not, in general, correspond to ordinary physical distances, and their derivatives with respect to cosmic time will not, in general, correspond to physical speeds. 
 
In conclusion, cosmic time $t$ is an unusual coordinate, and we must be careful not make unwarranted assumptions about how physical systems behave when described using a coordinate of this kind. 

\section{The Hubble relation}
\label{HubbleRelationSection}

Back to galaxies in the expanding universe. Figure \ref{ScaleFactor2}, with its snapshots of a universe that is undergoing scale-factor expansion, illustrates a systematic correlation: Galaxies that are further away from our spatial origin (arbitrarily chosen to be at the location of galaxy 1) move faster than their less distant kin.

The reason behind this is, of course, the unusual pattern of motions where distances change in proportion to one and the same {\em factor}. If I multiply a distance of 100 million light-years by a factor of 1.3 (corresponding to the scale-factor change between panels a and c of figure \ref{ScaleFactor2}), The 100 become 130 million light-years, and absolute difference of 30 million light-years. 

If I multiply 200 million light-years by the same factor, the absolute difference is twice as large, namely 60 million light-years. The larger the original distance, the larger the absolute difference --- and since these changes happen during the same period of time, we also have: the larger the original distance, the larger the {\em rate of change} of the distance.

We can make this more precise as follows: Consider any pair-wise distance $d_{ij}(t)$, which changes as specified in equation (\ref{ChangingPairwise}). Then we can calculate the rate of change, namely the derivative with respect to the time coordinate, 
as
\bea
\nonumber
v_{ij}(t) \equiv \dot{d}_{ij}(t) &=& \frac{\Dd}{\Dd t}\left(  \frac{a(t)}{a(t_0)}\; d_{ij}(t_0)  \right)=\frac{\dot{a}(t)}{a(t_0)}\; d_{ij}(t_0)\\[0.5em]
&=& \frac{\dot{a}(t)}{a(t)}\; \frac{a(t)}{a(t_0)}\;d_{ij}(t_0) =  \frac{\dot{a}(t)}{a(t)}\;d_{ij}(t).
\eea
The function
\be
H(t) \equiv  \frac{\dot{a}(t)}{a(t)}
\label{HubbleParameter}
\ee
is called the {\em \Index{Hubble parameter}}, and its value at the present time $t_0$,
\be
H_0\equiv H(t_0)= \left.\frac{\dot{a}(t)}{a(t)}\right|_{t=t_0}
\label{HubbleConstant}
\ee
is the {\em \Index{Hubble constant}}. The relation
\be
v_{ij}(t) = H(t)\cdot d_{ij}(t)
\label{HubbleRelation}
\ee
is called the {\em \Index{Hubble relation}}. Sometimes, instead of the Hubble constant $H_0$, astronomers will use the (dimensionless) {\em reduced Hubble constant} $h$ defined by
\be
H_0 = h\cdot 100\,\frac{\mbox{km/s}}{\mbox{Mpc}},
\label{ReducedHubbleConstant}
\ee
\index{Hubble constant!reduced}
where $1\;$ Mpc (``mega-parsec''), a distance measure commonly used in astronomy, corresponds to $3.26$ million light-years. Typical Hubble constant values are such that the reduced Hubble constant is around $h\approx 0.7$.

Naively, the Hubble relation is a relation between pair-wise relative velocities and pair-wise distances, valid at any specific time $t$. Whenever we can determine these distances and velocities, the expanding universe model predicts a clear relationship between the two -- a prediction to be tested against observational data. In particular, if we take one of the two galaxies to be our own, then equation (\ref{HubbleRelation}) is a relation between distant galaxies' distances from us and these galaxies' radial velocities; since, in an expanding universe, those velocities are away from us, they are commonly called \index{recession speed}{\em recession speeds}.

But taking into account the unusual properties of the cosmic time $t$ discussed in section \ref{CosmicTime}, we know to be be cautious. In particular, there is no reason to think that on large scales, the cosmic-time derivatives of the quantities $d_{ij}$ correspond to a relativistic generalisation of the concept of relative speed. Closer examination shows that they do not. There is, in fact, a sensible relativistic generalisation of the concept of relative speeds for the situation of two galaxies exchanging light signals in an expanding universe, and it gives a result that is very different from the cosmic-time derivative of the corresponding $d_{ij}$ (\citenp{Narlikar1994,BunnHogg2009}). In fact, the most obvious property unbecoming a relativistic relative speed, namely the fact that by the Hubble relation is that we can have $v_{ij}>c$ for sufficiently large distances $d_{ij}$, which taken at face value would mean superluminal motion. \index{superluminal motion!galaxies} This is a direct indication that $v_{ij}$ is no generalised relativistic relative speed.

In the direct cosmic neighbourhood of each free-falling galaxy in the Hubble flow, on the other hand, cosmic time $t$ is a good approximation for the usual time coordinate of special relativity and classical mechanics. There, the interpretation of the Hubble relation as linking our usual notion of distances and relative speeds is valid. This is true in the cosmic neighbourhood of our own galaxy, and in this approximation, we can link the Hubble relation to an actual observable: the redshift of light reaching us from another galaxy.

\subsection{Free-falling galaxies and the Doppler effect}
\label{HubbleDoppler}

All the galaxies in our simplified model are in free fall, so the crucial condition for applying the equivalence principle is fulfilled: at least in the direct vicinity of each galaxy, the laws of special relativity should hold -- limited by any tidal forces that might be present. We know from the Hubble relation (\ref{HubbleRelation}) that in the close vicinity of that galaxy, all distance changes happen rather slowly. In fact, by focusing on a sufficiently small region of space we can ensure that all the $d_{ij}$, and consequently all $v_{ij}$, are below some given limit. Given the observed value of the Hubble constant, of around 20 km/s per million light-years, even galaxies as far away as 140 million light-years will not reach recession speed values of more than about 1 percent of the speed of light in vacuum.

In this limit we can talk about motion in the usual way of classical physics --- we can talk about galaxies moving, and about light travelling from one galaxy to another along straight lines at the speed $c$, about $d_{ij}$ being an ordinary distance to be covered, and about $t$ behaving like an ordinary time coordinate. In this limit, we will also find that the light from distant galaxies is subject to the (classical) Doppler shift.

\begin{figure}[htbp]
\begin{center}
\includegraphics[width=\textwidth]{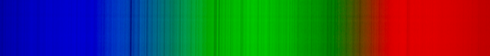}
\caption{Solar spectrum (measured in reflection by observing the Moon) showing dark absorption lines \index{solar spectrum} \index{spectrum!solar}}
\label{SpectralLines}
\end{center}
\end{figure}

Consider a simple (monochromatic) light wave, with its wave crests and troughs, propagating from one galaxy to another:
\begin{center}
\begin{tikzpicture}[scale=1.2]
\draw[yellow!80!red, ultra thick, domain=0:3,samples=200] plot (\x, {0.2*sin(360*\x)});
\draw [->,very thick] (3.1,0) -- (4.1,0);
\draw [gray,decorate,very thick, decoration={brace,mirror,amplitude=5pt},
   xshift=0pt,yshift=0pt] (0.75,-0.35)  -- (1.75,-0.35) 
   node [black,midway,below=4pt,xshift=-2pt] { wavelength $\lambda$};
\end{tikzpicture}
\end{center}
Since the galaxies are in relative motion, light emitted in one of the galaxies, and arriving at the other, will be subject to the (ordinary, non-relativistic) Doppler effect (the usual derivation for which is given in the first part of section \ref{SRDoppler}). Let us call $\lambda_e$ the wavelength of the light as measured at the time of emission $t_e$, by an observer moving along with the emitting galaxy, and $\lambda$ the wavelength of the same light, measured by us as the light reaches our own galaxy. In terms of these quantities, the redshift $z$ is defined as \index{z (wavelength shift)}
\be
z= \frac{\lambda - \lambda_e}{\lambda_e}.
\ee
The classical \Index{Doppler effect} links $z$ with the emitting object's \Index{radial speed} $v$ (that is, the component of its speed directly away from us or, for negative values, directly towards us) as
\be
z=\frac{v}{c}.
\ee
The wavelength shift $z$ can be measured very accurately. The light of stars and galaxies contains a wealth of \Index{spectral lines}: narrow wavelength regions in which the energy distribution of the light has a sharp maximum (\Index{emission lines}) or minimum (\Index{absorption lines}). Figure \ref{SpectralLines} shows an example of absorption lines.
In our particular situation, we substitute the speed at which the emitting galaxy is moving away from us in the Hubble flow.
This gives the {\em \Index{redshift-distance relation}}
\be
cz = H_0\cdot d,
\label{RedshiftDistanceApproximate}
\ee
linking distances $d$ and redshifts $z$. By introducing the {\em \Index{Hubble distance}} \index{distance!Hubble}
\be
D_H\equiv\frac{c}{H_0}
\label{HubbleDistance}
\ee
as a natural length scale for an expanding universe with Hubble constant $H_0$, this can also be written as
\be
d = D_H\cdot z.
\ee

Note that we are evaluating all quantities, and in particular the Hubble constant, at the present time. We assume that light travel times are too short to matter here, and that the Hubble constant does not change between the emission time $t_e$ and the time of reception of the light signal in question -- another consequence of including only comparatively nearby galaxies. The result -- a systematic redshift increasing with distance -- is our first acquaintance with the \Index{cosmological redshift}, \index{redshift!cosmological} if only in the limit of comparatively close galaxies. 

Some authors do not distinguish between the Hubble relation (\ref{HubbleRelation}) and the redshift-distance relation. It makes sense to clearly differentiate between the two, though, since the redshift-distance relation (\ref{RedshiftDistanceApproximate}) is an approximation that only holds for small distances, whereas the Hubble relation (\ref{HubbleRelation}) is an exact relation that holds on all length scales in an expanding universe.

\subsection{Measuring astronomical distances}
\label{MeasuringDistances}

Determining distances is one of the more difficult problems in astronomy. Our best current solution is the cosmic {\em \Index{distance ladder}}: a combination of various methods used by astronomers to successively determine distances within our cosmic neighbourhood and beyond (\citenp{deGrijs2011}). The term derives from the fact that the different methods for determining astronomical distances build one upon the other, representing the consecutive ``ladder steps''.

The first steps involve measurements to establish the distance scale within the solar system, notably the average Earth-Sun distance, which is known as the \Index{astronomical unit}. The traditional method for determining this basic scale made use of  \index{parallax} parallax measurements (corresponding to the way astronomical observations change as the observer changes location), notably during Venus transits in front of the Sun. Modern measurements are based on the light-travel time of radar signals sent to the nearest planets. \index{radar distance measurement} The next step involves measurements of stellar parallax, \index{stellar parallax} that is, the change in the stars' apparent positions as the Earth orbits the Sun. At the time of this writing, ESA's \Index{Gaia mission} is measuring highly accurate distances to more than a billion stars in this way. 

With accurate parallax measurements, one can hope to eliminate what used to be some extra steps of the ladder; in any case, extragalactic distances typically involve what are known as {\em \Index{standard candles}}:  objects whose total light output per unit time, that is, each object's {\em \Index{luminosity}}, is known, either because all objects of a certain type have the same luminosity, or because the luminosity can be derived from certain observable properties of the object.

Of great importance, both historically and for the modern distance scale, are \Index{Cepheid variables}. The luminosity of these comparatively rare, massive and bright stars changes periodically over time, with periods between days and months. The change is caused by an overall pulsation of the star: as the star gets bigger, it gets brighter. Size and time-scale are related, and so are the pulsation period and the star's brightness (both average and maximal/minimal). 

\begin{figure}[htbp]
\begin{center}
\includegraphics[width=0.9\textwidth]{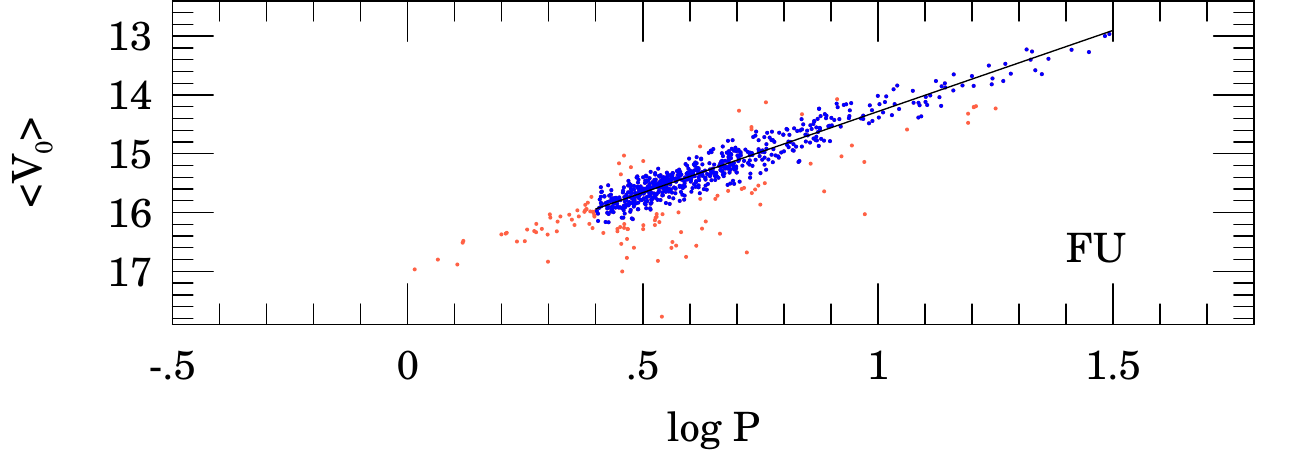}
\caption{Period-Luminosity relation (logarithm of period $P$ in days vs. average visual magnitude) for Cepheids in the simplest pulsation mode (fundamental mode) in the Large Magellanic Cloud. Lower part of figure 3 in \citenp{Udalski1999} }
\label{PLFigure}
\end{center}
\end{figure}

The relation between the two, the Cepheid's {\em \Index{period-luminosity relation}}, was first found empirically by the US astronomer Henrietta Swan Leavitt, who
\index{Leavitt, Henrietta Swan}
noticed in or around 1907 that those Cepheids with the longest period appeared to be the ones with the brightest peak brightness \citep{Leavitt1908}. She developed this observation into a period-luminosity relationship that allows astronomers to deduce distant Cepheid's luminosities from the period of their (regular) variations. A modern version of this relation, based on observations of Cepheids in the same galaxy already studies by Leavitt, namely the Large Magellanic Cloud, is shown in figure \ref{PLFigure}.

\begin{figure}[htbp]
\begin{center}
\includegraphics[width=0.55\textwidth]{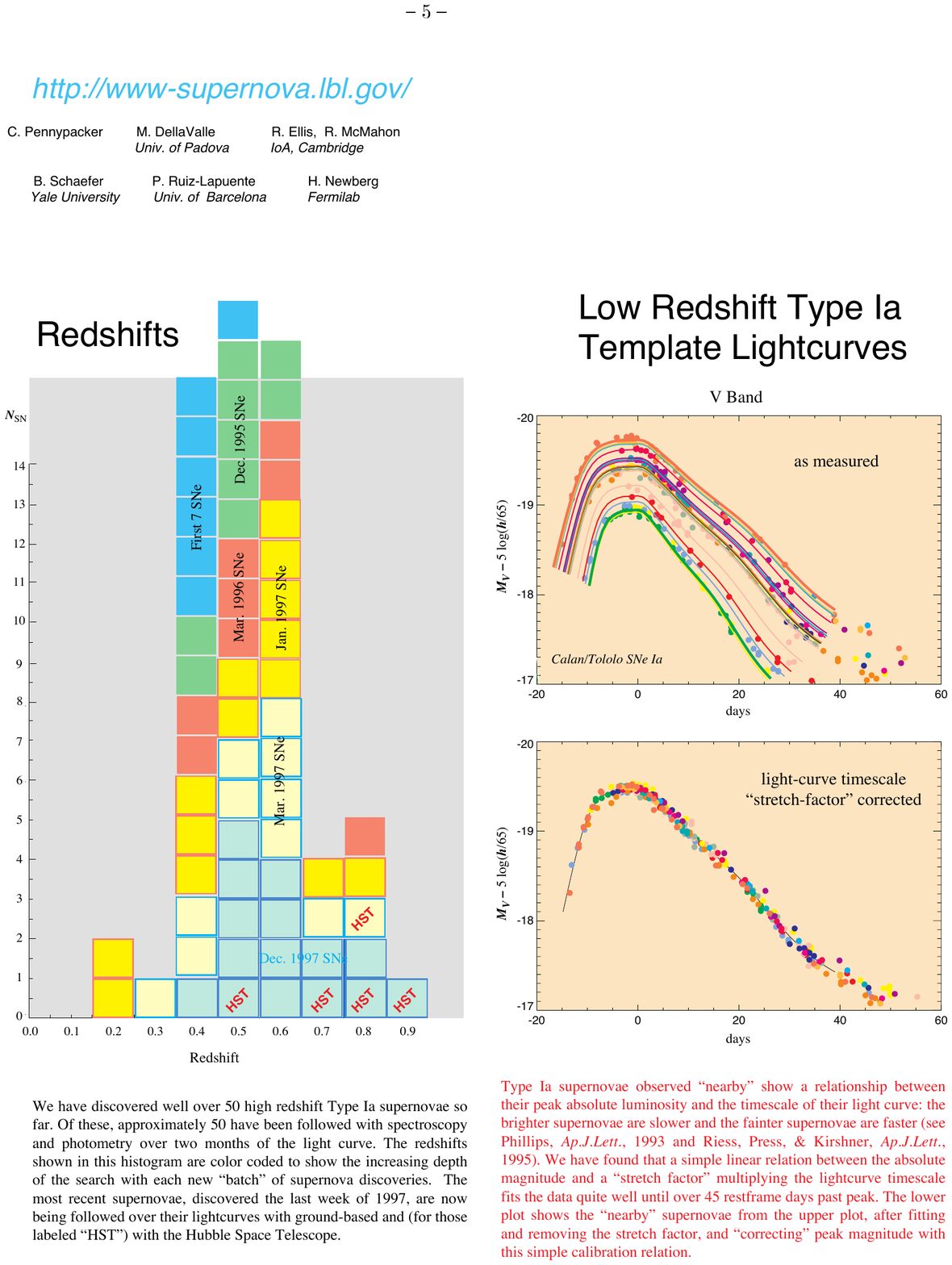}
\caption{Peak luminosities of uncalibrated supernovae type Ia light curves (top) are already fairly similar. Rescaling using the time scale of diminishing brightness gives calibrated light curves (bottom) that are highly accurate standard candles. Vertical axis shows visual magnitudes, shifted by a constant that depends on the reduced Hubble constant. Horizontal axis is days from peak brightness. Figure from \citenp{Perlmutter1997}
}
\label{SNLightCurve}
\end{center}
\end{figure}

Key standard candles for the largest extragalactic distances are \Index{supernovae of type Ia}, thought to be White Dwarf stars drawing hydrogen from a binary companion onto themselves and exploding once they have reached a certain critical mass. Peak luminosities of such explosions are fairly similar already (see top part of figure \ref{SNLightCurve}).
In addition, there is a relation between the peak luminosity and the overall width of the light curve (representing how fast brightness diminishes after the peak). This relation can be used to determine the peak brightness more accurately, resulting in calibrated supernova Ia light curves that are fairly accurate standard candles (bottom part of figure \ref{SNLightCurve}), which play a key role for modern cosmological observations.

Standard candles are useful for determining distances since the apparent brightness of an object of given luminosity is a direct measure of that object's distance from us. This is a matter of everyday experience -- we perceive a flashlight that is directly in front of us as much brighter than the same flashlight a hundred meters away.

The relation between luminosity, distance, and flux density can be more precise in the following way. We had already defined the luminosity, the amount of energy emitted by an object per unit time, as a physical measure of an object's intrinsic brightness. For the apparent brightness, we need to measure the amount of radiation we receive from a distant object.

But that amount will depend on our collecting area -- within any given period of time, larger telescopes collect more light than smaller telescopes. The physical measure of the {\em \Index{apparent brightness}} of an observed object is thus what is known in technical terms as the {\em \Index{irradiance}} or {\em (radiation) \Index{flux density}}: the amount of radiation energy received per unit time, per unit area. 

\begin{figure}[htbp]
\begin{center}
\includegraphics[width=0.4\textwidth]{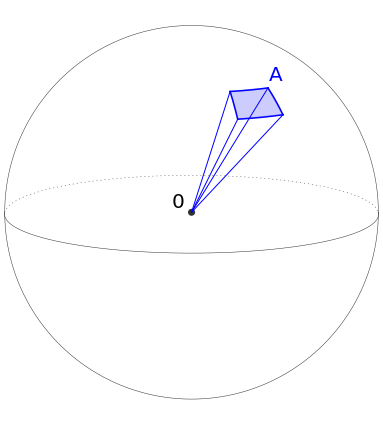}
\caption{Sphere centered around an object at 0, with the small collection area A at distance $d$ from the object marked in light blue}
\label{InverseSquare}
\end{center}
\end{figure}

In ordinary, flat space, the relation between flux density and luminosity is straightforward. Consider the light of an object that radiates isotropically in all directions, as many astronomical objects do. At a distance $d$ from that object, its light has spread out evenly over a spherical surface with the total surface area $4\pi d^2$. Assume that we collect light using a telescope with collection area $A$ pointed directly at the source, as shown in figure \ref{InverseSquare}.
Evidently, the fraction of light our telescope catches will be
\be
\frac{A}{4\pi d^2},
\ee
namely the ratio of our collecting area and the total area over which the light has spread out. Thus, luminosity $L$ (intrinsic brightness) and flux density $I$ (as a measure of apparent brightness) are related by the {\em \Index{inverse square law}}
\be
I = \frac{L}{4\pi d^2}.
\label{InverseSquareLaw}
\ee
This relation is the basis of standard candle distance measurements: Determine the luminosity $L$ from the properties of the standard candle, measure the flux density $I$ directly using a telescope, and you can solve equation (\ref{InverseSquareLaw}) for the distance. 

At least, this is true for our cosmic neighbourhood, up to distances of a few hundred million light-years. For very distant objects, there is a modified inverse square law, which we will derive below, in section \ref{ModifiedInverseSquareLaw}.

\subsection{Measuring the Hubble constant}
\label{MeasuringHubble}
\begin{figure}[htbp]
\begin{center}
\includegraphics[width=0.8\textwidth]{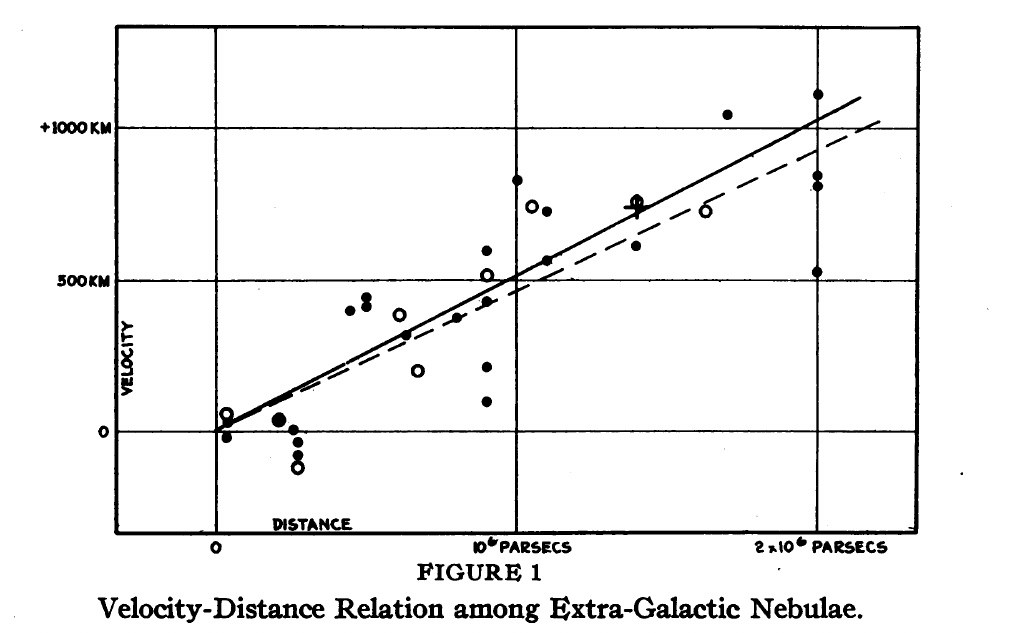}
\caption{The first Hubble diagram, figure 1 in \citenp{Hubble1929}. The y axis text is misleading; the units are not km, as written, but km/s}
\label{FirstHubblePlot}
\end{center}
\end{figure}

With these preparations, we are now in a good position to look at actual measurements of the Hubble constant, taken using comparatively close galaxies, so the approximations we made for this section will be valid. The first actual \Index{Hubble diagram} plotting distances against redshifts is due to Edwin Hubble \citep{Hubble1929}, and reproduced in figure \ref{FirstHubblePlot}.\footnote{An assessment of (partial) precursors can be found in \citenp{Trimble2013}.}

While there is considerable scattering, there is a clear linear trend --- which is, as we have seen, what one would expect in a universe that is expanding with a universal scale factor $a(t)$, where galaxies obey a \Index{Hubble relation}. This plot and others like it eventually convinced most astronomers that we live in an expanding universe, although Hubble himself remained strangely ambivalent about the matter.

From a modern perspective, the plot in figure \ref{FirstHubblePlot} has significant systematic errors. Perhaps the most fundamental one is that the stars originally lumped as Cepheids really belong to two different types, and have two different period-luminosity relations, as first pointed out by Walter Baade \index{Baade, Walter} at an IAU meeting in 1952 \citep{Hoyle1954,Baade1956}. \index{Hubble constant!measurements}

I will not give an account of the complete history of Hubble constant measurements. Instead, I fast-forward to a milestone: the \Index{Hubble Key Project}, which took advantage of the Hubble Space Telescope to calibrate the Cepheids \Index{period-luminosity relation} and, on that basis, other standard candle methods that reach out to much greater distances, including supernovae of type Ia and standard candles for the brightness of whole galaxies. The project's aim was to measure the \Index{Hubble constant} with an accuracy of 10\%. 

\begin{figure}[htbp]
\begin{center}
\includegraphics[width=0.95\textwidth]{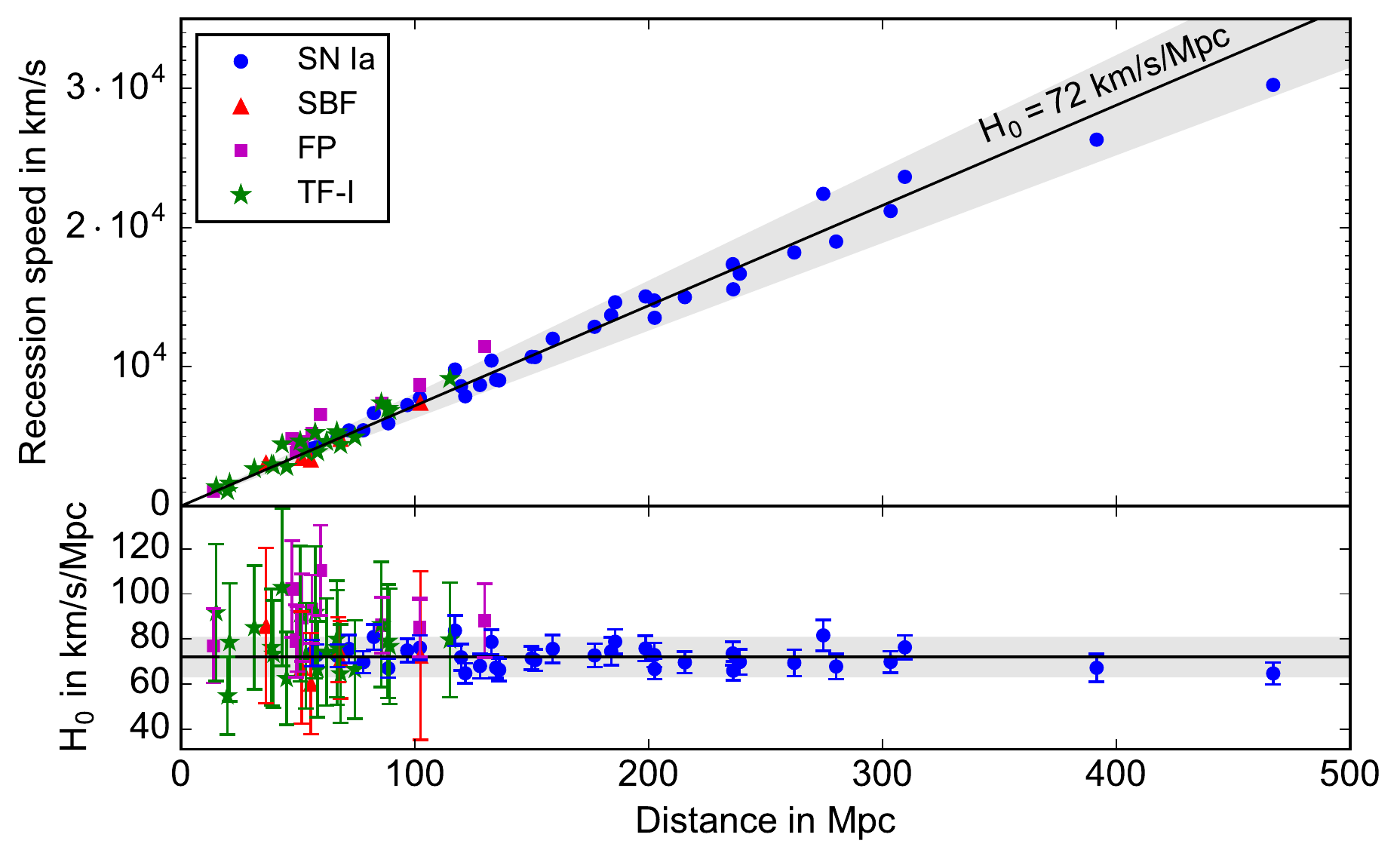}
\caption{Hubble diagram (top) and Hubble constant value (bottom) for distant standard candles of the Hubble Key Project, based on the data and results of \citenp{Freedman2001}. Standard candle types explained in table \ref{SCTypes}}
\label{HKPDistant}
\end{center}
\end{figure}

The Hubble Key Project results for the more distant \Index{standard candles} are shown in figure \ref{HKPDistant}, with a brief descriptions of the types of standard candles involved in table \ref{SCTypes}.
\begin{table}
\caption{Standard candle types used for the Hubble Key Project distance measurements in figure \ref{HKPDistant}. The table numbers in the rightmost column refer to tables in \citenp{Freedman2001}}
\footnotesize\vspace*{1em}
\bgroup
\renewcommand{\arraystretch}{1.2}
\begin{tabular}{|l|p{0.25\textwidth}|p{0.5\textwidth}|c|}
\hline
 & Type & Description & Table\\\hline\hline
Sn Ia & Supernovae Type Ia\index{supernovae of type Ia} & deducing absolute brightness from evolution of light-curve & 6\\\hline
SBF & Surface Brightness Fluctuations \index{surface brightness fluctuations} & link between distance and the fine-grainedness of a galaxy image &10 \\\hline
FP & Fundamental Plane \index{fundamental plane} & deducing \Index{elliptical galaxy} brightness from velocity dispersion and brightness profile & 9\\\hline
TF-I & I-Band Tully-Fisher \index{Tully-fisher relation} & deducing \Index{spiral galaxy} brightness from rotation speed & 8\\\hline
\end{tabular}
\egroup
\label{SCTypes}
\end{table}
Combining their measurements for the different standard candle methods they applied, the researchers obtained a value for the Hubble constant of
$H_0 = (72\pm 8)\,$km/s~Mpc${}^{-1}$ \citep{Freedman2001}.

The best current direct measurements of the Hubble constant have an accuracy of better than 3\%. However, there are now some discrepancies with determinations of the Hubble based on observations of the very early universe. The best value based on measurements by the Planck satellite gives a somewhat smaller Hubble constant of $H_0= (66.9 \pm 0.6)\,$km/s~Mpc${}^{-1}$, while the latest value based on supernovae of type Ia is $H_0=(73.2\pm 1.7)\,$km/s~Mpc${}^{-1}$ \citep{Riess2016}. It is not yet clear what the best explanation for this and similar discrepancies is. They could point towards new physics --- a deviation from the simple model presented here --- or they could mean that systematic uncertainties have been underestimated.

Given that we are now in the era of gravitational wave astronomy, I ought to mention that there is an interesting alternative way of measuring the Hubble constant. Gravitational waves are minute disturbances of the geometry of space, which propagate at the speed of light. Notably, such waves are produced wherever masses orbit each other. The waves are particularly strong when the orbiting masses are compact, and orbit each other closely and quickly. Since the gravitational waves carry away energy, the orbiting masses will move ever closer to each other, orbit each other ever more quickly, and thus produce ever stronger gravitational waves with ever smaller wavelengths until, at the end, the two masses merge. The run-up to this merger, with increasing gravitational wave amplitude and frequency, is governed by the basic laws of relativity. In fact, if one can detect the gravitational wave and measure its basic properties, including how its frequency changes over time, one can {\em derive} the wave's amplitude at the source. 

In other words: Pre-merger gravitational wave signals are ``standard sirens,'' whose amplitude can be determined from direct measurements! And just as with standard candles, the distance of standard sirens can be determined directly: by comparing the gravitational wave signal's reconstructed amplitude at the source with the amplitude that was measured when the signal arrived here on Earth.\footnote{
The relation, in this case, is not an inverse square law, but a $1/r$ law. This is because gravitational wave detectors directly measure the amplitude of a passing wave, not the intensity (which is proportional to the square of the amplitude).
} Crucially, this distance determination is independent of any other form of astronomical distance measurement --- it follows directly from the laws of general relativity! In those cases where one can also determine the redshift of the gravitational wave source, this allows for a gravitational-wave based measurement of the Hubble constant, as first pointed out by Bernard Schutz, who estimated that ten such measurements out to a distance of 100 Mpc (326 million light-years) could suffice to determine the Hubble constant with an accuracy of 3\% \citep{Schutz1986}.

The redshift, however, cannot be derived directly from the gravitational wave signal. That was one reason why the first detection of a merger event with an optical counterpart on August 17, 2017, was so exciting. This event, the merger of two neutron stars, dubbed GW170817, was detected not only by the LIGO and Virgo gravitational wave detectors, but also by many astronomical telescopes all over the globe and in space. The gravitational wave signal allowed a direct determination of the event's distance, while electromagnetic observations yielded precise measurements of the redshift $z$. By comparing the two, a direct value for the Hubble constant of $H_0 = (70\pm 12)\;$km/s Mpc${}^{-1}$ was derived \citep{2017Natur.551...85A}. 

Discrepancies aside, these values for the Hubble constant set the basic scales for our expanding universe. The \Index{Hubble time} has the value
\be
\label{HubbleTimeValue}
\tau_H \equiv \frac{1}{H_0} \sim 4\cdot 10^{17}\,\mbox{s} \sim 10^{10}\,\mbox{a},
\ee
which, as we shall see later on in section \ref{Singularities}, sets the scale for the age of the universe. The \Index{Hubble distance}
\be
\label{HubbleDistanceValue}
D_H\equiv \frac{c}{H_0} \sim 10^{26}\,\mbox{m} \sim 4000\,\mbox{Mpc} \sim 1.4\cdot 10^{10}\,\mbox{light-years}
\ee
sets the cosmic distance scale, and we will encounter both those scales at various times throughout the following sections.

\subsection{Approximating $a(t)$}

Later on, in section \ref{DynamicsCalc}, we will see how the function $a(t)$ can be determined, and how its properties depend on the content of our universe. But even now, we can write down an approximate solution. After all, mathematics tells us that certain functions (specifically, those that are infinitely differentiable) can be represented as polynomials with infinitely many terms, namely as a {\em Taylor series}. If we keep only the first few terms, we get an approximation of the function in question. If we zoom in on any point on the function's graph, then for all smooth functions, a small region around that point will begin to look more and more like a straight line (described by a linear). Zoom out again, and the first traces of curvature of the function's graph can be described quite well by a parabola (described by the linear plus a quadratic term).

Each such approximation works well in a small area around a chosen point. For $a(t)$, we concentrate on the present time $t_0$ and try to find an approximation that works for the immediate past and the immediate future. We approximate $a(t)$ as
\be
a(t) \approx c_1 + c_2(t-t_0) + c_3(t-t_0)^2,
\label{aApproximationInitial}
\ee
with three constants $c_1, c_2, c_3$. Setting $t=t_0$, the present-day value of the scale factor is $c_1$. Let us rename that constant and call it $a_0\equiv a(t_0)$.  Next, we calculate the Hubble constant, following the definition (\ref{HubbleConstant}). Applied to our approximate expression (\ref{aApproximationInitial}), the result is
\be
H_0 = \left.\frac{\dot{a}(t)}{a(t)}\right|_{t=t_0} = \frac{c_2}{c_1} = \frac{c_2}{a_0},
\ee
so that we have
\be
c_2 = a_0 H_0.
\ee
This leaves $c_3$, which is not fixed by any parameter we have yet introduced. The physical dimension of $c_3$ is that of $c_2$ but with an additional inverse time dimension, since $c_3$ is the coefficient for the term $(t-t_0)^2$ instead of $(t-t_0)$.  In order to keep the new parameter dimensionless, it is customary to define $c_3$ as
\be
c_3 = -\frac12\,a_0\,H_0^2 \,q_0,
\ee
where the dimensionless constant $q_0$ is called the {\em deceleration parameter} \index{q${}_0$} \index{deceleration parameter}
The minus sign is a historical legacy: When this parameter was initially defined, the universe was thought to contain ordinary matter which, as we will see later (and as is natural for a mutually attractive gravitational force), decelerates cosmic expansion. In this situation, the extra minus sign makes $q_0$ a positive parameter. When it was later found that our universe is instead accelerating, we were stuck with a negative $q_0$, and so far, nobody has seen fit to re-define this parameter as the acceleration parameter, with opposite sign. (Presumably for the same reason I am chickening out here, and going with the conventional, if awkward definition.)

All in all, we have the approximate expression for the scale factor that is
\be
a(t) \approx a_0\left[
1 + H_0(t-t_0) - \frac12 H_0^2 q_0(t-t_0)^2
\right].
\label{aApproximation}
\ee
It is possible, of course, to improve the approximation by including higher-order terms. A derivation up to fourth order can be found in \citenp{Visser2004}; we follow the more usual course of going no higher than the second-order term.

The approximation (\ref{aApproximation}), governing as it does the distances of nearby galaxies in the Hubble flow, is another example for the equivalence principle. As long as we only consider short time spans $t-t_0$, distances change approximately linearly, as they would if all galaxies were moving freely, without the influence of gravity --- exactly the behaviour prescribed by the equivalence principle, as long as we consider a spacetime region that is small enough. Higher-order terms encoding accelerations due to gravity become inevitable, though, as $t-t_0$ grows larger, in other words: as we consider larger and larger spacetime regions, tidal forces become ever more important.

\section{Consequences of scale-factor expansion}
\index{scale-factor expansion!consequences}
When it came to observable effects of the Hubble relation, we have so far relied local approximations: the (classical) Doppler effect applies to recession speeds only in a neighbourhood of our own galaxy --- even though that neighbourhood is large by everyday standards, extending out to objects at a distance of more than 500 Mpc, corresponding to a bit over 1.6 billion light-years.

In this section, we go further than that. We will again make local calculations; after all, the local approximations based on the equivalence principle and the Newtonian limit describe those situations where our usual physical intuition regarding distances, speeds, and various physical laws linking these and other entities apply. But we will find ways of generalizing our results to the whole of our universe --- either by integrating them up, or by deriving relations between locally defined quantities; if those hold at the location of one galaxy, then in a homogeneous universe, they will hold everywhere. 

There are additional consequences of scale-factor expansion that are not dealt with here: Information about how expansion influences number counts of objects, e.g. up to a certain redshift $z$, and a brief introduction to the Tolman surface brightness test relating different cosmic distance measures can be found in sections 1.11 and 1.7 of \citenp{Weinberg2008}, respectively.

\subsection{Diluting the universe}
\label{Diluting}
We begin, once more, on small scales, invoking both the equivalence principle and the Newtonian limit; recall that, in a cosmos expanding with a universal scale factor, restricting attention to a sufficiently small region will also limit the changes associated with the Hubble flow. Thus restricted, we are now, locally, in the domain of classical physics, at least to good approximation.

One of the most important classical laws of physics is the \Index{first law of thermodynamics}, \index{thermodynamics!first law} a particular way to define 
\Index{energy conservation}. Given a system with \Index{internal energy} $U$ and \Index{pressure} $p$, the internal energy will change as 
\be
\Dd U = \delta Q - p\,\Dd V \label{FirstLaw}
\ee
as the system's \Index{volume} changes by $\Dd V$, and heat $\delta Q$ is added to or withdrawn from the system (depending on the sign). 

Let us consider a small volume of space in our expanding universe that is expanding along with the galaxies. One example would be a cube, with one of our Hubble-flow galaxies in each of its eight corners.

In our expanding universe, the pattern of galaxies will be preserved. In particular, no galaxy will move into our cube from the outside, or leave the cube. Since the universe is homogeneous, the same must hold for anything that can move around --- gas, energy, heat; any net imbalance between adjacent regions would mean that one region would have more, another less, leading to inhomogeneities. In particular, our cube will not pick up net heat from the surrounding regions, or lose heat to them. Expansion must be adiabatic, $\delta Q=0$.

How does this picture change if we include the \Index{mass-energy equivalence} of special relativity? For one, the energy $U$ would be equivalent to a mass; also, we would need to consider an additional energy contribution in terms of rest energies of the particles involved. But since, following our homogeneity argument, the particle number in our cube will be constant (even separately, for all species of particles), extending the definition of $U$ to include rest energy contributions will not change the validity of our conservation equation (\ref{FirstLaw}). 

Thus, we can relate the inner energy $U$ to an energy density, and that energy to the mass density $\rho(t)$ in our cosmos as
\be
U=\rho c^2 V.
\ee
Inserting this expression, it is straightforward to rewrite  (\ref{FirstLaw}) as
\be
\Dd{}\rho = -(\rho + p/c^2) \frac{\Dd V}{V}.
\label{DensityChange1}
\ee
\index{density!change due to expansion}
Recall that we are not considering an arbitrary volume, but one that is linked to cosmic expansion. If all lengths scale $\sim a(t)$, then any volume will scale $\sim a(t)^3$. This is easily seen in the case of a cube, where the volume is the cube of the side length, and the side length scales with $a(t)$. 

In consequence, the volume changes as \index{volume!change due to expansion}
\be
V(t) = \left(\frac{a(t)}{a(t_0)}\right)^3\, V(t_0), 
\ee
so the differential change in volume, $\Dd V$, corresponding to a small change $\Dd a$ in the scale factor, is
\be
\frac{\Dd V}{V} = 3 \frac{\Dd a}{a}.
\ee
Insert this into (\ref{DensityChange1}) and you get the differential change in density, $\Dd\rho$, corresponding to the small scale factor change, namely
\be
\Dd{}\rho = -3(\rho + p/c^2) \frac{\Dd a}{a}.
\ee
A special case of this is how densities change with cosmic time, 
\be
\dot{\rho} = -3(\rho + p/c^2) \frac{\dot{a}}{a} = -3(\rho+p/c^2)\; H(t).
\label{dotRho}
\ee
This depends on two quantities describing the state of cosmic matter: the (mass) density $\rho$ and the pressure $p$. In order to find a solution to this equation, we will need additional information: we will need to know how density and pressure are related. This information is encapsulated in what is called the {\em \Index{equation of state}} of the matter in question,  $p=p(\rho)$.

In cosmology, it is usual to consider different equations of state, all of which are of the form 
\be
p = w\cdot \rho c^2
\label{LinearEOS}
\ee
for some specific constant $w$: \index{w (parameter for equation of state)}
\begin{center}
\begin{tabular}{lll}
Matter (``galaxy dust''): & $p = \phantom{-}\;\;0$,  & $w=\phantom{-}0$\\[0.4em]
Electromagnetic radiation: & $ p = \phantom{-}\rho c^2/3 $, & $w=\phantom{-}\frac{1}{3}$\\[0.4em]
Dark energy (scalar field): & $p = -\rho c^2 $, & $w=-1$.
\end{tabular}
\end{center}
\index{matter!equation of state}
\index{dark energy!equation of state}
\index{electromagnetic radiation!equation of state}
Matter, in this particular context, refers to our ``\Index{galaxy dust}'' of galaxy point particles. In a local frame, the speed values for such galaxies are much slower than the speed of light, and their momentum values much smaller than their energy divided by $c$. That is why it is an excellent approximation to set the pressure to zero, $p=0$.

For electromagnetic radiation filling an expanding universe --- for instance in the form of thermal radiation --- we can make no such approximation. From Maxwell's theory, physicists know, and have known for some time \citep[\S 792]{Maxwell1873}, that \Index{radiation pressure} and radiation energy density $\rho_E$ are linked as $p=\rho_E/3$.

The last example, dark energy, is the most unusual of the three. It does not correspond to any form of matter or energy we know from everyday experience, or even everyday physics. It does, though, correspond to a particularly simple kind of particle in elementary particle physics, namely a particle described by a so-called \Index{scalar field}. In cosmology, this particular ingredient will turn out to be needed to explain the observed expansion history --- the generalization of the Hubble relation that we derived in section \ref{HubbleRelationSection}. Still, at present nobody has a convincing physical theory of what dark energy actually is. At this point, the equation of state $p=-\rho c^2$ is pretty much the only thing that defines dark energy. Historically, dark energy was introduced in a somewhat different form: as an additional constant $\Lambda$, called the {\em \Index{cosmological constant}}, \index{constant!cosmological} introduced by Einstein \index{Einstein, Albert} in 1917 as an extension of the original form of his field equations. The dark energy density we shall be working with in this lecture is related to the value of the cosmological constant by
\be
\rho_{\Lambda} = \frac{c^2}{8\pi G}\, \Lambda.
\label{CosmologicalConstantDarkEnergy}
\ee
Back to our task of reconstructing dilution! For any equation of state of the form $p = w\cdot \rho c^2$, equation (\ref{dotRho}) is readily integrated by separation of variables. To this end, we rewrite the equation as
\be
\frac{\Dd\rho}{\rho} = -3(1+w)\;\frac{\Dd a}{a}.
\ee
Both sides are easy to integrate, and we obtain
\be
\ln\rho = -3(1+w)\ln a + C,
\ee
with an integration constant $C$. This equation can be rewritten as
\be
\rho(t) = C\cdot a(t)^{-3(1+w)}.
\ee
We can replace the constant $C$ by the present day, $t=t_0$, value $\rho(t_0)$ of the density and the present-day value $a_0\equiv a(t_0)$ of the cosmic scale factor, and thus obtain 
\be
\rho(t) = \rho(t_0)\cdot\left( \frac{a(t)}{a(t_0)}\right)^{-3(1+w)}
\label{DensityDilutionW}
\ee
\index{density!change due to expansion}
This is the explicit equation showing how the content of the universe is diluted as the scale factor changes over time.

For dark energy, the density turns out to be constant, \index{dark energy!constant density} since $1+w=0$! This is the motivation for interpretations of dark energy as some unusual property of empty space --- when expansion has doubled the ``amount of space'', the amount of dark energy has doubled, as well. One should be cautious with such interpretations, though. 

Matter --- consisting of our galaxy point particles --- dilutes as
\be
\frac{\rho_M(t)}{\rho_M(t_0)} =\left( \frac{a(t)}{a(t_0)}\right)^{-3}.
\ee
This is as expected: the number of galaxies within our ``co-moving cube'' remains constant, since the pattern inside the cube, defined by the galaxies and their locations, remains the same; each galaxy's mass remains unchanged, so the total mass due to galaxies inside the cube remains unchanged. On the other hand, as we have seen, volume scales as $V\sim a^3$, so the density should indeed scale $\sim a^{-3}$.

While this is not surprising, it is always good to have a cross-check to perform. The next case, that of electromagnetic radiation, is more illuminating (no pun intended).

\subsection{Redshifting photons}

\index{cosmological redshift}
The mass density $\rho_R$ of electromagnetic radiation, with its equation of state $p=\rho c^2/3$, scales as 
\be
\frac{\rho_R(t)}{\rho_R(t_0)} = \left( \frac{a(t)}{a(t_0)}\right)^{-4}.
\ee
What does that tell us about light? The easiest way to answer this is to remember that light is a mixture of light particles, called \Index{photons}:
\begin{center}
\begin{tikzpicture}[scale=0.8]
\draw [blue, very thick,  domain=0:6.3, samples=150,scale=0.2] 
 plot ({\x}, {sin(180*\x)});
 \draw [orange, very thick,  domain=0:6.3, samples=150,scale=0.2] 
 plot ({\x - 8}, {-3+sin(100*\x)});
 \draw [blue, very thick,  domain=0:6.3, samples=150,scale=0.2] 
 plot ({\x+4}, {10+sin(180*\x)});
 \draw [blue, very thick,  domain=0:6.3, samples=150,scale=0.2] 
 plot ({\x+10}, {2+sin(180*\x)});
 \draw [red, very thick,  domain=0:6.3, samples=150,scale=0.2] 
 plot ({\x-6}, {4+sin(80*\x)});
  \draw [green, very thick,  domain=0:6.3, samples=150,scale=0.2] 
 plot ({\x-7}, {8+sin(140*\x)});
   \draw [green, very thick,  domain=0:6.3, samples=150,scale=0.2] 
 plot ({\x+1}, {5+sin(140*\x)});
    \draw [red, very thick,  domain=0:6.3, samples=150,scale=0.2] 
 plot ({\x+2}, {-4+sin(80*\x)});
     \draw [yellow, very thick,  domain=0:6.3, samples=150,scale=0.2] 
 plot ({\x+9}, {-3+sin(120*\x)});
      \draw [yellow, very thick,  domain=0:6.3, samples=150,scale=0.2] 
 plot ({\x-12}, {3+sin(120*\x)});
\end{tikzpicture}
\end{center}
Each photon has a frequency $\nu$ and wavelength $\lambda$, with an energy $E=h\nu=hc/\lambda$ where $h$ is \Index{Planck's constant}, and photons travel at the vacuum speed of light $c$. Photons behave like particles, and we will take that to include that they do not vanish suddenly, or pop into existence.

The mass density of a bunch of photons is proportional to their energy density, and that is proportional to the sum of all the energy contributions from the various photons. As space expands, the same number of photons spread out over a larger volume $V\sim a^3$. If that were all, we would expect the photon mass density to vary as $\rho\sim a^{-3}$. The extra factor $a^{-1}$ must mean that the energies of all the separate photons decrease with time, as well, as $E\sim a^{-1}$. 

Given that for these photon energies, $E=hc/\lambda$, this means that each photon wavelength gets stretched in proportion to $a(t)$, namely as
\be
\lambda(t) = \lambda_0\cdot \frac{a(t)}{a(t_0)}.
\label{CosmologicalRedshift1}
\ee
\index{redshift!cosmological}
This is an alternative form of the \Index{cosmological redshift} we had encountered as an approximate \Index{Doppler shift} in section \ref{HubbleDoppler}: photon wavelengths increase in proportion to the universal scale factor, just like the distances between galaxies in the Hubble flow.

There are some aspects of this derivation one might well be skeptical about. Is photon number really conserved, or is that taking our physical intuition too far? Introducing photons in the first place means introducing some concepts from quantum mechanics. If we were to go further and include the quantum theory of electromagnetic fields, we enter a framework where photon number is not conserved, in general --- so which is the case in the particular situation we are considering here?

Other skeptical questions include: Given that the equivalence principle applies to limited space{\em time} regions, does our derivation really ensure that the cosmological redshift formula (\ref{CosmologicalRedshift1}) holds for all times $t$? Also, given that we started out with a formula for energy conservation, how come that individual photons are now apparently {\em losing} energy over time? 

Let us begin with the time limit. \index{light propagation!in expanding universe} Photons travel at the vacuum speed of light $c$, so after a longer time has passed, we will definitely have left the immediate vicinity of our original small region. Still, we can apply the formula (\ref{CosmologicalRedshift1}), as follows. Our aim is to deduce what happens to a photon that travels from one galaxy in the Hubble flow to another such galaxy (the latter galaxy, in all practical applications, is our own; our home base for observing the universe). Let us divide the time that passes between the cosmic time $t_e$ when the photon is emitted and the later time $t_r$ when the photon is received (observed) into many small steps, inserting $t_1, t_2, t_3,\ldots$ in between the emission and reception times. If we make the divisions sufficiently small, then each travel from $t_i$ to $t_{i+1}$ will occur in a small enough region, and during a small enough time interval, that we can apply equation (\ref{CosmologicalRedshift1}). During each of these portions of the photon's trajectory, its wavelength changes by the ratio $a(t_{i+1}) / a(t_i)$. In order to obtain the total change, we need to multiply all those different change factors:
\begin{center}
\begin{tikzpicture}[scale=1.2]
\foreach \i in {0,1,...,10} 
{
\draw[thick] (\i,-0.05) -- (\i,0.05);
}
\foreach \i in {1,2,...,9} 
{
\node at (\i,0.05) [anchor=south] {$t_{\i}$};
}
\foreach[evaluate=\i as \ni using {int(\i+1)}] \i in {1,2,...,8} 
{
\node at (\i+0.5,-0.3) [anchor=north] {$\cdot \frac{a(t_{\ni})}{a(t_{\i})}$};
}
\foreach \i in {0,1,...,9} 
{
\node at (\i+0.5,0.05) [anchor=north] {$\underbrace{\;\;\;\;\;\;\;\;}$};
}
\draw[->,thick] (0,0) -- (10.5,0);
\node at (0,0.05) [anchor=south] {$t_e$};
\node at (10,0.05) [anchor=south] {$t_r$};
\node at (0.5,-0.3) [anchor=north] {$\cdot \frac{a(t_{1})}{a(t_{e})}$};
\node at (9.5,-0.3) [anchor=north] {$\cdot \frac{a(t_{r})}{a(t_{9})}$};
\node at (-0.2,-0.45) [anchor=north] {\small $\lambda_e$};
\node at (10.5,-0.45) [anchor=north] {\small $ = \;\lambda_r$};
\end{tikzpicture}
\end{center}
But as one can see even in our comparatively coarse example with ten divisions, all the scale factor values cancel each other out, except the first and the last one. We find that equation (\ref{CosmologicalRedshift1}) indeed holds for arbitrary times $t$ and $t_0$, and that light emitted by a galaxy in the Hubble flow at cosmic time $t_e$, and received by another at a later cosmic time $t_r$, is redshifted as
\be
z = \frac{\lambda_r-\lambda_e}{\lambda_e} =  \frac{a(t_r)}{a(t_e)} - 1.
\label{CosmologicalRedshift2}
\ee
So what about individual photons losing energy, even though we began by assuming energy conservation? 
\index{energy conservation!and cosmological redshift}
This is a somewhat subtle point. All our calculations in a small spacetime region in free fall are approximate, limited by the influence of tidal accelerations -- cosmic time $t$ is only approximately the time coordinate of special relativity, distances behave only approximately as they do in special relativity, and so on. If we want to go beyond these approximations and obtain exact results, we will need to take the limit and go to a region that is {\em infinitesimally} small. An equation like the one describing the relation between the density and the scale factor, equation (\ref{DensityDilutionW}), remains valid in that infinitesimal limit, since the density of an infinitesimally region is well-defined --- which is why, in an inhomogeneous situation, we are able to talk about the density of a continuum at a specific location.

But an equation like the one we have just derived for the redshift of a photon, (\ref{CosmologicalRedshift2}), is not preserved if we go to the infinitesimal limit. The limit of restricting ourselves to an infinitesimal region, $t_r\to t_e$, is the limit where $z\to 0$, that is, where the photon is not redshifted at all. Equation (\ref{CosmologicalRedshift2}), as it stands, describes a non-infinitesimal region. Which is fine, but it also means that we have to face up to cosmic time $t$ being an unusual coordinate. The two galaxies emitting and receiving the light are not at rest relative to each other, yet cosmic time links their proper time coordinates to create the global time coordinate we call cosmic time. And for galaxies that are not at rest relative to each other, there will be a Doppler shift. 

In special relativity, when light is emitted by an observer, and received by an observer who is in motion relative to the first, of course the second observer will measure a different wavelength, and thus energy, for each individual photon (cf. the brief account of the Doppler effect in special relativity in section \ref{SRDoppler}). In that situation, we would never wonder how the photon in question had ``lost'' or ``gained'' energy. We would see quite clearly what had happened, namely that we are dealing with more than one frame of reference. The initial energy was measured in one inertial frame, and the receiver was in another inertial frame. Emitter and receiver are in relative motion; hence, the Doppler effect and the fact that the receiver measures a different energy.

In our cosmological situation, we know that at least in the small-region approximation, our galaxies are in motion; at large scales, we know that cosmic time $t$ is not an ordinary, special-relativistic time coordinate, but something more unusual. Given those two circumstances, even with our rather limited understanding, the most parsimonious explanation is that the apparent loss of photon energy is due to the fact that emitter and receiver are in relative motion, or in what passes for relative motion in the more general framework of an expanding universe. A more thorough examination, including a generalization of the special-relativistic notion of relative speed by introducing a concept known as parallel transport, confirms \index{cosmological redshift!as Doppler shift} this \citep{Narlikar1994,BunnHogg2009}.

Still, with the remaining doubts about photon number conservation, and the question of whether or not equation (\ref{CosmologicalRedshift2}) remains valid beyond the (trivial) infinitesimal limit $z\to 0$, it makes sense for us to look at a more rigorous derivation of how photons propagate, and are redshifted.

\subsection{Light propagation in an expanding universe}
\label{LightPropagationSection}
\index{light propagation!in expanding universe|(}
Consider the propagation of light that was emitted by one galaxy at time $t_e$, and received in another galaxy at a later time $t_r$.  We think of light as a point-like pulse with a well-defined location at each moment in time, and follow the same strategy as before: We make our calculations locally, in freely falling systems, considering only a small spacetime region for each and exploiting the fact that cosmic time $t$ is an inertial coordinate there, at least approximately. A similar calculation for a rubber band model of the expanding universe can be found in \citenp{Price2001}.

Specifically, let us look at the distance between the emitting and receiving galaxy at fixed time $t_r$, and divide that distance into $N$ equal pieces of length $\Delta r$. Assume that our light signal reaches the center of the $i$th such piece at time $t_i$. If the length of the $i$th piece at time $t_r$ was $\Delta r$, then the length of the same piece, as measured in our local free-falling system at the time $t_i$ when the light signal was passing through, will be
$$
\Delta r\frac{a(t_i)}{a(t_r)},
$$
since that is how lengths change in our expanding universe. Also, since we are in free fall in a reasonably small neighbourhood, light propagation is governed by special relativity: in an inertial system, light travels in vacuum at the constant speed $c$ (cf.\ the basic principles of special relativity as recounted in the appendix in section \ref{Simultaneity}). Thus, in our particular situation, light will take the cosmic time interval
\be
\Delta t_i = \frac{\Delta r}{c}\cdot\frac{a(t_i)}{a(t_r)}
\ee
to pass that particular segment of its trajectory. And remember that wherever I have talked about approximations in this section so far, the deviations from the laws of special relativity are proportional to the size of the small free-falling region we are considering, since these deviations are due to tidal forces. If we go to the limit of infinitesimally small regions, our approximate statements about the lengths of certain segments, and the propagation speed of light being $c$, will become exact.

Naively, one might think to add up all these travel times to obtain the total interval of cosmic time it has taken for light to travel from the emitting to the receiving galaxy, similar to here:
\begin{center}
\begin{tikzpicture}[scale=1.8]
\foreach \i in {0,1,...,6} 
{
\draw[thick] (\i,-0.05) -- (\i,0.05);
}
\foreach \i in {1,2,...,5} 
{
\node at (\i,0.05) [anchor=south] {$t_{\i}$};
}
\foreach[evaluate=\i as \ni using {int(\i+1)}] \i in {1,2,...,5} 
{
\node at (\i,-0.3) [anchor=north] {$ + \frac{\Delta r}{c}\! \cdot\! \frac{a(t_{\i})}{a(t_{r})}$};
}
\foreach \i in {1,2,...,5} 
{
\node at (\i,0.05) [anchor=north] {$\underbrace{\;\;\;\;\;\;\;\;\;\;\;\;\;\,}$};
}
\draw[thick] (0,0) -- (6,0);
\node at (0,0.05) [anchor=south] {$t_e$};
\node at (6,0.05) [anchor=south] {$t_r$};
\node at (0.1,-0.3) [anchor=north] {$\frac{\Delta r}{2c}\!\cdot \!\frac{a(t_{e})}{a(t_{r})}$};
\node at (5.75,-0.34) [anchor=north] {$+\frac{\Delta r}{2c}$};
\node at (-0.6,-0.4) [anchor=north] {\small $T =$};
\node at (0.25,0.05) [anchor=north] {$\underbrace{\;\;\;\;}$};
\node at (5.75,0.05) [anchor=north] {$\underbrace{\;\;\;\;}$};
\end{tikzpicture}
\end{center}
But things aren't that simple -- the time $t_i$ implicitly depends on how long it has taken light to reach that particular point of its trajectory, which is the kind of travel we hope to calculate!

The solution is to add up the small contributions in another way. Starting from 
\be
\Delta t_i = \frac{\Delta r_i}{c}\cdot \frac{a(t_i)}{a(t_r)},
\ee
where $\Delta r_i=\Delta r/2$ for the first and last bit, and $\Delta r_i=\Delta_r$ for all other segments,
we move all the $t_i$-dependent factors to the left-hand side of the equation, which gives
\be
\frac{\Delta t_i}{a(t_i)} = \frac{\Delta r_i}{c\cdot a(t_r)}.
\ee
Summing over all the segments involved, we get 
\be
\sum_{i=1}^N \frac{\Delta t_i}{a(t_i)} =  \frac{1}{c\cdot a(t_r)}\; d_{r,e}(t_r),
\ee
where on the right-hand side we have used the fact that, by definition, the length segments add up to the distance of the emitting and receiving galaxies at the time $t_r$ of reception.

If we make $N$ infinitely large, and our segments infinitesimally small, the left-hand side becomes an integral, and we end up with\footnote{Why does this appear to have a minus sign, compared to the usual derivation using the metric and radial coordinate? Because with the radial coordinate, we specifically look at light travelling inwards towards smaller radius values, so $r_r-r_e$ is equal to minus the comoving distance.} 
\be
d_{r,e}(t_r) = c\cdot a(t_r)\cdot \int\limits^{t_r}_{t_e}\frac{\Dd t}{a(t)}.
\label{LightPropagation}
\ee
This is the key equation for light propagation in an expanding universe, linking light travel time $t_r-t_e$ to the distance $d_{r,e}(t_r)$ at the time of reception. Of course, in order to evaluate the integral, we will need the explicit form of the universal scale factor as a function of cosmic time, $a(t)$. This requires additional calculations, which I will present in section \ref{DynamicsCalc}.
\index{light propagation!in expanding universe|)}

For signals we receive at the present time, $t_r = t_0$, it is again straightforward to see how this distance is related to the look-back time if we consider our series expansion for $a(t)$ given in equation (\ref{aApproximation}). Inserting that approximation, we have
\be
d_{r,e}(t_0) = c\cdot \int\limits^{t_0}_{t_e}\frac{\Dd t}{1+H_0(t-t_0)-\frac12H_0^2q_0(t-t_0)^2}.
\ee
We can simplify this integral in two ways. For the denominator, we make use of the geometric series approximation
\be
\frac{1}{1+X} = 1-X+\dots;
\label{GeometricSeries}
\ee
this formula gives a good approximation whenever $X$ is small, in our case: whenever $t_0-t$ is considerably smaller than the Hubble time $\tau_H=1/H_0$.
In addition, we choose another integration variable, namely $t'=t_0-t$. Our expression for the distance then becomes
\bea
\nonumber
d_{r,e}(t_0) &=& c\cdot \int\limits^{t_0-t_e}_{0}\frac{\Dd t'}{1-H_0t'-\frac12H_0^2q_0(t')^2} \approx c\cdot\int\limits^{t_0-t_e}_0 \Dd t'\left(
1+H_0t'
\right)\\[0.5em]
&=& c(t_0-t_e) +\frac12 cH_0(t_0-t_e)^2,
\label{LookBackDistance}
\eea
where we have again only kept terms up to second order. This expression shows that, at least for shorter distances, the proper distance $d_{r,e}$ is approximately equal to the look-back time $t_0-t_e$ times the usual vacuum speed of light, just as in special relativity. For larger distances, and thus larger look-back times $t_0-t_e$, additional effects come into play, and the relation becomes more complicated. 

In fact, the \Index{look-back time}, as the time it takes light from a distant object to reach us, is one possible definition of a distance in an expanding universe. Our formula shows that for smaller distance values, the look-back time and the proper distance $d_{r,e}$ are approximately the same, while for larger distances, the two definitions diverge.

\subsection{Cosmological redshift revisited}
\label{CosmologicalRedshiftRevisited}

\index{redshift!cosmological|(}
Even though we do not yet have the knowledge to find explicit solutions to the light propagation equation (\ref{LightPropagation}), since we do not yet know the explicit form of $a(t)$, there are some conclusions we can draw already.  Imagine two light signals leaving a distant galaxy at $r=r_e$ at consecutive cosmic times $t_e$ and $t_e+\delta t_e$, arriving at a second galaxy at time $t_r$ and $t_r+\delta t_r$. Then by our light propagation equation (\ref{LightPropagation}), we have
\be
\int\limits^{t_r}_{t_e}\frac{c\;\Dd t}{a(t)} = \frac{d_{r,e}(t_r)}{a(t_r)} \;\;\;\;\;\mbox{and} \;\;\;\;\;
\int\limits^{t_r+\delta t_r}_{t_e+\delta t_e}\frac{c\;\Dd t}{a(t)} = \frac{d_{r,e}(t_r+\delta t_r)}{a(t_r+\delta t_r)}.
\ee
But the expressions on the right-hand sides of these two equations are the same --- after all, that is how distances change in a universe expanding with a universal scale factors, cf.\ equation (\ref{ScaleFactorExpansionEquation}). That means we have
\be
\left[\;\int\limits^{t_r+\delta t_r}_{t_e+\delta t_e}-  \int\limits^{t_r}_{t_e} \right] \frac{c\;\Dd t}{a(t)} = 0,
\ee
or, exploiting the facts that we can split integration ranges into separate segments, and combine adjacent segments, and that an integral changes sign if you reverse upper and lower limit,\footnote{
Written out, this means that we have additivity,
$$
\int\limits_a^{b}f(x) \Dd x = \int\limits_a^{c}f(x) \Dd x+ \int\limits_c^{b}f(x) \Dd x,
$$
and a sign change when we reverse the boundary values,
$$
\int\limits_a^{b}f(x) \Dd x = -\int\limits_b^{a}f(x) \Dd x.
$$
}
\bea
\nonumber\left[\;\int\limits^{t_r+\delta t_r}_{t_r} + \int\limits^{t_r}_{t_e+\delta t_e}-  \int\limits^{t_r}_{t_e} \right] \frac{c\;\Dd t}{a(t)} =
\left[\;\int\limits^{t_r+\delta t_r}_{t_r} - \int\limits_{t_r}^{t_e+\delta t_e}-  \int\limits^{t_r}_{t_e} \right] \frac{c\;\Dd t}{a(t)} && \\[0.5em]
=\left[\;\int\limits^{t_r+\delta t_r}_{t_r} - \int\limits_{t_e}^{t_e+\delta t_e} \right] \frac{c\;\Dd t}{a(t)} &=& 0.
\eea
However, for very small integral ranges $\delta t$, there is an approximate equality
\be
\int\limits_{\bar{t}}^{\bar{t}+\delta t} f(t)\,\Dd t \approx f(\bar{t})\cdot\delta t,
\label{IntegralSmallLimit}
\ee
which in our case translates to 
\be
\frac{\delta t_r}{a(t_r)} = \frac{\delta t_e}{a(t_e)}.
\label{DeltaStretchExpansion}
\ee
Our two light signals, one closely following the other, could be one of several things: real light pulses, but also consecutive crests (or troughs) of elementary light waves of frequency $f \propto 1/\delta t$. In the latter case, we would have 
\be
\frac{f_r}{f_e} = \frac{a(t_e)}{a(t_r)}, 
\ee
so the wavelength of the light changes as
\be
\frac{\lambda_r}{\lambda_e} = \frac{a(t_r)}{a(t_e)},
\ee
reproducing formula (\ref{CosmologicalRedshift1}) for the cosmological redshift, namely
\be
z = \frac{\lambda_r-\lambda_e}{\lambda_e} =  \frac{a(t_r)}{a(t_e)} - 1.
\label{CosmologicalRedshiftNew}
\ee
Note that while there was still some approximation involved in the derivation, it is of a different kind than before. Our only assumption now, necessary for evaluating the integral as in (\ref{IntegralSmallLimit}), was that $a(t)$ should be approximately constant during the time interval between the first and the second signal (crest, trough\dots). For all realistic $a(t)$ we are dealing with in cosmology, with the exception of the so-called inflationary phase in the very early universe, that approximation is valid.

Also note that not only does the derivation in this section validate our earlier results -- it extends them: formula (\ref{DeltaStretchExpansion}) is valid not only for wave crests or troughs, but more generally for light-like signals following each other. The arrival time difference of these signals will be larger than the emission time difference. The same will apply to {\em all} consecutive signals that carry information about what is happening in a distant region of the cosmos. Formula (\ref{DeltaStretchExpansion}) describes not only a cosmological redshift, but, more generally, {\em cosmological time dilation}. 
\index{time dilation!cosmological}
The corresponding effects have been observed in the light-curves of more vs.\ less distant supernovae of type Ia \citep{Leibundgut1996}.
\index{redshift!cosmological|)}

The most common application is the case where we receive the light of a distant object at the present time, $t_r = t_0$. In that case,
we can make equation (\ref{CosmologicalRedshiftNew}) more concrete by using the series approximation for the scale factor $a(t)$ around the present time $t_0$ which is given in equation (\ref{aApproximation}) in terms of the constants $a_0, H_0, q_0$. In order to get everything nice and polynomial, we will again use the series approximation (\ref{GeometricSeries}). The series expansion for $a(t)$ is in terms of $t_0-t_e$, called the \Index{lookback time}, namely the time that has passed between the emission of the light and us receiving it. Putting everything together, we get
\be
z \approx H_0(t_0-t_e) + \frac12H_0^2\,(q_0+2)(t_0-t_e)^2.
\ee
This can be inverted to give the lookback time as a polynomial of the redshift $z$, again keeping only terms up to second order, namely
\be
H_0(t_0-t_e) \approx z -\frac12(q_0+2) z^2.
\label{LookbackRedshift}
\ee
Using this in our approximate relation (\ref{LookBackDistance}) between the present-day distance $d_{r,e}(t_0)$ and the look-back time, we can derive an approximate redshift-distance relation, which turns out to be
\be
d_{r,e}(z) = \frac{c}{H_0}\left(
z - \frac12(q_0+1)z^2\right) = D_H\left(z-\frac12(q_0+1)z^2\right),
\ee
where the last expression again uses the Hubble distance $D_H\equiv c/H_0$. The term linear in $z$ is the linear redshift-distance relation we had already derived as equation (\ref{RedshiftDistanceApproximate}), while the term that is second order in $z$ is the lowest-order correction.

\subsection{Redshift drift}
Consider once more the cosmological redshift formula  \index{redshift!drift}
\be 
z(t_r) = \frac{a(t_r)}{a(t_e(t_r))} - 1,
\label{CosmoRedshifttr}
\ee
here rewritten to make explicit that, given a fixed reception time and a fixed distant galaxy in the Hubble flow, both the redshift of that galaxy's light arriving at our own location at time $t_r$ and the time at which that light was emitted are functions of $t_r$, namely $z(t_r)$ and $t_e(t_r)$. 

We can differentiate both sides of this equation to see how these quantities change as the reception time changes. The result ist
\be
\frac{\Dd z}{\Dd t_r} = \frac{\dot{a}(t_r)}{a(t_e(t_r))} - \frac{a(t_r)}{a^2(t_e(t_r))}\dot{a}(t_e(t_r))\frac{\Dd t_e}{\Dd t_r}.
\ee
We can replace the $\dot{a}$-terms using the Hubble parameter $H(t)$ defined in equation (\ref{HubbleParameter}), and obtain
\be
\frac{\Dd z}{\Dd t_r} = \frac{{a}(t_r)}{a(t_e(t_r))}\left[H(t_r)  - H(t_e(t_r))\frac{\Dd t_e}{\Dd t_r}\right].
\ee
But in equation (\ref{DeltaStretchExpansion}) during our derivation of the cosmic redshift formula, we saw that the ratio of infinitesimal changes in the emission and reception times is proportional to the ratio of scale factor values at those times,
\be
\frac{\Dd t_e}{\Dd t_r}  = \frac{a(t_e(t_r))}{a(t_r)}.
\ee
Finally, we can make use of the redshift formula (\ref{CosmoRedshifttr}) to rewrite all scale factor ratios in terms of the redshift $z(t_r)$. The result, now once more dropping the explicit $t_r$-dependence-notation and noting that the Hubble parameter at the reception time is the Hubble constant $H_0$, is
\be
H(t_e)  = H_0 (1+z) -  \frac{\Dd z}{\Dd t_r}.
\ee
On the right-hand side, we have $H_0$, which can be measured as described in section \ref{MeasuringHubble}, and we have the redshift, which is straightforward to measure. If we could measure the derivative, as well, and observe how the redshift of that distant galaxy changes a little bit over time, then using this formula, we could directly reconstruct the value $H(t_e)$ of the Hubble parameter at the time the light was emitted! By performing such measurements for galaxies at different redshifts, we could reconstruct the change of the Hubble parameter over time, and in this way, reconstruct the expansion history of the universe!

Excitingly, such measurements are just about within reach for the next generation of telescope, namely 40-meter-class telescopes such as ESO's Extremely Large Telescope currently under construction in Chile. With patience, namely investing 4000 hours of observing time over 20 years, the effect could be measured for extremely distant cosmic clouds of hydrogen at redshifts $2<z<5$ \citep{Liske2008}.

\subsection{The geometry of space}
\label{Geometry}
We have seen in section \ref{MeasuringDistances} how astronomers measure distances, using standard candles and the inverse-square law (\ref{InverseSquareLaw}). Which parts of the derivation of the inverse-square law needs to be changed in an expanding universe? In the lecture itself, I immediately assumed Euclidean geometry for my calculations, mentioning in passing that this does indeed turn out to be the geometry of our particular expanding universe. In the written version, I will be more general, although not to the extent of writing down explicit proofs. These proofs, none of which require familiarity with the concept of a space(time) metric, can be found in chapter 26 of \citenp{Schutz2004}, on which this section is largely based.

To begin with, we should be cautious when it comes to the geometry of space as defined by constant values of the cosmic time $t$. As we saw in section \ref{CosmicTime}, we should be very careful about which of the concepts of special relativity, regarding both time and space, we can legitimately apply in an expanding universe.

Fortunately, the fact that our simple model universe is homogeneous and isotropic restricts the possibilities to a remarkably small number, namely three. The geometric arguments are beyond the scope of this lecture, but they, too can be formulated without recourse to the concept of a metric. The three kinds of three-dimensional spaces that are compatible with homogeneity and isotropy are (1) ordinary Euclidean space, (2) a {\em spherical} space \index{spherical space} that is the three-dimensional analogue of a two-dimensional spherical surface, and (3) a {\em hyperbolic} space that is the three-dimensional analogue of a two-dimensional saddle-shaped surface. \index{geometry of space} \index{hyperbolic space}

What is important for a closer look at luminosity and apparent brightness is to know the surface area of a sphere, defined as the surface of constant distance from a fixed origin. In our case, we look at a fixed cosmic time $t$; the sphere we are interested in is the surface defined by the location of all Hubble-flow galaxies $i$ that have the distance distance $d_{i,0}(t)=d$ from the origin $0$. That surface area, it turns out, is\footnote{In the usual coordinates of standard cosmology, where $r$ is the comoving radial coordinate defined so that the surface area of a sphere is
always $4\pi a(t)^2 r^2$, one introduces the adapted radial coordinate $\chi$ by \index{radial coordinate!adapted}
\be
\Dd\chi \equiv \frac{\Dd r}{\sqrt{1-Kr^2}}.
\label{FLRWAdaptedRadialDefined}
\ee
By rescaling $r$ and $a(t)$ simultaneously, one can always obtain one of the three cases $K=-1,0,+1$. Our global distance coordinate is $d(t)=a(t)\cdot \chi$, corresponding to what is usually called proper distance between galaxies in the Hubble flow.
}
\be
\label{CurvedArea}
A[d(t)] = 4\pi\cdot \left\{    
\begin{array}{ll}
d(t)^2 & \mbox{Euclidean space}\\[0.5em]
a(t)^2\cdot \sin^2[d(t)/a(t)] & \mbox{spherical space}\\[0.5em]
a(t)^2\cdot \sinh^2[d(t)/a(t)] & \mbox{hyperbolic space}.
\end{array}
\right.
\ee
\index{sphere!surface area}
Note that, for small distances $d(t)/a(t)$, all three formulae have the same linear limit, due to $\sin x\approx x$ and $\sinh x\approx x$ for $x\ll 1$. This is a good cross-check on our practice of treating space within a small region around a free-falling galaxy as Euclidean space.

There is a closely related batch of formulae which relates the arc length for a circular arc with radius $d(t)$ and opening angle $\Delta\phi$ as \index{arc length}
\be
\label{CurvedArcLength}
l[d(t),\Delta\phi] = \Delta\phi\cdot \left\{    
\begin{array}{ll}
d(t) & \mbox{Euclidean space}\\[0.5em]
a(t)\cdot \sin[d(t)/a(t)] & \mbox{spherical space}\\[0.5em]
a(t)\cdot \sinh[d(t)/a(t)] & \mbox{hyperbolic space}.
\end{array}
\right.
\ee
In fact, one can get from (\ref{CurvedArcLength}) to (\ref{CurvedArea}) and back quite easily, by looking at the limit of very small angles $\Delta\phi$. In that case, the surface patch of the sphere is very small, and can be approximated as flat. (On Earth, we do the same every time we print a city map onto a flat piece of paper; and yes: this is exactly analogous to our strategy of using the equivalence principle for describing small regions of spacetime.) Then, with an angle $\Delta\phi$ and another angle $\Delta\alpha$ whose arc is perpendicular to that of $\Delta\phi$ where the two intersect on the sphere's surface, we can build a very small rectangle, whose surface area will be given by $l[d(t),\Delta\phi]\cdot l[d(t),\Delta\alpha]$, or
\be
\Delta\alpha\cdot\Delta\phi\cdot \left\{    
\begin{array}{ll}
d(t)^2 & \mbox{Euclidean space}\\[0.5em]
a(t)^2\cdot \sin^2[d(t)/a(t)] & \mbox{spherical space}\\[0.5em]
a(t)^2\cdot \sinh^2[d(t)/a(t)] & \mbox{hyperbolic space},
\end{array}
\right.
\ee
which in turn is proportional to the solid angle $\Delta\Omega\equiv \Delta\phi\cdot\Delta\alpha$. The $d(t)$-dependence is the same for all solid angle patches that lie on the same sphere with radius $d(t)$, so if we add up numerous small patches that, between them, cover the whole sphere, the solid angle patches by themselves will add up to the full solid angle $4\pi$. Thus, the total area will indeed turn out to be given by (\ref{CurvedArea}).

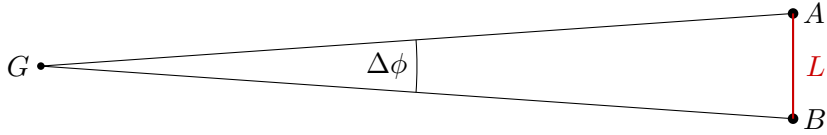
\begin{figure}[htbp]
\begin{center}
\begin{tikzpicture}
\fill (0,0) circle (0.05);
\def\ll{10}
\draw (\ll,0.7) -- (0,0) -- (\ll,-0.7);
\fill (\ll,0.7) circle (0.07);
\fill (\ll,-0.7) circle (0.07);
\node at (\ll,0.7) [anchor=west] {$A$};
\node at (\ll,-0.7) [anchor=west] {$B$};
\draw[red!80!black,thick] (\ll,-0.7) -- (\ll,0.7);
\draw (0,0) ++ (4.01399: 0.5*\ll) arc(4.01399:-4.01399: 0.5*\ll);
\node at (\ll+0.3,0) [red!80!black] {$L$};
\node at (-0.3,0) {$G$};
\node at (0.5*\ll - 0.4, 0) {$\Delta\phi$};
\end{tikzpicture}
\caption{The geometry of two Hubble-flow objects $A$ and $B$, separated perpendicularly to the line of sight by a proper distance $L$, emitting light towards an observer within our own galaxy $G$ at some past time $t_e$. The diagram shows the geometry at time $t_e$}
\label{AngularDistanceGeometry}
\end{center}
\end{figure}

The arc length equations (\ref{CurvedArcLength}) can be used to define an interesting measure of distance in cosmology, called the {\em angular diameter distance} \index{distance!angular diameter} \index{angular diameter distance}. Assume we have two objects A and B in the Hubble flow, equally distant from us, separated by a proper distance $L$ perpendicularly to the line of sight, and that both objects emit light towards Earth at some past time $t_e$. The geometry at the time of light emission is as shown in figure \ref{AngularDistanceGeometry}.
Note that while the triangle is drawn with straight lines, geometry need not be flat, and in fact, the distance $d_{ang}\equiv \overline{AG}=\overline{BG}$, the distance $L$ and the angle $\Delta\phi$ shown in the figure will not, in general, be related following the rules governing triangles in Euclidean space. For small angles $\Delta\phi$, the straight distance $L$ is approximately equal to the corresponding arc length, and for $\Delta\phi$ measured in radians, we have
\be
L = \Delta\phi\cdot d_{ang},
\label{angularDistanceAngle}
\ee
For a configuration of known size $L$, we can measure the apparent size $\Delta\phi$ from observations and use (\ref{angularDistanceAngle}) to derive the configuration's angular diameter distance $d_{ang}$.

Following (\ref{CurvedArcLength}), the angular diameter distance is related to the proper distance as
\be
\label{AngularDistanceAndD}
d_{ang} =  \left\{    
\begin{array}{ll}
d(t_e) & \mbox{Euclidean space}\\[0.5em]
a(t_e)\cdot \sin[d(t_e)/a(t_e)] & \mbox{spherical space}\\[0.5em]
a(t_e)\cdot \sinh[d(t_e)/a(t_e)] & \mbox{hyperbolic space}.
\end{array}
\right.
\ee
The same expression can be re-written in a different form, which we will make use of later. By the definition of scale factor expansion, equation (\ref{ScaleFactorExpansionEquation}), we have $d(t_e)/a(t_e) = d(t_0)/a_0$, with proper distance and scale factor evaluated at the present time $t_0$. Also, the formula (\ref{CosmologicalRedshift2}) for the cosmological  redshift relates the redshift $z$ of light received from a distant object to the scale factor values at the time $t_e$ of emission and the time $t_0$ of reception of the light as $1+z = a_0/a(t_e)$. As a result, we have
\be
\label{AngularDistanceAndDt0}
d_{ang} =  \left\{    
\begin{array}{ll}
d(t_0)/(1+z) & \mbox{Euclidean space}\\[0.5em]
a_0/(1+z)\cdot \sin[d(t_0)/a_0] & \mbox{spherical space}\\[0.5em]
a_0/(1+z)\cdot \sinh[d(t_0)/a_0] & \mbox{hyperbolic space}.
\end{array}
\right.
\ee

The most important application of the concept of angular distance concerns measurements of the fluctuations in the cosmic microwave background (CMB), the radiation that reaches us from the hot, dense Big Bang phase of our universe, 13.8 billion years ago.
\index{cosmic background radiation} From models of the universe, one can derive a maximum size for such fluctuations, given by how far a sound wave in the early universe's plasma can travel from the time of the Big Bang to the time the cosmic microwave radiation is set free. In that case, the radius of the corresponding arc segment in terms of proper distance can be calculated directly, as long as we know the cosmic time now, and the cosmic time the CMB was produced. The angular size of the largest fluctuations can be measured on the sky, and the proper distance of these fluctuations from us can be derived from reconstructions of cosmic history --- we can tell at what age, and at what redshift, the cosmic microwave background was set free. The only unknown, then, is which of the three formulae in (\ref{AngularDistanceAndD}) best describes the relationship between these quantities, in other words: these measurements and deductions constrain the geometry of the universe. The evidence shows that the space of our expanding universe is Euclidean, within the limits of accuracy of these measurements \citep{Boomerang2000,WMAP2013,Planck2016}.

\subsection{Luminosity, apparent brightness, and distance}
\label{LuminosityDistanceSection}
\label{ModifiedInverseSquareLaw}

After these geometric preparations, let us return to the questions of luminosity, apparent brightness, and distance. Just as we argued in section \ref{MeasuringDistances}, in an expanding universe, light sent out isotropically from a distant source will spread out uniformly over the surface of a sphere; when we measure that light, the fraction we capture is the ratio of our telescope area (pointed directly at the distant source!) and the total surface area of the sphere upon which the light has spread out.

Consider light that was emitted at some previous time $t_e$ and is just now reaching us at the present time $t_0$. By (\ref{CurvedArea}), we know how the sphere's surface area is related to the proper distance $d_{i,0}(t_0)$ between the light source and us, the observers, evaluated at the moment our telescope receives the light.

But that is not all. Both luminosity and flux density are measured in terms of energy per unit time. In the case of an expanding universe, this quantity is affected by the cosmological redshift, and by cosmological time dilation. Each individual photon will undergo a redshift, so by equation (\ref{CosmologicalRedshift1}), its energy $E_e$ at the time of emission and $E_0$ at the time $t_0$ of reception will be related by
\be
E_0 = \frac{a(t_e)}{a_0}\cdot E_e.
\ee
But these photons, with the energy of each individual photon diminished, are flying towards our detector in a steady stream. In this respect, they are like the light pulses we considered in section (\ref{CosmologicalRedshiftRevisited}), and their rate of arrival is afflicted by the cosmological redshift: the distance between one photon and the following increases in proportion to the universal scale factor. A stream of $n_e$ photons per second sent out by the emitting galaxy will arrive in our telescope as a diminished stream of particles, at the rate
\be
n_0=  \frac{a(t_e)}{a_0}\cdot n_e.
\ee
The relation between an object's luminosity and its apparent brightness, described as the flux density measured by a distant observer, is given by the combination of these three effects: the homogeneous spreading-out, individual photon redshift and a diminution of the flux rate by cosmic time dilation. The result is
\be
I = \frac{L}{4\pi}\cdot\left(\frac{a(t_e)}{a_0}\right)^2\cdot \left\{    
\begin{array}{ll}
d(t_0)^{-2} & \mbox{Euclidean space}\\[0.5em]
a_0^{-2}\cdot \sin^{-2}[d(t_0)/a_0] & \mbox{spherical space}\\[0.5em]
a_0^{-2}\cdot \sinh^{-2}[d(t_0)/a_0] & \mbox{hyperbolic space}.
\end{array}
\right.
\ee
There are several ways to simplify this expression. We can eliminate the sine/hyperbolic sine terms by using the angular distance. Also, we can replace the scale factor ratio by the redshift $z$, a quantity that has the advantage of being directly observable, as per the redshift formula (\ref{CosmologicalRedshift2}), 
\be
1+z = \frac{a_0}{a(t_e)}.
\ee
When we apply both simplifications, the result is
\be
I = \frac{L}{4\pi\cdot d_{ang}^2(t)\cdot (1+z)^4}.
\ee
Motivated by the simple inverse square formula, the quantity
\be
d_L(t) \equiv d_{ang}(t)\cdot (1+z)^2
\label{LuminosityDistance}
\ee
is called the {\em \Index{luminosity distance}}. \index{distance!luminosity} In terms of the luminosity distance, the relation simply becomes
\be
I =\frac{L}{4\pi\,d_L(t)^2},
\label{LuminosityDistanceInverseSquare}
\ee
\index{inverse square law}
just like in an ordinary, non-expanding, Euclidean-space universe.

\subsection{Luminosity distance and distance moduli}

Astronomy has a traditional system of expressing apparent brightness in terms of \Index{apparent magnitude} --- a logarithmic system that originated at a time when astronomers still relied chiefly on that logarithmically sensitive measuring instrument, the human eye. \index{magnitude!apparent} The apparent magnitude $m_2$ of an object is defined relative to that of another object with apparent magnitude $m_1$ via the respective \index{flux} fluxes $F_2$ and $F_1$, that is, the amounts of energy received from each object per unit time, as
\be
m_2-m_1 = -2.5\log_{10}\left(\frac{F_2}{F_1}\right).
\ee
Since the flux only enters as a ratio, flux densities, measured with the same telescope, can be substituted for the fluxes. Magnitude scales typically have reference objects to define what $m=0$ means, so that stand-alone apparent magnitude values can be given.

(There is a whole area of complications that arises since real astronomical measurements involve measuring not the total flux, but the flux in a certain wavelength band only, using appropriate filters. We neglect this complication; strictly speaking, we will always talk about the total flux, or {\em bolometric flux}, to use the technical term. A helpful introduction to these complications, and the so-called {\em k correction}, can be found in \citenp{Hogg2002}.)

The \Index{absolute magnitude} $M$ of an object is defined as that object's apparent magnitude $m$ when observed from a distance of 10 pc = 32.6 light-years. \index{magnitude!absolute}

Using this definition, and applying the formula (\ref{LuminosityDistance}) for the luminosity distance, a distant object's absolute magnitude $M$ and its apparent magnitude $m$ as measured by an observer are related by
\be
m - M = -2.5\log_{10}\left(\frac{I(d_L)}{I(10\;\mbox{pc})}
\right) = -2.5\log_{10}\left(\frac{10\;\mbox{pc}}{d_L}\right)^2
=5\log_{10}\left(\frac{d_L}{10\;\mbox{pc}}\right).
\ee
The quantity on the right is called the {\em \Index{distance modulus}} corresponding to the luminosity distance $d_L$,
\be
\mu(d_L) \equiv 5\log_{10}\left(\frac{d_L}{10\;\mbox{pc}}\right) = 5\log_{10}\left(\frac{d_L}{1\;\mbox{Mpc}}\right)+25,
\label{DistanceModulus}
\ee
sometimes designated as DM, where the re-write in terms of Mpc is to better accommodate typical extragalactic distance scales.
 
Recall the initial diagram figure \ref{InitialDiagram} we are working to understand? The y axis represents exactly this: distance moduli for the luminosity distances of the objects in question, derived by comparing standard candles' known luminosities $L$ and measured flux intensities $I$. 

We have already made substantial strides in understanding the original diagram, but there is still one ingredient missing: we need to look at what determines the functional form of the universal scale factor $a(t)$. 

\section{Dynamics: how $a(t)$ changes over (cosmic) time}
\label{DynamicsCalc}

So far, $a(t)$ was an unknown function, and all we did was calculate various consequences of an expansion with a given universal scale factor $a(t)$, or
a series approximation thereof. Now, it's time to go a step further and look at what determines the functional form of $a(t)$. \index{scale factor!dynamics|(} \index{cosmic scale factor!dynamics|(} 
This involves exploring the {\em dynamics} of our expanding universe --- how the matter content influences cosmic expansion. \index{Newtonian cosmology}
\index{cosmology!Newtonian}
Newtonian cosmological calculations meant as a {\em replacement} of the models of general relativity go back to the 1930s \citep{Milne1934,McCreaMilne1934,McCrea1955}. Newtonian calculations as an aid to understanding can be found in a number of text books, e.g. \citenp{Rindler1977,Harrison2000,Lotze2002,Weinberg2008}. There are also texts that argue (as I do in these lecture notes) that the conjunction of equivalence principle, Newtonian approximation, and the universality of the scale factor, makes the Newtonian derivation an exact one, and thus much more than a mere analogy or approximation. The earliest example I have found is \citenp{Callan1965}; a more recent article with goals very similar to these lecture notes is
 \citenp{Jordan2005}.

\subsection{Friedmann's equations}
\label{FriedmannEqSection}
\begin{figure}[htbp]
\begin{center}
\begin{tikzpicture}
\draw[fill, white!70!black] (0,0) circle (2);
\draw[black] (0,0) circle (2);
\pgfmathsetmacro{\angle}{20}
\pgfmathsetmacro{\x}{2*cos(\angle)}
\pgfmathsetmacro{\y}{2*sin(\angle)}
\draw[fill, magenta!80!black] (\x,\y) circle (0.07);
\draw[->, thick] (0,0) to ({0.98*\x},{0.98*\y});
\node at ({\x+0.1},{\y+0.17}) [anchor=west,magenta!80!black] {\small test};
\node at ({\x+0.1},{\y-0.17}) [anchor=west,magenta!80!black] {\small galaxy};
\node at (0.0,-0.3) {\small $0$};
\node at (0.7,0.6) {\small $r(t)$};
\fill[black] (0,0) circle (0.07);
\end{tikzpicture}
\caption{Matter within a distance $r(t)$ around a free-falling (Hubble flow) galaxy located at the spatial origin $0$ of our chosen coordinate system}
\label{GravitySphere}
\end{center}
\end{figure}
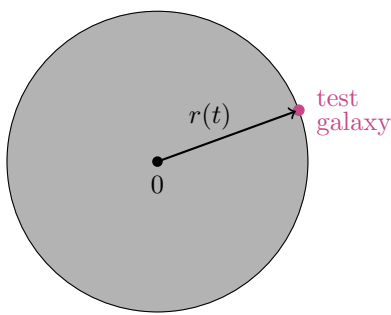

In order to calculate $a(t)$, we once more focus our attention on a small spacetime region around a free-falling, Hubble flow galaxy, which we place in the origin $0$ of our coordinate system. The situation is sketched in figure \ref{GravitySphere}.
We now consider a test galaxy at proper distance $r(t)$ from the first galaxy. Making $r(t)$ sufficiently small, we can consider this situation using classical mechanics and Newtonian gravity, although we will expand our description as discussed in section
\ref{EinsteinsEquations} to include not only mass density, but density and an isotropic pressure term as well.

In the Newtonian picture, and in a spherically symmetric situation (since we are in an isotropic universe), the only gravity acting upon our test galaxy are those at distances smaller than or equal to $r(t)$; none of the mass, energy or pressure outside this sphere has any influence.

(Are you skeptical, because that would mean extending the Newtonian region too far outwards, beyond the expected validity of the Newtonian limit? Very well; in that case, please refer back to section \ref{EinsteinsEquations} of these lecture notes; section 4 in \citenp{Baez2005} shows how the same result can be derived from their exact simple form of Einstein's equations.)

Also in the Newtonian picture, the gravitational acceleration due to the mass inside the sphere with radius $r(t)$ on the test galaxy is the same as in the case where
the same mass is concentrated in the origin. (This, too, is derived in  section 4 of \citenp{Baez2005}, but since in this case, the Newtonian argument applies only to a very small region, the additional support should not be necessary.)

Hence, the acceleration acting to change the test galaxy's radial position is
\be
\ddot{r}(t) = -\frac{GM(r)}{r(t)^2}.
\ee
The effective mass inside the sphere is
\be
M(r) = V(r)\cdot( \rho + 3p/c^2),
\ee
where $V$ is the volume of the sphere, $\rho$ the mass density (including contributions from all available types of energy), and $p$ the pressure, as per our discussion in section \ref{EinsteinsEquations} about sources of gravity.

Since we are in a small region, geometry is approximately Euclidean, and the volume of the sphere is given by the usual formula,
\be
V(r) = \frac{4}{3}\pi\,r^3,
\ee
and since the test galaxy is in the Hubble flow, we have
\be
r(t) = \frac{a(t)}{a(t_0)}\cdot r_0,
\ee
for $r_0$ the proper radius of the sphere at some reference time $t_0$. Putting everything together, it turns out that the situation-specific parameters, notably $r_0$ and $t_0$, cancel out, and that we are left with a second-order equation for $a(t)$, namely
\be
\frac{\ddot{a}(t)}{a(t)} = -\frac{4\pi G}{3}\left(\rho+3\frac{p}{c^2}\right).
\label{SecondFriedmann}
\ee
\index{Friedmann equation!second (order)}
This is known as the {\em second (order) Friedmann equation}, after the Russian cosmologist Alexander Friedmann, the first to write down general cosmological models for an expanding universe. 

Note that this is a purely local equation: We can make our region infinitesimally small and still obtain the same relation between the local quantities $\rho$ and $p$ and the universal function $a(t)$. Also note that while we have restricted our attention to a small spacetime region in order to apply our equivalence principle and Newtonian limit approximations, there is nothing special about the cosmic time $t$ we examined. The same reasoning can be applied to a small spacetime region at {\em any} cosmic time, showing that equation (\ref{SecondFriedmann}) holds for {\em all} moments $t$ of cosmic time. 

That (\ref{SecondFriedmann}) is a second order equation is no surprise. It continues a general trend in physics: dynamical laws concern second time derivatives, accelerations of some sort. 

Given the equation we had derived for how density changes over (cosmic) time, equation (\ref{dotRho}), we can in fact integrate the second order Friedmann equation. To do so, we use equation (\ref{dotRho}) to replace the pressure term on the right-hand side of the Friedmann equation, which yields
\be
\frac{\ddot{a}(t)}{a(t)} = \frac{4\pi G}{3}\left( \dot{\rho}\frac{a}{\dot{a}} +2\rho\right)
\ee
or, multiplying both sides with $a\dot{a}$,
\be
\ddot{a}\dot{a} = \frac{4\pi G}{3}\left( \dot{\rho} a^2 +2\rho a \dot{a}\right).
\ee
This is readily integrated as
\be
\frac{1}{2}(\dot{a})^2 = \frac{4\pi G}{3} \rho a^2 + C.
\label{PreFirstFriedmann}
\ee
Until now, we have not addressed the question of the uniqueness of $a$. In fact, by going through the equations we have developed so far, you will see that the only thing that matters are ratios of scale factor values at different time. That means we have the freedom of rescaling $a$ with a positive constant. We choose to give $a$ the physical dimension of a length, so that both sides of (\ref{PreFirstFriedmann}) have the dimension of speed to the second power. Also, we rescale $a$ by a constant factor so that the integration constant $C$ becomes $C=-Kc^2$, with $c$ the vacuum speed of light and $K$ now restricted to the three possible values $-1,0, +1$. 

The result is the standard form of the {\em first (order) Friedmann equation}, \index{Friedmann equation!first (order)}
\be
\frac{(\dot{a})^2+Kc^2}{a^2} = \frac{8\pi G}{3}\,\rho.
\label{FirstFriedmann}
\ee
\index{scale factor!dynamics|)} \index{cosmic scale factor!dynamics|)}

In fact, we were a little less free in our choices than I have led you to believe. The geometrical formulae for the area of a sphere, and the length of a circular arc, as given in section \ref{Geometry}, already presume this particular rescaling of the scale factor. This is due to one of the relations in an expanding universe that I will not derive here, but which follows directly from general relativity (and, I have no doubt, also from a closer look at the interplay between light and geometry in an expanding universe in the manner of the arguments in this lecture): the connection between the parameter $K$ and the geometry of space, namely that $K=0$ corresponds to Euclidean space, $K=+1$ to spherical and $K=-1$ to hyperbolic space. \index{geometry of space} \index{space geometry}

The general-relativistic models of a homogeneous, isotropic universe, expanding with a universal scale factor, the dynamics of the expansion governed by the Friedmann equation(s), are called {\em \Index{Friedmann-Lema\^{\i}tre-Robertson-Walker models}}, abbreviated FLRW, \index{FLRW models} after the four scientists who made key contributions to their development: the Russian mathematician Alexander Friedmann, \index{Friedmann, Alexander} who discovered that general relativity allowed for different types of matter-filled expanding universe \citep{Friedmann1922,Friedmann1924}; the Belgian astronomer Georges Lema\^{\i}tre \index{Lema\^{\i}tre, Georges} who discovered expanding universe solutions independently, proposing a link with astronomical observation and, later, pioneering the notion of a dense primordial state of the universe \citep{Lemaitre1927,Lemaitre1931a,Lemaitre1931b}; the mathematician and physicist Howard P. Robertson \index{Robertson, Howard P.} from the US, who also discovered expanding universes independently \citep{Robertson1927} and who, like the British mathematician Arthur Geoffrey Walker, \index{Walker, Arthur Geoffrey}, put the models on a solid mathematical foundations, showing that these are indeed the only possibilities to describe isotropic and homogeneous universes in Einstein's theory \citep{Walker1935}.

Synonyms of ``FLRW models'' are ``FLRW spacetimes,'' ``FLRW solutions'' (of Einstein's equations) or ``FLRW universes''. What we will be developing in the following sections is called FLRW cosmology. Variations omitting one or more names, such as \Index{Robertson-Walker spacetimes} or \Index{Friedmann-Robertson-Walker models}, can be found throughout the literature.

\subsection{Filling the universe with different kinds of matter}

Recall that, in the context of how expansion dilutes the contents of the cosmos, we had already introduced three simple kinds of matter, with different dilution behaviour as the scale factor grows, to wit:
\begin{center}
\bgroup
\renewcommand{\arraystretch}{1.2}
\begin{tabular}{l|c|l}
Name  & index & scaling behavior\\\hline\hline
 matter \index{matter} & $M$ & $\rho_M(t) \sim a(t)^{-3}$\\\hline
 radiation \index{radiation} & $R$ & $\rho_R(t) \sim a(t)^{-4}$\\\hline
 dark energy \index{dark energy} &$\Lambda$ & $\rho_{\Lambda}(t) \sim $ const.
\end{tabular}
\egroup
\end{center}
In the following, we assume that there is no significant interaction between these different kinds of content, in the sense that no significant fraction of one type turns into content of another, and that no significant new energy terms arise merely because these three types of content are now present together in all regions of the cosmos. 

Under these circumstances, the mass density (again including energy terms) will simply be the sum of the partial densities,
\be
\rho = \rho_M + \rho_R + \rho_{\Lambda},
\ee
and the overall pressure the sum of partial pressures.

A considerable simplification stems from the fact that the Hubble constant $H_0$ and the gravitational constant $G$ can be combined into a constant with the physical dimension of a density, which in cosmology is called the {\em \Index{critical density}} \index{density!critical} and made to absorb as many numerical factors as possible,
\be
\rho_{c0} \equiv \frac{3H_0^2}{8\pi G}.
\label{CriticalDensity}
\ee
Using this critical density, we can bring the Friedmann equations into a particularly simple dimensionless form. This process is a bit tedious, as it involves a considerable number of steps of substitution and redefinition, but the result is worth it.

First, let us make use of the fact that we know how each density changes over cosmic time from the various special cases of the scaling equation (\ref{DensityDilutionW}), namely
\bea
\nonumber
\rho(t) &=& \rho_M(t)+\rho_R(t)+\rho_{\Lambda}(t)\\[0.5em] 
&=&
 \rho_M(t_0) \cdot \left(\frac{a(t)}{a(t_0)}\right)^{-3}+\rho_R(t_0)\cdot  \left(\frac{a(t)}{a(t_0)}\right)^{-4}
 +\rho_{\Lambda}(t_0),
\eea
where the reference time $t_0$ is taken to be the present time. The second simplification is to express all the present-time densities as multiples of the critical density (\ref{CriticalDensity}). To this end, we introduce dimensionless parameters $\Omega$ via \index{$\Omega$ parameters (density)}
\be
\rho_M(t_0) \equiv \Omega_M\cdot \rho_{c0},\;\;\;\;
\rho_R(t_0) \equiv \Omega_R\cdot \rho_{c0},\;\;\;\;
\rho_{\Lambda}(t_0) \equiv \Omega_{\Lambda}\cdot \rho_{c0},
\label{OmegaParameters}
\ee
and finally an $\Omega$ term of a different kind for the integration constant,
\be 
\label{OmegaK}
Kc^2 \equiv - \Omega_K\cdot [a_0\cdot H_0]^2 \;\;\; \Leftrightarrow \;\;\; K \equiv \Omega_K \left[\frac{a_0}{D_H}\right]^2,
\ee
where for the second version we have used the Hubble distance $D_H\equiv c/H_0$. Note that, with these definitions, the dark energy term $\Omega_{\Lambda}$ is related to the cosmological constant $\Lambda$ introduced in (\ref{CosmologicalConstantDarkEnergy}) as
\be
\Omega_{\Lambda} = \frac{c^2}{3H_0^2}\;\Lambda .
\ee
There is a direct connection between these $\Omega$ parameters and the deceleration parameter $q_0$ we had introduced in our series expansion of $a(t)$, equation (\ref{aApproximation}). In that series expansion, the second-order term was
\be
\left.\frac{\ddot{a}}{a}\right|_{t=t_0} = -q_0H_0^2.
\label{SeriesSecond}
\ee
Using the second Friedmann equation (\ref{SecondFriedmann}) to express the left-hand side of (\ref{SeriesSecond}) in terms of density and pressure, using the appropriate equations of state (\ref{LinearEOS}) to eliminate the pressure terms, and expressing the remaining densities by the critical density (\ref{CriticalDensity}) and $\Omega$ parameters (\ref{OmegaParameters}), we have
\bea
\nonumber
\left.\frac{\ddot{a}}{a}\right|_{t=t_0} &=& -\frac{4\pi G}{3}\left.\left( \rho + 3\frac{p}{c^3}\right)\right|_{t=t_0}
= -\frac{4\pi G}{3}\left.\left( \rho_M + 2\rho_R - 2\rho_{\Lambda}\right)\right|_{t=t_0}\\[0.5em]
&=& -H_0^2\left(\frac12\Omega_M + \Omega_R - \Omega_{\Lambda} \right),
\eea
so that
\be
q_0=\frac12 \Omega_M +\Omega_R - \Omega_{\Lambda}.
\ee

\subsection{Dimensionless Friedmann equation and light-travel distance}
\label{LightTravelDistance}

With the $\Omega$ parameters, we can also rewrite the relation between the look-back time for light reaching us from a distant object and the redshift $z$ of that light. In order to do so, let us begin by abbreviating the cumbersome scale factor ratio, defining \index{x (scale factor ratio)}
\be
x(t)\equiv \frac{a(t)}{a_0}.
\ee
The present-day value of this function, $x(t_0)$, is equal to one.

For $t_0 > t$, this function $x(t)$ is directly related to the redshift $z$ associated with light that is emitted by a distant galaxy at time $t$ and received by us at time $t_0$; by our redshift formula (\ref{CosmologicalRedshift2}), 
\be
x(t) = \frac{1}{1+z}. \label{ScaleFactorRatioX}
\ee
Last but not least, recall the definition of the \Index{Hubble parameter} $H(t) = \dot{a}(t)/a(t)$, which can be rewritten as
\be
H(t) = \frac{\dot{x}}{x}.
\ee
Drawing all this together, we can rewrite the Friedmann equation (\ref{FirstFriedmann}) as \index{Friedmann equation!first (order)}
\be
\left(\frac{\dot{a}(t)}{a(t)}\right)^2 = H(t)^2 = H_0^2 \left[\Omega_M \,x^{-3} + \Omega_R\, x^{-4} + \Omega_{\Lambda} + \Omega_K\, x^{-2}   \right].
\label{FirstFriedmannSimple}
\ee

Separation of variables allows us to rewrite equation (\ref{FirstFriedmannSimple}) as
\be
t-t_0=\int\limits^t_{t_0} \Dd t' =\frac{1}{H_0} \int\limits^x_1 \frac{\Dd x'}{x'\,\sqrt{\Omega_M \,x'{}^{-3} + \Omega_R\, x'{}^{-4} + \Omega_{\Lambda} + \Omega_K\, x'{}^{-2}}}.
\label{FriedmannDimensionless}
\ee
If you can calculate the integral on the right-hand side and solve the resulting equation for $x(t)$, you have derived the functional form of $a(t)$ in terms of the five parameters $H_0, \Omega_M, \Omega_R, \Omega_{\Lambda},\Omega_K$. 

In the general case, the right-hand side integral, cannot be calculated analytically in terms of elementary functions. (It is what mathematicians call an {\em \Index{elliptic integral}}.) But it is straightforward to perform the integration numerically, since in the general case, the integral is well-behaved.\footnote{By definition, the scale factor $x$ cannot be negative; this drastically limits the possibility of poles (where the denominator in the integrand goes to zero). The denominator can only become zero if all $\Omega$s except for $\Omega_{\Lambda}$ are themselves zero; even then, the integral is well-defined, as long as one does not set the integration limit $x$ to zero. As we shall see in section \ref{Explicit}, this is because in that particular case, expansion is exponential, as documented in eq.\ (\ref{adS}), so even if we go back into the past, the limit $x=0$ is not reached in finite time.} We will have a look at some simple limiting cases, which can be integrated directly, in section \ref{Explicit}, p.\ \pageref{Explicit}ff.

Alternatively to the equation (\ref{FriedmannDimensionless}) in terms of the scale factor $x$, we can rewrite the integral directly in terms of the redshift variable $z$, using the relation (\ref{ScaleFactorRatioX}). Substitution of variables yields
\be
t=t_0- \int\limits_0^z \frac{\Dd z'}{H_0 (1+z')\,\sqrt{\Omega_M \,(1+z')^3 + \Omega_R\, (1+z')^{4} + \Omega_{\Lambda} + \Omega_K\, (1+z')^{2}}}.
\label{FriedmannDimensionlessRedshift}
\ee
\index{Friedmann equation!in terms of redshift}
The time difference $t_0-t$ is once more the \Index{look-back time}, and the formula is a more exact version of the approximate formula (\ref{LookbackRedshift}) we derived earlier.

This version of the equation can be used for calculating directly the emission time of a light signal reaching us with a specific redshift $z$. In particular, we can define the \Index{light-travel distance} \index{distance!light-travel} $d_{LT}$ as the light-travel time from a distant source to us, multiplied with the speed of light $c$. This light-travel distance is then related to the redshift $z$ of that light by 
\be
d_{LT} \equiv c(t_0-t) = D_H \int\limits_0^z \frac{\Dd z'}{(1+z')\,\sqrt{\Omega_M \,(1+z')^3 + \Omega_R\, (1+z')^{4} + \Omega_{\Lambda} + \Omega_K\, (1+z')^{2}}},
\label{LightTravelTime}
\ee
where the basic length scale is given by the Hubble distance $D_H\equiv c/H_0$. Alternatively, one can talk about this distance in terms of travel time, such as stating that a particular object is so far away that its light has taken $X$ years to reach us.

Before we study simple explicit solutions to the scale factor equation (\ref{FriedmannDimensionless}) in section \ref{Explicit}, we will first look at some general properties of model universes governed by that equation.

\subsection{Densities and geometry}

Since this equation (\ref{FriedmannDimensionless}) is valid for all cosmic times, it also holds true at the present time, where $x=1$ and $H(t)=H_0$. This implies a restriction on the values of the $\Omega$s, namely \index{Friedmann equation!present-day}
\be
\Omega_M + \Omega_R + \Omega_{\Lambda} + \Omega_K = 1,
\label{FriedmannSimplePresent}
\ee
the present-day version of the first order Friedmann equation.
Recall that $K$, and with it $\Omega_K$, stands for the overall geometry of space in our expanding universe. This equation shows how that geometry is linked to the content of our cosmos; taking into account the three cases $K=-1,0,+1$ as discussed in section \ref{Geometry}, the link is as follows:
\index{geometry of space!and critical density} \index{critical density!and space geometry}
\be
\Omega_M + \Omega_R + \Omega_{\Lambda} \left\{
\begin{array}{ll}
>1 & \mbox{spherical space}\\[0.5em]
=0 & \mbox{Euclidean space}\\[0.5em]
<1 & \mbox{hyperbolic space}.
\end{array}
\right.
\ee
Since the $\Omega$s are all relative to the critical density $\rho_{c0}$, this clarifies the critical density's role as the density value separating the different spatial geometries. 

Sometimes, the above classification is erroneously linked to the global structure of space, in particular whether space is of finite size (``\Index{closed universe}''), or infinite (``\Index{open universe}''). \index{universe!closed vs.\ open}\index{homogeneous spaces}The (countably) infinite varieties of homogeneous spherical space will always be of finite size, and all are derived from the standard \Index{three-sphere} \citep{Gausmann2001}. The situation is more complicated for locally Euclidean \index{Euclidean space} and for hyperbolic spaces: \index{hyperbolic space} for the former, there are 18 possibilities for the large-scale structure of space, in 10 of which space has a finite volume \citep{Riazuelo2004}; these were first classified in the context of crystallography as early as 1885! The hyperbolic case, which features an (uncountably) infinite number of spaces, some that are finite and some that are infinite in volume, is much more complicated, and the search for a complete classification an active area of mathematical research, linked to fundamental topics such as the \Index{Poincar\'e conjecture} \citep{Cornish1998}. The more complicated spaces are connected in multiple ways. The simplest two-dimensional example of a multiply connected space is a \Index{torus}, as shown in figure \ref{Torus}:
\begin{figure}[htbp]
\begin{center}
\includegraphics[width=0.55\textwidth]{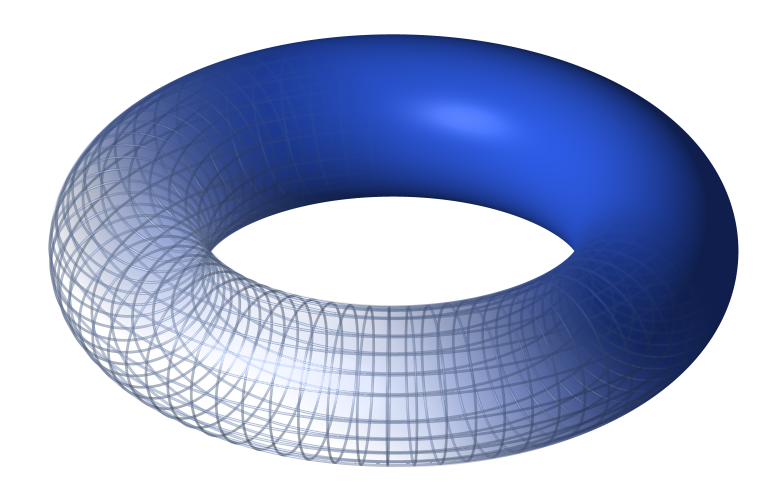} \hspace*{3em}
\begin{tikzpicture}[scale=1.3]
\fill [black!10] (0,0) -- (0,3) -- (3,3) -- (3,0) -- (0,0);
\draw [->>, very thick] (0,0) -- (0,1.8);
\draw [very thick] (0,1.8) -- (0,3);
\draw [->,very thick] (0,3) -- (1.6,3);
\draw [very thick] (1.6,3) -- (3,3);
\draw [->>, very thick] (3,0) -- (3,1.8);
\draw [very thick] (3,1.8) -- (3,3);
\draw [->,very thick] (0,0) -- (1.6,0);
\draw [very thick] (1.6,0) -- (3,0);
\draw[opacity=0] (0,-0.4)--(0,0);
\end{tikzpicture}
\caption{Two representation of a torus: embedded (left) and mathematical (right). Left image: User LucasVB via Wikimedia Commons, in the public domain.}
\label{Torus}
\end{center}
\end{figure}
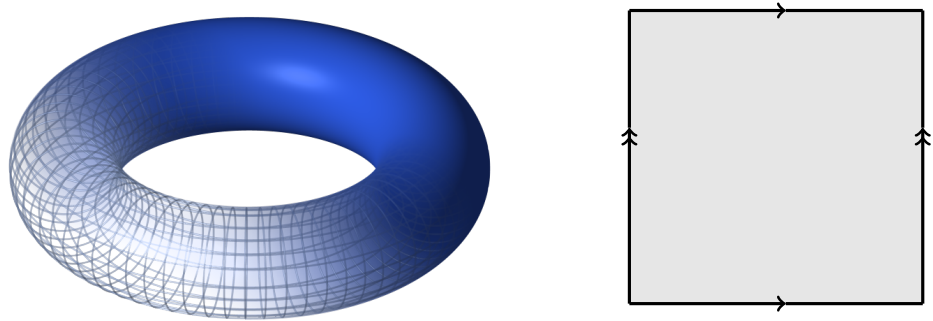
Most people's mental image of a torus is probably that on the left, reminiscent of a donut or the inner tube of a tire. This embedding illustrates nicely the non-trivial global properties of the torus, what mathematicians call its \Index{topology}. For instance, the fact that the torus encloses a hole means that not all curves on the torus are contractable. Imagine a loop of string, such as a shoelace, in the ordinary Euclidean plane. You can always shrink the size of that loop to zero by shortening the string. On the surface of a torus, that is not always possible -- notably, when such a loop circles the tube that is the embedded torus, it cannot be contracted beyond a certain minimum size while remaining on the surface of the torus. 

One aspect of the torus embedding on the left of figure  \ref{Torus} is misleading: the embedding does not preserve the surface's geometry -- the lengths of lines, as measured on the embedded torus, do not correspond (and are not proportional) to the lengths of lines in the two-dimensional, locally flat space the embedded torus is meant to represent.

Mathematically, the easiest way to picture a torus is the one on the right. This is a flat rectangle, on which the usual rules of Euclidean geometry apply. In order to complete the torus, identify the two pairs of opposite edges, one pair marked with a single arrow, the other with a double arrow (orientation should be chosen so each pair of arrowheads is aligned). In this flat representation, straight lines on the torus are shown as straight lines, and all lengths of line segments are represented faithfully.

A good way of picturing the identification is as in old, two-dimensional computer games, where a spaceship (or other object of interest) exiting the screen towards the left would re-enter the screen immediately from the right; a spaceship exiting the screen via the top edge would re-appear from the bottom edge, and vice versa for both sets of directions. 

We can go directly from the mathematical torus on the right-hand side of figure \ref{Torus} to the embedding on the left-hand side by making the identifications manifest. The first identification can be made without distorting the rectangle, and gives a cylinder; closing the cylinder will give the usual torus on the left, but will distort the rectangle, which is why the embedding does not preserve the lengths of lines on the torus.

The mathematical definition of identified edges clearly preserves the local geometry; triangles drawn onto the paper will still have 180${}^{\circ}$ as the sum of their three inner angles, and circles with radius $r$ the circumference $2\pi r$. Both are diagnostics that clearly show that this is locally Euclidean geometry.
Similarly, the more complicated spaces belonging to expanding universes will locally have a spherical, or Euclidean, or hyperbolic geometry, but will be multiply connected. 

\subsection{Different eras of cosmic history; the Big Bang singularity}
\label{Singularities}

The general solution of the Friedmann equation (\ref{FriedmannDimensionless}) for the scale factor $a(t)$ is not readily written down in terms of elementary functions --- although easy enough to solve numerically. We can, however, make some fairly general statements about solutions.

The first batch of general statements follows directly from the scaling behaviour of the different content types, which we had derived in section \ref{Diluting} from the equations of state. If $a$ grows ever larger, then if dark energy is present, it will come eventually to dominate any further evolution, simply because the other densities have been diluted away, $\rho_M\sim a^{-3}$ and $\rho_R\sim a^{-4}$. If $a$ was very small in the past, then if there is a radiation component, it will have been dominant back then because of $\rho_R\sim (1/a)^4$. In between, there may have been a phase where matter was the dominant contribution to the overall density, at a time when the density due to radiation had already been diluted, but before dark energy becomes dominant. 

\begin{figure}[htbp]
\begin{center}
\begin{tikzpicture}[scale=1.0]
\begin{loglogaxis}[xlabel=Relative scale factor $a(t)/a(t_0)$,ylabel=Density/present density, width=11cm, height=8cm,grid=major, ymin=1e-14,xmin=1e-6,xmax=100]
\addplot[color=green!70!black, line width = 1pt, domain=0.000001:100] {0.317/x^3};
\addplot[color=red, line width = 1pt, domain=0.000001:100] {0.00005/x^4};
\addplot[color=blue, line width = 1pt, domain=0.000001:100] {0.683 }; 
\end{loglogaxis}
\node[red] at (7,1) {Radiation};
\node[blue] at (1.4,2) {Dark energy};
\node[green!70!black] at (6,3.3) {Matter};
\end{tikzpicture}
\caption{Different types of content dominate at different scale factors: fairly generic progression of eras, using the estimates for our own universe given
in table \ref{ParameterEstimates} on page \pageref{ParameterEstimates}}
\label{DifferentEras}
\end{center}
\end{figure}
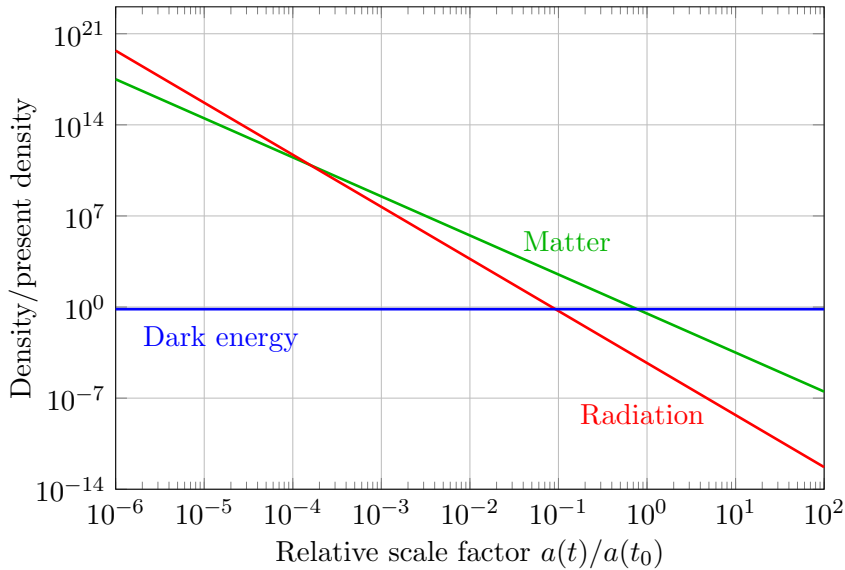

This progression of various eras of dominance is fairly generic whenever all three components are present; an example, using the best estimates for density values in our own universe (cf.\ table \ref{ParameterEstimates} on page \pageref{ParameterEstimates}), is shown in figure \ref{DifferentEras}.

In this diagram with two logarithmic scales, the line representing the matter density has slope $-3$, the radiation line $-4$, while dark energy density remains, of course, constant. From left to right, we see the {\em \Index{radiation era}} up to when the universe was about $10^{-4}$ its present size where radiation dominates, followed by the {\em \Index{matter era}}, while shortly before the present scale factor is reached, dark energy becomes dominant. 

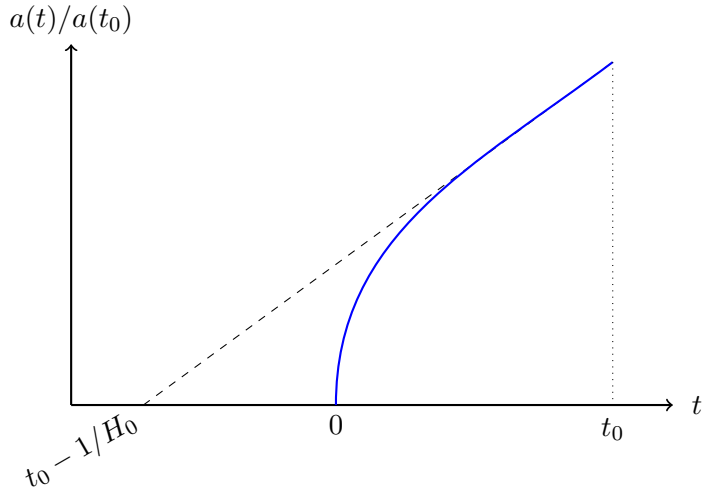
\begin{figure}[htbp]
\begin{center}
\begin{tikzpicture}[scale=1.6]
\draw[thick,->] (0,0) -- (5,0);
\draw[thick,->] (0,0) -- (0,3);
\node at (0,3) [anchor=south] {$a(t)/a(t_0)$};
\node at (5.2,0) {$t$};
\def\steigung{0.57735}
\draw[dashed] (0.6,0) -- (4.5,{(4.5-0.6)*\steigung+0.6});
\draw[blue, thick] (2.2,0) to [out = 90, in = 217] (4.5,{(4.5-0.6)*\steigung+0.6});
\draw[dotted] (4.5,{(4.5-0.6)*\steigung+0.6}) -- (4.5,0);
\node at (0.5,0.1) [anchor=north east,rotate=30] {$t_0-1/H_0$};
\node at (4.5, -0.2) {$t_0$};
\node at (2.2,0) [anchor=north] {$0$};
\end{tikzpicture}
\caption{Generic evolution for a universe not dominated by dark energy: a singular beginning}
\label{Earlya}
\end{center}
\end{figure}

So far, though, these different eras are defined in terms of the scale factor value, not in terms of cosmic time. That is why, as a next step, we turn to the second-order form (\ref{SecondFriedmann}) of the Friemann equation, using the equations of state of our content components to rewrite the right-hand side in terms of mass density alone, as 
\be
\frac{\ddot{a}(t)}{a(t)} = -\frac{4\pi G}{3}\left(\rho_M +2\rho_R -2\rho_{\Lambda}\right).
\label{InitialAcceleration}
\ee
This shows that, as long as the density of dark energy does not dominate everything else as $\rho_{\Lambda} > \rho_R+\rho_M/2$, we have $\ddot{a} < 0$, corresponding to a deceleration. In such situations, tracing the evolution of $a(t)$ back in time as in figure \ref{Earlya}, we will generically find some point in the past with $a(t)=0$. As the figure shows, the value of the Hubble constant $H_0$, which describes the present-day slope of $a(t)/a(t_0)$, can be used to estimate how far back in time that point $a(t)=0$ can be. If $a(t)$ were a linear function, then we could easily extrapolate. The time at which $a(t)=0$ would be
\be\label{HubbleTime}
\tau_H\equiv 1/H_0,
\ee 
namely the \Index{Hubble time}, before the present. Since the actual evolution is not linear, but instead has $\ddot{a}<0$, the linear extrapolation gives an upper limit of how much time has passed since $a(t)=0$.

At the time where $a(t)=0$, weird things happen. Since the densities of matter and radiation are proportional to $a^{-3}$ and $a^{-4}$, respectively, they become infinitely large at $a(t)=0$. In our galaxy dust picture, all finite distances between other galaxies and us (and some infinite distances, as well!) were zero at that time. With these disquieting properties, the point at which $a(t)=0$ is a so-called {\em \Index{singularity}}, more specifically the {\em \Index{initial singularity}}, {\em \Index{Big Bang singularity}}, or just {\em \Index{Big Bang}} for short. The time of that singularity is commonly chosen as the zero point of cosmic time $t=0$.

Our argument for why the Big Bang singularity must occur is a very crude version of much more rigorous {\em \Index{singularity theorems}}, notably the cosmological singularity theorem first formulated by Stephen Hawking \index{Hawking, Stephen}and Roger Penrose, \index{Penrose, Roger} which show that Big Bang singularities occur under much more general circumstances than in simple homogeneous models. 

The Big Bang singularity is a problem for our models; in fact, singularities and infinities like this commonly indicate the limits of the model in question. A widely accepted limit that would apply at such small scales in any case would be that neither our simplified description nor the more complete description in the framework of general relativity takes into account quantum effects, which one should expect to become important at such small length scales and such high energies. Unfortunately, no consistent and complete theory of {\em \Index{quantum gravity}} has been found yet, although physicists have been searching for one for the past 70+ years. 

The good news is that, even without understanding the Big Bang singularity at $t=0$, it is possible to model what happens directly afterwards, at a time when the universe was still very crowded, when the scale factor was smaller than a thousandth of its present value, and densities for matter and radiation were correspondingly high. The key factor in this is that, at present, the universe contains about a billion times more photons than matter particles such as protons, neutrons, and electrons. When we extrapolate back to smaller scale factor values, when these photons were much more energetic (cosmic redshift!), we end up with a very hot soup of elementary particles, dominated by these photons --- the {\em \Index{Big Bang phase}}, sometimes called the {\em Big Bang} for short, as well. The choice of terms here requires some caution: a cosmologist talking about the Big Bang could be referring to this [fairly well understood] phase of the early universe, or to the [highly problematic] Big Bang singularity --- learn to deduce from the context what is meant! 

The physics of the Big Bang phase is beyond the scope of this lecture, and was covered at this summer school by Matthias Bartelmann's lecture, a video recording of which is available at
\begin{center}
\href{http://youtu.be/m35fXJoQLA0}{http://youtu.be/m35fXJoQLA0}
\end{center}

The distinction between the Big Bang and the Big Bang phase is an important one to make to both students and to the general public. Naturally, the Big Bang itself, as the beginning of everything, holds a special fascination for anyone interested in the big questions. It's important to be honest at this point, and state clearly that cosmologists do not have reliable models for that singular beginning. We do not even know whether it {\em is} an abrupt beginning, or merely the end of an earlier phase. But it is equally important to communicate that this ignorance about the Big Bang singularity does not mean we can say nothing worthwhile about the beginning of the universe as we know it. 

On the contrary: When it comes to the Big Bang phase, our knowledge becomes more and more secure with increasing cosmic time, and after a bit less than a second of cosmic time, we have reached the realm of standard, experimentally testable physics. Our models of the Big Bang phase and its aftermath have a lot to say about how the universe became what it is today -- from the abundances of light elements to the existence and properties of the \Index{cosmic background radiation} to the evolution the first stars, the first galaxies, and the large-scale structure of the observable universe. 

Let us briefly talk about what $a(0)=0$ does and doesn't mean. Taken at face value, it does mean that all particles that are at finite distance from us at present had  distance zero from us (and from each other) at the Big Bang singularity at $t=0$. In particular, everything we can observe around us, in other words: everything in the observable universe was, at that time zero, compressed into an arbitrarily small volume. (The concept of the observable universe will be made more precise in section \ref{Horizons}.) 

This does not necessarily mean that the {\em whole} universe was compressed into a mathematical point. In those models where the universe is infinite, there are Hubble flow galaxies that are infinitely far away from us at the present moment. Multiplying an infinite present distance with a zero-valued scale factor is an ill-defined problem. So is the Big Bang singularity a ``spacetime point'', or in some way {\em like} a spacetime point, or not? 

There is a way of understanding light propagation in relativistic spacetimes that shows the situation to be more complicated. There exists a way of mapping even an infinite universe into a finite, so-called {\em \Index{conformal diagram}} (also known as a \Index{Penrose diagram} or, more rarely, \Index{Carter-Penrose diagram}). Such maps show that when it comes to causality --- which regions of spacetime can, in principle influence which other regions? --- the initial singularity has more in common with infinite space than with an ordinary point in space  (section 5.3 in \citenp{HawkingEllis1973}). For matter emerging from an ordinary single point of space, one might assume that all this matter could have ``inherited'' common properties, such as a common temperature, from its common origin. But for matter shortly after the Big Bang, this kind of inheritance is impossible. Notably, a common origin of this particular kind is insufficient to explain the fact that early on, all baryonic matter in the observable universe appears to have had nearly the same temperature (chapter 5 in \citenp{Earman1995}). This is known as the \Index{horizon problem}, and one of the stated motivations of modifying the expansion history of the very early universe by introducing an early \Index{inflationary phase} of exponential expansion prior to cosmic time $t=10^{-30}\;\mbox{s}$.

Again, we will not go into the history of the very early universe in these lecture notes. We do take away the message that, in order to avoid confusion, if we feel the need to talk about how dense and compressed matter was in the early universe, we had better avoid the singularity at $t=0$. It is perfectly correct to say that, within these models, all the matter within the observable universe (alternatively, ``all the matter we see around us'') was once compressed into a volume the size of a basketball, or a cricket ball, or the pin of a needle, and so on. But talking about it all being compressed ``into a single point'' or ``a region of volume zero'' skips over significant complications, doesn't add anything significant in terms of understanding, and should not detract from the facts that (a) within the general-relativistic models, the Big Bang singularity is not part of spacetime, (b) the occurrence of the singularity a strong indication that our model breaks down at cosmic time $t=0$, and (c) that at the present time, nobody has a solid alternative model for the very early universe. 

On to less problematic matters: Now that we have chosen the zero point of cosmic time $t=0$ to correspond to the Big Bang singularity, at which $x=0$, we can use equation (\ref{FriedmannDimensionless}) to obtain the age of the universe as 
\be
t_0=\frac{1}{H_0} \int\limits^1_0 \frac{\Dd x'}{x'\,\sqrt{\Omega_M \,x'{}^{-3} + \Omega_R\, x'{}^{-4} + \Omega_{\Lambda} + \Omega_K\, x'{}^{-2}}}.
\ee
The time scale for the \Index{age of the universe} \index{universe!age} is set by the \Index{Hubble time} $\tau_H \equiv 1/H_0$ (cf. equation \ref{HubbleTime}).

\subsection{Expansion and collapse}
\label{ExpansionCollapse}

\index{fate (of the universe)!collapse vs.\ eternal expansion|(}
Will an expanding universe keep expanding forever? Or might it collapse back onto itself, all pairwise distances now {\em shrinking} in proportion to the universal scale factor, just as they grew in unison before, ending in what has been called a {\em Big Crunch}, \index{Big Crunch} as the opposite of the Big Bang (or, more facetiously, the \Index{Gnab Gib}, [\citenp{Adams1980}])?

For an answer, let us revisit the dimensionless Friedmann equation, 
\be \tag{\ref{FirstFriedmannSimple}}
\left(\frac{\dot{a}(t)}{a(t)}\right)^2 = H(t)^2 = H_0^2 \left[\Omega_M \,x^{-3} + \Omega_R\, x^{-4} + \Omega_{\Lambda} + \Omega_K\, x^{-2}   \right].
\label{FriedmannSimple2}
\ee
For an expansion phase $\dot{a}(t)>0$ to make the transition into a contraction phase with $\dot{a}(t)<0$, there must be a moment in between where $\dot{a}(t)=0$, and hence the whole equation (\ref{FriedmannSimple2}) must be zero. Let us assume that if at all, this happens only after the early Big Bang phase, so that we can neglect the radiation term. Since $x\ne 0$, we can rewrite the condition that (\ref{FriedmannSimple2}) be zero as
\be
\Omega_M + \Omega_{\Lambda}x^3 + \Omega_K\, x =0,
\ee
and since we know how to express $\Omega_K$ in terms of the other two terms as per equation (\ref{FriedmannSimplePresent}), the necessary condition for re-collapse is
\be
\Omega_{\Lambda}x^3 + (1-\Omega_M-\Omega_{\Lambda} ) x + \Omega_M=0.
\label{GeneralRecollapse}
\ee
In the absence of any dark energy, $\Omega_{\Lambda}=0$, this is readily solved as
\be
x = \frac{\Omega_M}{\Omega_M -1},
\ee
and since $x$ must always be positive, this is only a solution if $\Omega_M>1$, in other words: if the present-day matter density is larger than the present-day critical density. In older text books, written before it was discovered that $\Omega_{\Lambda}$ played a significant role in our own universe, this circumstance was often linked to space geometry: if only matter is present, a spherical geometry corresponds to a re-collapsing universe, while Euclidean and hyperbolic universes expand forever.

We can at least make some statements about the general case of equation (\ref{GeneralRecollapse}). At the present day, $x=1$, and we must have $\Omega_M + \Omega_{\Lambda} + \Omega_K = 1 >0.$ And if $\Omega_{\Lambda}<0$, corresponding to a type of dark energy that seeks to decelerate the universe, then for suitably large $x$, the cubic term will dominate, and the left-hand side of (\ref{GeneralRecollapse}) will be negative; if it was positive before, and negative then, it must have been zero some time between. Will such large $x$ values be reached, or could the evolution approach some constant $x$ value, asymptotically? No, because then the acceleration would need to go to zero, which is not compatible with the acceleration equation (\ref{InitialAcceleration}). Such universes will always re-collapse.

For $\Omega_{\Lambda}>0$, the situation is less clear, but it is clear that equation (\ref{GeneralRecollapse}) can only be zero for some (necessarily positive!) $x$ if the linear coefficient is negative, $1-\Omega_M-\Omega_{\Lambda} =\Omega_K<0$. So in this case, as in the case without dark energy, re-collapse is only possible if the space geometry is spherical. This is a necessary, but not a sufficient condition --  for instance for $\Omega_M=0$, there will be no collapse even in a spherical space: in that case, the only solution for the turning point is
\be
x=\sqrt{\frac{\Omega_{\Lambda}-1}{\Omega_{\Lambda}}} < 1,
\ee
which, in a currently expanding universe, is a point in the past, when the universe was smaller. For there to be a turning point, we must have had a universe that was initially collapsing, but then turned and is now expanding, and will keep expanding forever.
\index{fate (of the universe)!collapse vs.\ eternal expansion|)}

\subsection{Explicit solutions}
\label{Explicit}
After these general considerations, let us turn to special cases of the Friedmann equation (\ref{FriedmannDimensionless}), which allow for simple solutions. We concentrate on solutions with an initial singularity, and change the integration limits to include the singularity $a(0)=0$. Then we have
\be
t=\int\limits^t_0 \Dd t' = \int\limits^x_0 \frac{\Dd x'}{H_0 x'\,\sqrt{\Omega_M \,x'{}^{-3} + \Omega_R\, x'{}^{-4} + \Omega_{\Lambda} + \Omega_K\, x'{}^{-2}}}.
\label{CalculateTFromX}
\ee
\index{universe!radiation-dominated}
We begin with a universe that contains only radiation --- as we have seen in section \ref{Singularities}, this is a good approximation for the very early universe, shortly after the Big Bang. Given the scaling of the term containing $\Omega_K$, we can neglect that term (and thus any non-Euclidean geometry of the universe) when we restrict ourselves to the early universe with small values of $a$. This leaves us with 
\be
t= \int\limits^x_0 \frac{\Dd x'\,x'}{H_0\sqrt{ \Omega_R}} = \frac{1}{2H_0\sqrt{\Omega_R}}\, x^2, 
\ee
which is readily solved as
\be
a(t) = a(t_0)\cdot\sqrt{\sqrt{\Omega_R}\cdot 2H_0 t} \sim t^{1/2}.
\ee
If this is not just an approximation, with $\Omega_K$ neglected, but instead a generic Euclidean model universe with $\Omega_K=0$, then by the present-day Friedmann equation (\ref{FriedmannSimplePresent}), which states that all the $\Omega$s must sum up to one, 
we must have $\Omega_R=1$, and our evolution is given by
\be
a(t) = a(t_0)\cdot\sqrt{2H_0 t}.
\label{RadiationEvolution}
\ee
For consistency, if \index{model universe!pure radiation} we insert the present age $t_0$ of the universe, we must recover $a(t=t_0)=a(t_0)$. This means that the age of a pure radiation universe must be given by
\be
t_0 = \frac{1}{2H_0},
\ee
half the Hubble time, in other words: half as old as for a linearly expanding universe. In particular, this means that we can write the scale factor in the particularly simple form
\be
a(t) = a(t_0)\cdot \sqrt{t/t_0}.
\ee

Next, consider a Euclidean ($\Omega_K=0$), matter-only universe. \index{model universe!Einstein--de-Sitter (EdS)} This solution, which is called the Einstein--de-Sitter (EdS) universe, \index{Einstein--de-Sitter universe} must have $\Omega_M=1$ in order to satisfy the present-day Friedmann equation (\ref{FriedmannSimplePresent}). Integrating up, we obtain 
\be
a(t)=a(t_0)\cdot \left(3/2\cdot H_0t  \right)^{2/3} \sim t^{2/3}.
\ee
Again, consistency demands, and direct calculation of $H(t)$ confirms, that the age $t_0$ of such a universe must be
\be
t_0 = \frac{2}{3H_0},
\ee
and the simplest form for the scale factor is
\be
a(t) = a(t_0)\cdot (t/t_0)^{2/3}.
\label{aEdS}
\ee

Our last elementary case is a universe with Euclidean space ($\Omega_K=0$) that only contains dark energy, so $\Omega_{\Lambda}=1$. In this case, there will be no finite time at which $a(t)=0$, so we revert to the original integration limits of (\ref{FriedmannDimensionless}) to obtain
\be
t-t_{0} =  \int\limits^x_1 \frac{\Dd x'}{H_0 \,x'} = \frac{1}{H_0} \; \ln(x) 
\ee
so that
\be
a(t) = a(t_0)\cdot\exp\left[H_0(t-t_0) \right].
\label{adS}
\ee
\index{de Sitter solution} \index{model universe!de Sitter (dS)}
This type of exponential solution, albeit disguised by another choice of coordinates, was first found by the Dutch astronomer Willem de Sitter \index{de Sitter, Willem} \citep{deSitter1917}. In modern cosmology, it is important for the latest stages of evolution, when dark energy dominates, but also for a very early, hypothetical stage known as {\em \Index{inflation}}, postulated to explain, among other things, the homogeneity of the universe as well as the absence of certain exotic particles.

The de Sitter scale factor can also be written in terms of the cosmological constant. In the de Sitter universe, the critical density (\ref{CriticalDensity}) is equal to the dark energy density, which can be expressed in terms of the cosmological constant as in equation (\ref{CosmologicalConstantDarkEnergy}). Equating the two, we have
\be
H_0 = c\sqrt{\Lambda/3},
\ee
and can re-write the de Sitter scale factor as 
\be
a(t) = a(t_0)\cdot\exp\left[\sqrt{\Lambda/3}\cdot c\,(t-t_0) \right].
\label{adS2}
\ee
This concludes our gallery of selected explicit solutions; a more complete overview can be found in chapter 23 of \citenp{dInverno1992} or chapter 9 of \citenp{Rindler1977}.

With the arguments concerning the different eras, and these simple solutions, we can make a reasonable estimate for how $a(t)$ looks for our own universe. Our cosmos contains radiation (in particular the cosmic background radiation), matter in the form of galaxies, and dark energy. Thus, we would expect a beginning that looks like $a(t)\sim t^{1/2}$ while radiation dominates, exponential expansion at the later stages, and some kind of interpolation linking the two.

\begin{table}[htp]
\caption{Parameter estimates for our universe}
\begin{center}
\small
\begin{tabular}{l@{$\;=\;$}l}
$H_0$ & $67.74$ km/s/Mpc \\
            & $1/(4.6\cdot 10^{17}\;\mbox{s}) = 1/(1.4\cdot 10^{10}\;\mbox{a})$ \\[0.2em]
$\Omega_{\Lambda}$ & $0.69$\\[0.2em]
$\Omega_M$ & $0.31$\\[0.2em]
$\Omega_R$ & $0.0001$
\end{tabular}
\end{center}
\label{ParameterEstimates}
\end{table}%
\index{cosmological parameter values} \index{parameter values, cosmological} \index{Hubble constant!value} \index{$\Omega$ parameters (density)!values}
The parameter values applying to our specific universe need to derive from observation. A set of realistic estimates combining the measurements of ESA's \Index{Planck satellite} with other observations (notably of the redshift-distance relation, cf.\ section \ref{RDRSection}) can be found in table \ref{ParameterEstimates} \citep{Planck2016}.\footnote{Rightmost column in table 4.
The density parameter for radiation can be derived from $\Omega_M$ and the redshift $z_{eq}$ specified in the table, which indicates the redshift at which matter and radiation density were equal. Given the scaling behaviour for radiation and matter discussed in section \ref{Diluting}, we must have $\Omega_R=\Omega_M/(1+z_{eq})$, which yields the value for $\Omega_R$ quoted in table \ref{ParameterEstimates}. } The matter part $\Omega_M$ can be split in baryonic matter (such as protons, neutrons, electrons and the nuclei, atoms and molecules they form) with 
$\Omega_b= 0.05$ and a dark matter part with $\Omega_{dm}= 0.26$. 

The critical density (\ref{CriticalDensity}) in such a universe amounts to 
\be
\rho_{c0} = 8.6\cdot 10^{-27}\,\frac{\mbox{kg}}{\mbox{m}^3} = 4.8\,\frac{\mbox{GeV}/c^2}{\mbox{m}^3},
\index{critical density!value}
\label{CriticalDensityValue}
\ee
which corresponds to about five proton masses per cubic meter. 

\begin{figure}[htbp]
\begin{center}
\includegraphics[width=0.99\textwidth]{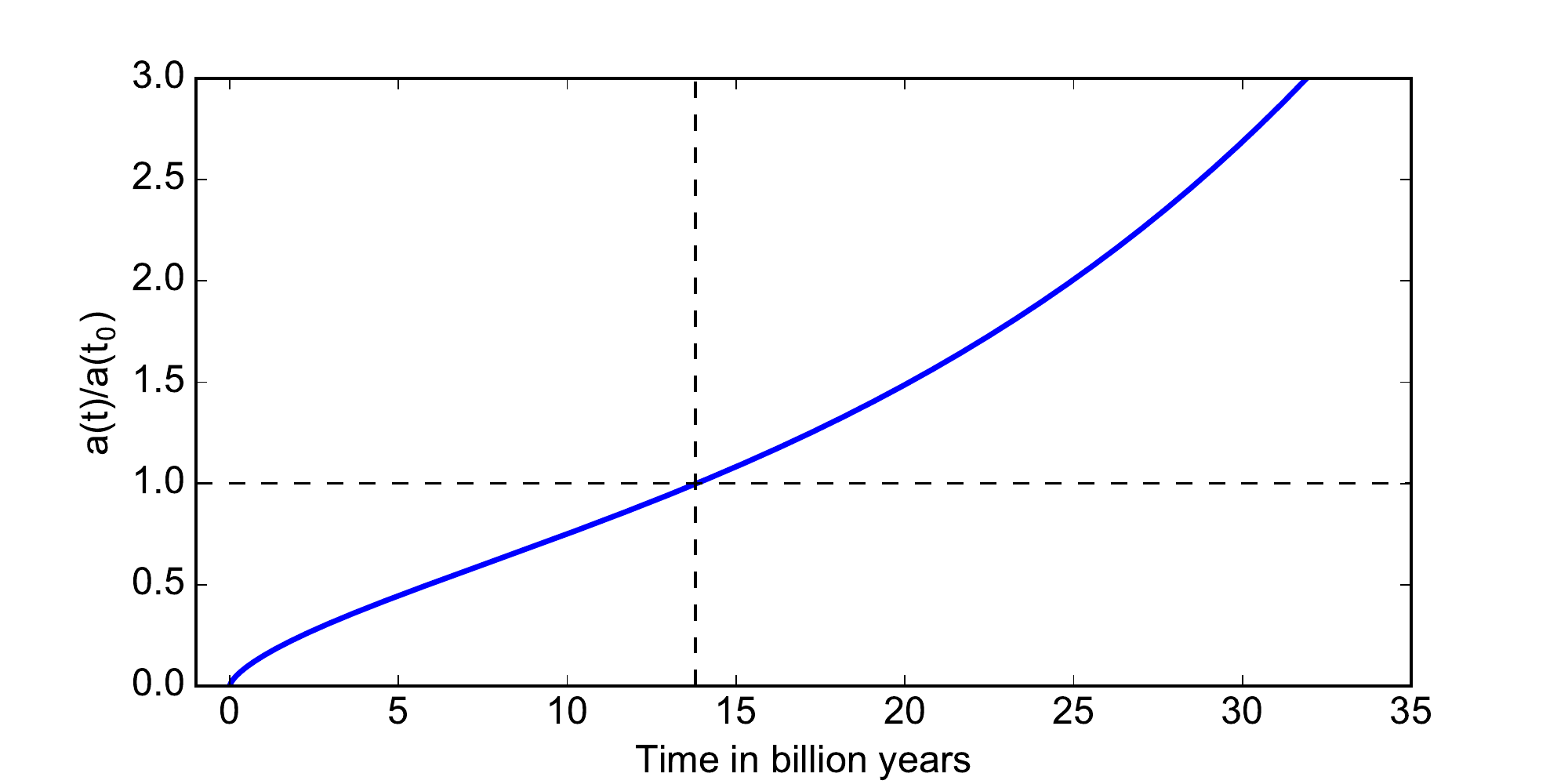}
\caption{Expansion $a(t)$ of our universe, using the parameters given in table \ref{ParameterEstimates}. The dotted lines correspond to the present time, $t=13.8$ billion years}
\label{OurExpansion}
\end{center}
\end{figure}

A numerical solution using these parameter values is shown in figure \ref{OurExpansion}.
The fate of our particular universe appears to be for distant galaxies to drift ever further apart. We shall see below in section \ref{Fate} that this will severely restrict future astronomer who, from a certain point in time onwards, will see hardly any distant galaxies any more, severely restricting their capacity to perform research in cosmology.

This shape of the curve of $a(t)$ conforms rather well with our expectations from studying the different eras. The expansion history shown in figure \ref{OurExpansion} forms the backbone of the modern \Index{$\Lambda$CDM models} of the universe (pronounced ``Lambda-CDM''). In this abbreviation, $\Lambda$ stands for the presence of dark energy (or, equivalently, a cosmological constant) which, as we shall confirm in our final evaluation of the redshift-distance diagram in section \ref{RedshiftDistanceRevisited}, is the dominant ingredient in our universe. 

CDM stands for ``\Index{cold dark matter}'' and indicates that the dominant matter component is in the shape of so-called \Index{dark matter} --- not the ordinary matter made of atoms (which in cosmology is called {\em \Index{baryonic matter}}), \index{matter!dark} \index{matter!baryonic} but matter that neither emits nor absorbs electromagnetic radiation. Dark matter thus remains unobservable by our telescopes, and can only detected indirectly by its gravitational action on the visible constituents of the universe. The contribution of baryonic matter to the $\Omega_M$ given in table \ref{ParameterEstimates} on page \pageref{ParameterEstimates} is a mere $\Omega_b=0.05$, that of dark matter $\Omega_{cdm} = 0.26$. (Upon closer inspection, even 90\% of the baryonic contribution is not in the shape of the galaxies I had introduced as prototypical matter constituents, but in the form of warm, \Index{intergalactic plasma}, cf.\ \citenp{Fukugita2004}.) \index{plasma, intergalactic}

In addition to the FLRW model universes described in this lecture, $\Lambda$CDM models encompass models for the hot, early universe. These describe both the creation of the cosmic microwave background (CMB) radiation \index{cosmic background radiation} and the primordial fusion reactions which created the first light elements (\Index{Big Bang nucleosynthesis}, BBN), as well as models for the formation of \Index{large-scale structure} in the universe. For the latter, the adjective ``cold'' in cold dark matter becomes important. As opposed to hot dark matter, particles of cold dark matter move slowly compared to the speed of light; this has significant consequences for the way the large-scale structure in our universe has formed over the past billions of years.

\section{Consequences of cosmic evolution}

Now that we have some notion of the different $a(t)$, we can revisit the topic of consequences of cosmic expansion and introduce some consequences that have an interesting connection with the specific form of $a(t)$.

\subsection{Horizons}
\label{Horizons}
Our light propagation formula 
\be \tag{\ref{LightPropagation}}
d_{r,e}(t_r) = c\cdot a(t_r)\cdot \int\limits^{t_r}_{t_e}\frac{\Dd t}{a(t)},
\ee
contains fundamental information about the causal structure of our model universe. In particular, no light signal that has reached us by now, at the present time $t_0$, can have begun its journey earlier than the time of the Big Bang, $t=0$. By the light propagation formula, such light cannot have started at a (proper) distance greater than
\be
d_{PH} = c\cdot a(t_0)\cdot \int\limits^{t_0}_{0}\frac{\Dd t}{a(t)}.
\ee
This defines our {\em \Index{particle horizon}}. \index{horizon!particle}  Light from galaxies, or other objects, that lie within the sphere defined by this distance value, has had sufficient time to reach us; in consequence, we can observe such objects. Light from objects outside the horizon has not yet had sufficient time to reach us, so we cannot observe these objects. 

Inside $d_{PH}$ is what is, at this moment in cosmic time, our {\em \Index{observable universe}} -- at least in theory: in practice, there can be additional restrictions. Notably, since our universe was in a hot, dense state during the Big Bang phase 13.8 billion years ago, the farthest we can look into the distance, and thus back in time, is to the end of the hot dense phase. The \Index{cosmic background radiation} is radiation from that particular transition from a hot dense opaque universe to a transparent universe, and marks the current boundary of the observable universe.

But back to particle horizons: For instance, in the Einstein--de-Sitter universe, where only matter density $\Omega_M=1$ contributes and space is Euclidean, given the scale factor (\ref{aEdS}), we have \index{Einstein--de-Sitter solution!particle horizon}
\be
d_{PH,EdS} = c\cdot (t_0)^{2/3}\cdot \int\limits^{t_0}_{0}t^{-2/3}\Dd t = 3\,c\,t_0 = 2\frac{c}{H_0} = 2\,D_H,
\ee
where in the last step, we have substituted the Hubble distance $D_H=c/H_0$.

A complementary concept is that of the cosmological {\em \Index{event horizon}}. \index{horizon!event} In a universe that is expanding sufficiently fast, there could be a distance beyond which light emitted at the present time $t_0$ will never reach us, no matter how long we wait. Given this definition, the proper distance to this horizon is 
\be
d_{EH} = c\cdot a(t_0)\cdot \int\limits^{\infty}_{t_0}\frac{\Dd t}{a(t)},
\label{CosmicEHFormula}
\ee
at least for infinitely expanding universes, marking the distance below which light emitted that far away will just about reach us, even though it takes an arbitrary long time to do so. (In a re-collapsing universe, one would need to adjust the upper limit so as to reflect the finite maximum time available for that light to reach us.)

For the Einstein--de-Sitter universe, we find
\be
d_{EH,EdS} = c\cdot (t_0)^{2/3}\cdot \int\limits^{\infty}_{t_0}t^{-2/3}\Dd t =c\cdot (t_0)^{2/3} \lim_{t\to\infty} t^{1/3},
\label{deSitterEH}
\ee
which is infinite. In such a universe, there is no finite event horizon --- as long as we wait sufficiently long, we will be able to see objects at an arbitrary distance.

Matters are different in an exponentially expanding de Sitter universe: for the scale factor (\ref{adS}), we have the event horizon \index{de Sitter universe!event horizon}
\be
d_{EH,dS} =  c\cdot \int\limits^{\infty}_{t_0} \exp[-H_0(t-t_0)]\,\Dd t = -\frac{c}{H_0} \bigg[
\exp[-H_0(t-t_0)]
\bigg]^{\infty}_{t_0}=\frac{c}{H_0} = D_H.
\ee
In a de Sitter universe, light signals emitted at the present time $t_0$ from a (proper) distance greater than the Hubble distance, which in that particular case is equal to
\be
D_H = \frac{1}{\sqrt{\Lambda/3}},
\ee
will never reach us.

\subsection{Do galaxies, humans, atoms expand?}
\label{BoundSystems}
\index{bound systems!and cosmic expansion|(} \index{cosmic expansion!and bound systems|(}
Another consequence of cosmic expansion concerns the interaction between expansion and bound systems. This issue frequently comes up whenever cosmic expansion is not described as a form of motion, but instead as a distance change of a completely new kind, with phrases such as ``space itself is expanding'' or  ``space is growing larger'' between galaxies. Based on such (misleading) models, it is quite natural to ask: What is happening with bound systems under such circumstances? Are galaxies, humans, atoms expanding, too --- are they growing larger in size? 

If yes, that could have significant consequences. A cartoon version of what might happen is the statement ``If all meter sticks grew in length along with cosmic expansion, how could we detect cosmic expansion in the first place?'' Meter sticks, however, are an overly simplified way of looking at this question; as long as our (idealized) clocks do not change, radar-ranging using light signals will tell us local distances, and enable us to detect the effects of cosmic expansion. The question of how universal expansion of all objects would affect length measurements depends on the less straightforward question of how expansion affects clocks.

But instead of talking about clocks, let us stay with meter sticks and other bound systems. The following treatment is, once more, simplified; more complete and more advanced treatments can be found in \citenp{Pachner1963,Pachner1964,Cooperstock1998,Carrera2010,Price2012,Giulini2014}.

When we talk about atoms, humans, planetary systems and even a single galaxy, we are automatically talking about small scales, where the special-relativistic notions of time and space constitute an excellent approximative description. Furthermore, the recession velocities associated with cosmic expansion are so small on these scales that we can make use of the concepts and laws of classical mechanics, including Newton's law of gravity. In short, we are in a situation very similar to that we analyzed in section \ref{FriedmannEqSection} when we derived Friedmann's equations. As an example, we consider a bound system consisting of a central mass $M_{cen}$, or central charge $Q_{cen}$, at the origin $0$ and an orbiting test particle with mass $m$ at a distance $r(t)$: a planet orbiting a central mass, or an electron orbiting an atomic nucleus. 

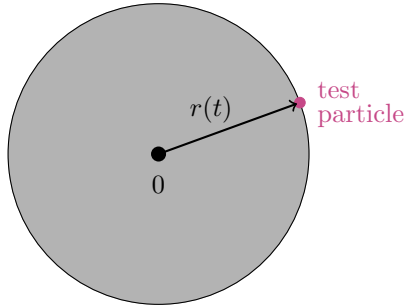
\begin{figure}[htbp]
\begin{center}
\begin{tikzpicture}
\draw[fill, white!70!black] (0,0) circle (2);
\draw[black] (0,0) circle (2);
\pgfmathsetmacro{\angle}{20}
\pgfmathsetmacro{\x}{2*cos(\angle)}
\pgfmathsetmacro{\y}{2*sin(\angle)}
\draw[fill, magenta!80!black] (\x,\y) circle (0.07);
\draw[->, thick] (0,0) to ({0.98*\x},{0.98*\y});
\node at ({\x+0.1},{\y+0.17}) [anchor=west,magenta!80!black] {\small test};
\node at ({\x+0.1},{\y-0.17}) [anchor=west,magenta!80!black] {\small particle};
\node at (0.0,-0.4) {\small $0$};
\node at (0.7,0.6) {\small $r(t)$};
\fill[black] (0,0) circle (0.1);
\end{tikzpicture}
\caption{Matter within a distance $r(t)$ of the origin $0$, here taken to be the center (central mass, central charge) of a bound system}
\label{BoundSphere}
\end{center}
\end{figure}
From the central object, our test particle will feel some force $F_{cen}$: A central mass will exert a Newtonian force on our test particle, and a central charge will exert a Coulomb force. In an idealized, perfectly homogeneous universe, there would be an additional force just as we derived in \ref{FriedmannEqSection}: While none of the homogeneously distributed matter outside the sphere with radius $r(t)$ will contribute, the homogeneous matter inside $r(t)$, represented by the gray area in figure \ref{BoundSphere}, will attract the test particle with a mass
\be
M_{cos}= \frac{4}{3} \pi r(t)^3\cdot[\rho(t) + 3p(t)/c^2)].
\ee
Here, we have already included the extra pressure term we learned about in section \ref{EinsteinsEquations}, and where the density $\rho$ includes contributions from the energy density. For short, we will refer to the mass of the homogeneous matter, and to the force on the test particle resulting from it, as ``cosmological'', $M_{cos}$ and $F_{cos}$ respectively, in order to distinguish it from the central force $F_{cen}$. \index{force!cosmological}

That mass $M_{cos}$ will act as if it were the mass of a point particle at the origin, so the gravitational force from the homogeneous matter that our test particle feels will be radial, and of the strength
\be
F_{cos} = -\frac{4 G}{3}\pi r(t) \cdot m[\rho(t) +3p(t)/c^2].
\label{HomBound}
\ee
Which kinds of cosmic matter contribute to this force $F_{cos}$? Ordinary matter, while assumed to be homogeneous on large scales, is most certainly {\em not} homogeneously distributed on the scales of atoms, humans, or planetary systems. When we talk about atoms and their typical length scales, all of the ordinary matter except for the atomic nucleus, whose influence is accounted for in $F_{cen}$, will be outside, and not contribute to $F_{cos}$. 

Does dark matter contribute? If dark matter is in the form of particles, as is commonly believed, that would depend on the particle mass. Going by the parameter estimate based listed in table \ref{ParameterEstimates} on page \pageref{ParameterEstimates} and the critical density value in (\ref{CriticalDensityValue}), the present mass density of dark matter amounts to
\be
\rho_{dm} = \Omega_{dm}\,\rho_{c0} = 2.2\cdot 10^{-27}\,\frac{\mbox{kg}}{\mbox{m}^3} = 1.2\,\frac{\mbox{GeV}/c^2}{\mbox{m}^3}.
\ee
No dark matter particle has yet been detected, and estimates for their masses depend on the chosen hypothetical model. For so-called axions, \index{axion}
their (current) masses could be in the range of $0.1\,\mbox{meV}/c^2$, corresponding to $\sim 10^{13}$ such particles per cubic meter. But the typical volume of an atom is on the order of $10^{-29}\,\mbox{m}^3$ at most. The probability to find a single one of these \Index{dark matter particles} inside a randomly placed atom would be $10^{-16}$. One might try to calculate the minute influence these particles might have on an atom when they happen to pass through, but numbers are such that approximating dark matter as a homogeneous continuum on the scale of atoms makes no sense. The numbers are, of course, much worse, for heavier dark matter candidates, such as \Index{WIMPs} (for ``weakly interacting massive particle'') with expected masses of $10\,\mbox{GeV}/c^2$ and more \citep{2016ChPhC..40j0001P}. On the other hand, over planet- or solar-system-size volumes, we will find so many even of these heavy  particles that a homogeneous distribution should be a good approximation.

Inserting the respective equations of state (\ref{LinearEOS}) for dark matter and dark energy into equation (\ref{HomBound}) and, in a second step, the definition
of the critical density (\ref{CriticalDensity}), we find that
\be
F_{cos} = \frac{4\pi G}{3} rm[2\rho_{\Lambda} - \rho_{dm} ] = m\, r\, H_0^2 [\Omega_{\Lambda}-\Omega_{dm}/2].
\label{HomPre}
\ee
On atomic or molecular scales, as per the argument given above, we will neglect the dark matter contribution; on planetary scales and above we will include dark matter. We can describe both scenarios in a unified way by defining 
\be
\Omega_{cos} = \left\{   \begin{array}{ll}
\Omega_{\Lambda} & \mbox{on atomic/molecular scales}\\[0.5em]
\Omega_{\Lambda} - \frac12\Omega_{dm} & \mbox{on scales greater than that of planets};
\end{array}
\right.
\ee
with this definition,
\be
F_{cos} = m\, r\, H_0^2 \,\Omega_{cos}.
\label{HomBoth}
\ee
In our own universe, $\Omega_{cos}>0$ as dark energy is the dominant contribution (cf.\ table \ref{ParameterEstimates}), so the cosmological force is repulsive. 

In addition to the cosmological force, there is the central force as an additional attractive force acting on the test particle, pulling it towards the origin; we introduce a unified description of a gravitationally bound system and an electrostatically bound system by writing that force as
\be
F_{cen} = - \frac{mC}{r^2},
\ee
where
\be
C \equiv \left\{\begin{array}{ll}
\frac{1}{4\pi\epsilon_0} \frac{Qq}{m} & \mbox{electrostatic force}\\[0.5em]
GM & \mbox{gravitational force,}
\end{array}
\right. 
\label{UnifiedForces}
\ee
and where we have encoded the attractive nature of the force in the minus sign.

Since the cosmological force $F_{cos}$ is repulsive, it serves to counteract, albeit very weakly, the attractive binding force of our system. The relative strength of cosmological force and central force varies with distance -- as the test particle moves outward from the origin, the homogeneous force will increase, and the binding force decrease.

It is instructive to calculate the distance at which both forces balance out precisely, $F_{cos} + F_{cen} = 0$. The result is 
\be
r_{balance} = \left(\frac{C}{\Omega_{cos}\,H_0^2} \right)^{1/3}.
\ee
Let us look at gravitationally bound systems first. In that case, with $C=GM$ and using the values from table \ref{ParameterEstimates} for $H_0$, as well as a $\Omega_{cos}$ that includes the dark matter contribution, we have
\be
r_{balance} = \left(\frac{GM}{\Omega_{cos}\,H_0^2}\right)^{1/3} = \left(\frac{M}{M_{\odot}}\right)^{1/3}\; 386\; \mbox{light-years}.
\ee
Here are a few examples for typical masses of astronomical objects, order of magnitude only: \index{planetary systems!and cosmic expansion} \index{galaxies!and cosmic expansion}
\begin{center}
\bgroup
\renewcommand{\arraystretch}{1.2}
\begin{tabular}{|c|c|c|c|}
\hline
central object & mass [$M_{\odot}$] & system size [ly] & $r_{balance}$ [ly] \\\hline\hline
Jupiter-like planet & $0.001$ & $<10^{-8}$ &  10 \\\hline
star / planetary system & 1 & $<10^{-3}$ & $10^2$\\\hline
galaxy & $10^{12}$ & $10^5$ & $10^6$\\\hline
galaxy cluster & $10^{14} - 10^{15}$ & $10^7 - 10^8$ & $10^7$\\\hline
\end{tabular}
\egroup
\end{center}
Stars and their planetary systems and galaxies are safely bound: their real size is much smaller than the $r_{balance}$ corresponding to their mass. But for large galaxy clusters, their real size has the same order of magnitude as their $r_{balance}$; these are indeed, as we would expect, the largest bound systems in the universe, and loosely bound at that. Systems larger than that are pushed apart by the acceleration due to dark energy. Considerations like these can help us understand which cosmic systems remain bound, and which ones join the Hubble flow!

Next, for an ``atom'' bound by the \Index{Coulomb force}. In that case, both the central charge $Q$ and the test particle charge $q$ are equal to the elementary electric chage $e$. \index{charge!elementary}The test particle mass is the electron mass $m_e$. With all this, the constant $C$ from equation (\ref{UnifiedForces}) is
\be
C=\frac{1}{4\pi\epsilon_0} \frac{Qq}{m} =\frac{1}{4\pi\epsilon_0} \frac{e^2}{m_e}. 
\ee
Inserting these values, we find that the attractive Coulomb force and the repulsive cosmological force balance at the distance \index{atoms!and cosmic expansion}
\be
r_{balance} = \left(\frac{e^2}{4\pi\epsilon_0m_e\Omega_{\Lambda}H_0^2} \right)^{1/3} = 28\;\mbox{au}.
\ee
This corresponds to the average distance of Neptune from the Sun, and is much, much larger than the real size of any atom. Atoms, too, are safely bound, and will hardly be affected by dark energy.

Next, let us analyze circular orbits, which are readily derived by balancing the centripetal acceleration
\be
a_{cp} = \frac{v^2}{r} = \omega^2\,r,
\ee
where $\omega$ is the angular frequency, related to the orbital period $T$ as $\omega=2\pi/T$,
with the acceleration created by the forces acting on our test particle, namely
\be
a_{forces} = \frac{C}{r^2} -  r\,\Omega_{\Lambda}\,H_0^2. 
\label{CombinedBoundAcceleration}
\ee
The result is a slightly modified version of Kepler's third law: orbital radius and angular frequency are related as
\be
\omega^2 = \frac{C}{r^3} - \Omega_{\Lambda}\,H_0^2 = \frac{C}{r^3} \left( 1 - \frac{r^3\Omega_{\Lambda}\,H_0^2}{C}  \right).
\ee
For a gravitationally bound system, the correction term is 
\be
 c_{corr} = \frac{r^3\Omega_{\Lambda}\,H_0^2}{C}  = 2\cdot 10^{-8}\cdot \left(\frac{r}{\mbox{ly}}\right)^3\,\left(\frac{M}{M_{\odot}}\right)^{-1}.
\ee
Here are, again, a few examples for typical masses and sizes of astronomical objects:
\begin{center}
\bgroup
\renewcommand{\arraystretch}{1.2}
\begin{tabular}{|c|c|c|c|}
\hline
central object & mass [$M_{\odot}$] & system size [ly] & $c_{corr}$ \\\hline\hline
star / planetary system & 1 & $<10^{-3}$ & $10^{-8}$\\\hline
galaxy & $10^{12}$ & $10^5$ & $10^{-5}$\\\hline
galaxy cluster & $10^{14} - 10^{15}$ & $10^7 - 10^8$ & $10^{-1} - 10^1$\\\hline
\end{tabular}
\egroup
\end{center}
Again, it is only on the level of loosely bound galaxy clusters that this correction becomes important. 

All in all, we find that local physics, on scales much smaller than that of galaxy clusters, is influenced only very slightly by cosmic expansion. Crucially, any influence is via the acceleration caused by the matter and dark energy content of the universe. 

This is in line with a very general property of time evolution in physics. How a system changes over time depends both on the dynamics and on initial conditions. Dynamical effects stem from interactions within the system, or with the external world. In classical physics, these interactions manifest themselves via forces that cause accelerations. Distinct from these effects are the \Index{initial conditions}, typically initial positions and initial speeds. Within a system, up to a certain point, initial conditions can be chosen freely. I can decide how hard to throw a ball, and choose the direction of my throw. The ball's trajectory depends both on the initial conditions I have chosen and on the gravitational force acting on the ball.

What we have described as cosmic expansion corresponds, in part, to a specific choice of initial conditions, in part to gravitational interactions within the system. Once we decide to describe a bound, local system, the initial conditions are set --- and they are different from the initial conditions for objects in the Hubble flow. Only the gravitational interactions are the same, and, as is usual for dynamics, manifest themselves in acceleration terms. These will inevitably influence our bound system (if only a tiny little bit, as we have seen).

While we have, so far, restricted our attention to small scales, there is a Gedankenexperiment on cosmological scales that illustrates the important difference between consequences of expansion depending directly on $\dot{a}(t)$ and the real, dynamical changes governed by $\ddot{a}(t)$. It is known as the \Index{tethered galaxy problem} \citep{Peacock2002,DavisLineweaverWebb2003,Whiting2004,Davis2005,Clavering2006,GronElgaroy2007,BarnesEtAl2006}, although as we see, that is slightly misleading; I will instead call this the ``attached galaxy problem.''

The problem can be formulated as follows: Consider a small galaxy that is not initially part of the Hubble flow; instead, initially, that galaxy is at rest relative to our own. In the thought experiment, the small galaxy is ``attached'' to our own using a thin, solid rod.\footnote{In the original formulation, the small galaxy is, instead, ``tethered'' to our own. But a tether will only prevent movement of the two galaxies away from each other; a rod will keep the distance constant. In the usual terminology of cosmology, we initially hold our proper distance to that galaxy constant.} The galaxy needs to be small so we consider it as a test particle, and assume the fixation does not affect our own galaxy's motion. \index{attached galaxy problem} If we now release that galaxy (``break the rod''), what will happen? How will the galaxy move once it is freed from the cumbersome constraint?

\begin{figure}[htbp]
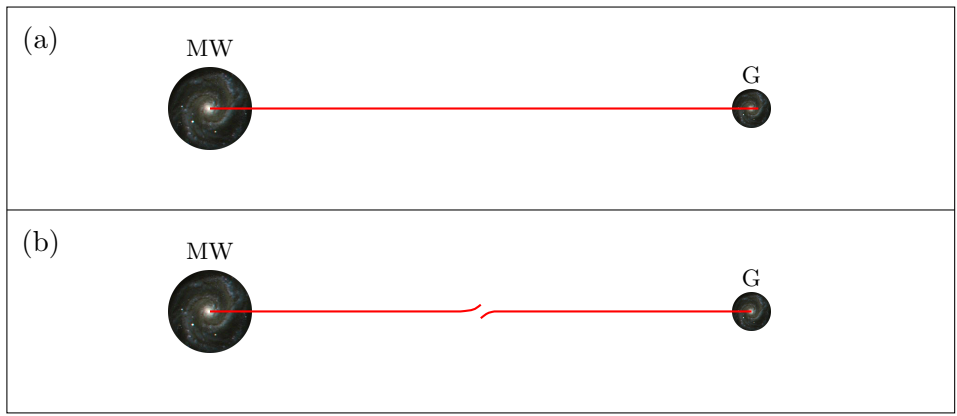

\begin{center}
\begin{tikzpicture}[scale=0.9]
\def\ss{3.0}
\node at (-4,0) {\includegraphics[width=1.1cm]{galaxy-small.png}};
\node at (-4,0.9) {\footnotesize MW};
\node at (4,0) {\includegraphics[width=0.5cm]{galaxy-small.png}};
\node at (4,0.5) {\footnotesize G};
\draw[red,thick] (-4,0) -- (-0.3,0) .. controls(-0.1,0.02) .. (0,0.1);
\draw[red,thick] (4,0) -- (0.2,0) .. controls (0.1,-0.02) .. (0,-0.1);
\node at (-6.5,0.5*\ss-0.5) {(b)}; 
\draw (-7,1.5*\ss) --  (7,1.5*\ss) -- (7,-0.5*\ss) -- (-7,-0.5*\ss) -- cycle;
\draw (-7,0.5*\ss) -- (7,0.5*\ss);

\node at (-6.5,1.5*\ss-0.5) {(a)}; 
\node at (-4,\ss) {\includegraphics[width=1.1cm]{galaxy-small.png}};
\node at (-4,\ss+0.9) {\footnotesize MW};
\node at (4,\ss) {\includegraphics[width=0.5cm]{galaxy-small.png}};
\node at (4,\ss+0.5) {\footnotesize G};
\draw[red,thick] (-4,\ss) -- (4.1,\ss);
\end{tikzpicture}

\caption{Setup for the attached galaxy problem: If a small (test particle) galaxy G is (a) initially kept at constant distance from our own galaxy MW by a thin rod, and (b) the rod is then broken, how will G move relative to us? }
\label{TetheredGalaxy}
\end{center}
\end{figure}

If you base your intuition on some notion of ``expanding space'', in particular the misguided notion that ``new space is created'' between the galaxies, you might assume that the galaxy will be quickly joining the Hubble flow as soon as the impediment, the rod, is removed. After all, there is the exact same amount of space between MW and G that would be between MW and a galaxy in the Hubble flow that happened to be at the present location of G. You would likely come to a similar conclusion if your mental picture involved cosmic expansion ``carrying away'' galaxies, just like an ordinary (viscous) fluid would carry away objects immersed in it.

But this is not what happens. Instead, the galaxy slowly accelerates -- away from us if dark energy dominates, and towards us if the ordinary attractive gravity of matter dominates. (The latter case clearly illustrates that it is not the flow itself that counts, but the acceleration -- otherwise, even in the latter case of attractive gravity, the galaxy would be carried away with the others.)

In this situation, too, our classical intuition about dynamics serves us well. In classical physics, the Hubble flow itself is irrelevant. The pattern of motion is a combination of forces (which cause accelerations) and initial conditions for locations and velocities. If we choose the moment where the rod is broken as our reference time, then clearly the Hubble flow is part of the initial conditions for all of the participating galaxies at that moment. Our recently attached galaxy G has different initial conditions. Its motion only changes as G accelerates. On small scales, that acceleration can be calculated classically, as we did when deriving the Friedmann equations in section \ref{FriedmannEqSection}, or the effect of cosmic expansion on a bound system in this section. On larger scales, somewhat more complex calculations following the prescriptions of general relativity are required.

What happens will depend on the dynamics involved, in other words: on the $\Omega$ parameters of the cosmos, a galaxy G in this position will join the Hubble flow, eventually ending up alongside the Hubble flow galaxies; in the general case, however, a galaxy G in such a situation will not join the Hubble flow.\footnote{The details depend on your definition of joining the Hubble flow, cf. \citenp{BarnesEtAl2006}.}

Returning to bound system one last time, it should be noted that there are (speculative) hypothetical scenarios that lead to much more dramatic consequences for bound systems in the future \citep{Caldwell2003}. As we saw in section \ref{Diluting}, the dilution behaviour of different kinds of matter content (including dark energy) depends crucially on the equation of state. In all cases of interest to standard cosmology, that equation has the form (\ref{LinearEOS}),
\be\tag{\ref{LinearEOS}}
p = w\cdot \rho c^2,
\ee
and we have seen in equation (\ref{DensityDilutionW}) that depending on the value of $w$, the corresponding density will change as
\be \tag{\ref{DensityDilutionW}}
\rho(t) = \rho(t_0)\cdot\left( \frac{a(t)}{a(t_0)}\right)^{-3(1+w)}.
\label{DensityDilutionRevisited}
\ee
On the other hand, both in the second-order Friedmann equation (\ref{SecondFriedmann}) and in the corresponding equation (\ref{HomBound}) for the additional force acting within a bound system due to homogeneous cosmological terms, the gravitational attraction is proportional to
\be
\rho + 3p/c^2.
\label{GravSourceBoundRip}
\ee
Under these circumstances, an ingredient with $w<-1$ would have a dramatic effect. By (\ref{GravSourceBoundRip}), such an ingredient, which has been dubbed {\em \Index{phantom energy}}, would exert a repulsive force on bound systems, similar to the influence of dark energy we have already explored in this section. But unlike dark energy, the density of this hypothetical ingredient would not remain constant, but grow larger with cosmic expansion, as per equation (\ref{DensityDilutionRevisited}). In consequence, the radii $r_{balance}$ where the repulsive force and a bound system's attractive force balance would grow ever smaller, and one after the other, bound systems on ever smaller scales would be ripped apart: galaxy clusters, galaxies, stars, planets, humans, atoms, atomic nuclei. Within a finite time, it turns out, the repulsive influence would become infinitely strong. The universe would end in a \Index{Big Rip}. At present, however, there is no evidence whatsoever that such phantom energy exists in our cosmos.
\index{bound systems!and cosmic expansion|)}\index{cosmic expansion!and bound systems|)}

\subsection{The fate of our own universe}
\label{Fate}
\begin{figure}[htbp]
\begin{center}
\includegraphics[width=0.9\textwidth]{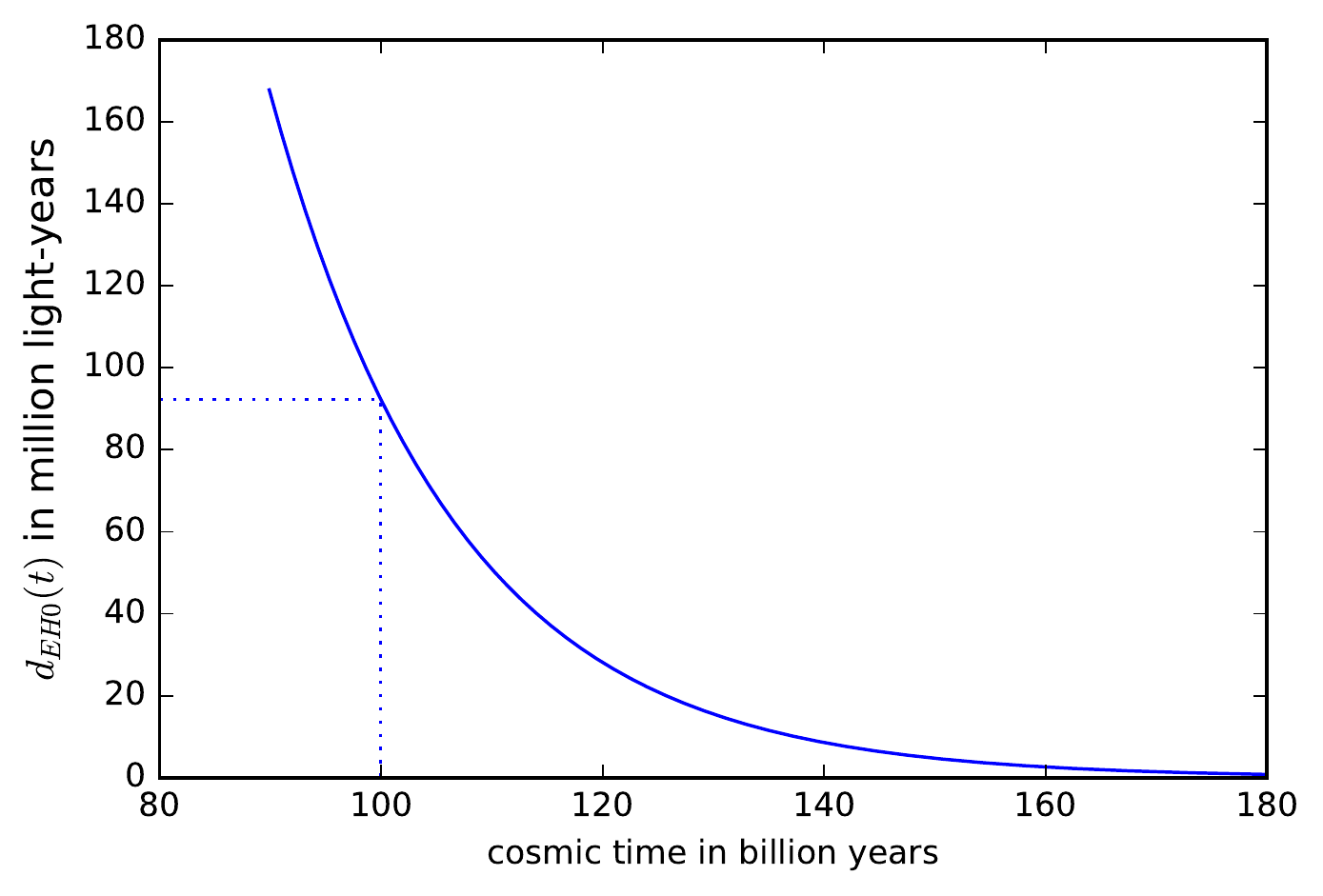}
\caption{The future cosmic event horizon at some future cosmic time $t$, expressed as a present-day proper distance $d_{EH,0}$. At each time, the plot shows
which present-day objects at a given proper distance will be beyond the cosmic event horizon at that future time}
\label{CosmicEHPlot}
\end{center}
\end{figure}

\index{fate (of the universe)|(}
Given that our own universe is currently dominated by dark energy, and will continue to be so dominated, expansion is set to continue indefinitely -- barring effects so unusual that they are not included in the current cosmological standard models. By that token, the event horizon for the de Sitter universe is relevant for our distant future, making for a lonely long-term fate. We can explore the specifics as follows (in a way similar to that of the article of \citenp{2000ApJ...531...22K}):

As a first step, let us extrapolate into the future, to the time $t_{100}$ when the scale factor will be 100 times larger than it is today. We can calculate that time using equation (\ref{CalculateTFromX}), namely as
\be
t_{100} = \frac{1}{H_0} \;\int\limits_0^{100} \frac{\Dd x'}{x'\;\sqrt{\Omega_Mx'{}^{-3} +\Omega_{\Lambda} }}.
\ee
Using the values given in table \ref{ParameterEstimates} on page \pageref{ParameterEstimates}, this can be evaluated numerically as
\be
t_{100} = \frac{6.4}{H_0} = 9\cdot 10^{11}\;\mbox{a}.
\ee
At that time, by the dimensionless Friedmann equation (\ref{FirstFriedmannSimple}), the Hubble constant will be
\be
H_{100}\approx H_0\cdot\sqrt{\Omega_{\Lambda}},
\ee
the matter term $\Omega_M$ being suppressed by a factor $10^{-6}$. By the same reasoning, scale factor evolution will be described in very good approximation by an exponentially expanding universe, equation (\ref{adS}), namely
\be
a(t) = a_{100}\cdot\exp[H_{100}(t-t_{100})] = 100\cdot a_0\cdot\exp[ H_0 \sqrt{\Omega_{\Lambda}} (t-t_{100})].
\label{Lateat}
\ee
On this basis, we can look at the cosmic event horizon. In equation (\ref{CosmicEHFormula}), we had written down the event horizon at the present time $t_0$. At any later time $t$, the event horizon, as a proper distance measured at that future time $t$, is
\be
d_{EH}(t) = c\cdot a(t)\cdot \int\limits^{\infty}_{t}\frac{\Dd t'}{a(t')}.
\ee
But what we are really interested in is the proper distance {\em as measured today} that value corresponds to. That will tell us which of the  objects in the present-day universe we will be able to see in the future, and which not. Since proper distances are directly proportional to the scale factor, the conversion is simple -- a factor $a(t_0)/a(t)$ will do the trick, so
\be
d_{EH,0}(t) = \frac{a_0}{a(t)}\,d_{EH}(t) = 
 c\cdot a_0\cdot \int\limits^{\infty}_{t}\frac{\Dd t'}{a(t')},
\ee
which is the event horizon at time $t$, expressed as a present-day proper distance. Restricting ourselves to $t>t_{100}$ and using (\ref{Lateat}) to describe cosmic evolution during that epoch, we find
\be
d_{EH,0}(t) = \frac{c}{100H_0\sqrt{\Omega_{\Lambda}}} \;\exp\left[ -H_0\sqrt{\Omega_{\Lambda}}(t-t_{100}) \right].
\ee
As time passes, the event horizon shrinks further and further, separating us from ever closer regions in the present-day universe. Figure \ref{CosmicEHPlot} shows a plot of $d_{EH,0}(t)$ against cosmic time. \index{horizon!event!of our own universe} \index{event horizon!our own universe}
This plot should be read as follows: Pick some future time $t$ on the x axis, say: a cosmic time of 100 billion years (as shown by the dotted lines). The value of the curve shows you that those galaxies in the Hubble flow that currently, at the present time, have a proper distance of 92 million light-years or more from us will be beyond our event horizon at a cosmic time of 100 billion years. 92 million light-years is not such a lot in present cosmic terms! Once the universe has reached about ten times its present age, most of what we see today will be inaccessible to us. 

Another consideration is that of the redshift experienced by light reaching us from various distant objects. With $a(t)$ of the form 
(\ref{Lateat}), the scale factor will grow by an additional factor 1000 every $1.2\cdot 10^{11}$ years or so. As distances and light travel times grow, there will be a point where even the most highly energetic light from a specific distant object in the Hubble flow will be shifted to wavelengths much too large to be detected by our most sensitive radio telescopes -- much sooner so for the cosmic microwave background, the thermal radiation reaching us from the end of the Big Bang phase that has already been shifted from visible light into the microwave range. 

In the long term, our universe will be a lonely place. Only the systems that are still bound to us, as per the considerations in section \ref{BoundSystems}, will still be in our cosmic vicinity, and visible to us. Astronomers of the far future will need to rely on ancient writings for their cosmological considerations -- the evidence, in the form of the cosmic background radiation, and of the distance-redshift relation, will have become inaccessible.

\index{fate (of the universe)|)}

\subsection{The generalized redshift-distance relation}
\label{RDRSection}
\index{redshift-distance relation!generalized|(}
Now that we have considered the past, evolution and future of the universe, let us return to our initial aim of understanding the distance-redshift relation. To that end, we take the Friedmann equation (\ref{FriedmannDimensionless}), revisit our light propagation formula 
 (\ref{LightPropagation}), apply it to light from distant sources reaching us at the present time $t_0$, and rewrite the result directly in terms of the basic parameters $H_0,\Omega_M,\Omega_R,\Omega_K$ and $\Omega_{\Lambda}$.
 
First, some conventions. Let us denote by $d_e$ the present-day proper distance of a light source whose light reaches us right now at time $t_0$ after leaving the light source at an earlier time $t_e$. Using this notation, our light propagation formula now reads
 \be
d_e = c\cdot a(t_0)\cdot \int\limits^{t_0}_{t_e}\frac{\Dd t}{a(t)}.
\ee
Assume that we are in a universe that is not re-collapsing. Then $a(t)$ is a monotonically increasing function of $t$, and thus can be inverted. We use this fact to rewrite the previous equation, now using $a\equiv a(t)$ as the independent variable, with particular values $a_e\equiv a(t_e)$ and $a_0\equiv a(t_0)$.  Applying this change of variable to the integral, the result is
\be
d_e = c\cdot a_0\cdot\int\limits^{a_0}_{a_e}\frac{\Dd a'}{a'{}^2 H(a')},
\label{LightPropagationa}
\ee
where $H(a')$ is the Hubble parameter, now rewritten in terms of the new independent variable $a'$. Next, for another change of variables. In a universe that is not re-collapsing, the redshift $z$ of light we receive from a distant object is directly related to the scale factor value at the time that light was emitted by the cosmological redshift relation (\ref{CosmologicalRedshift2}). 

For any object whose light we are receiving, the redshift $z$ we measure for that light and the value $a$ of the scale factor at time of emission are related as 
\be
z+1 = \frac{a_0}{a}. 
\ee
This provides the basic for our next change of variables, introducing $z$ as the new independent variable. The redshift $z=0$ corresponds to an object with distance zero, which is right were we are. In terms of this new variable, the light propagation formula (\ref{LightPropagationa}) now reads
\be
d_e(z) = c\cdot\int\limits^{z}_0\frac{\Dd z'}{H(z')},
\label{LightPropagationz}
\ee
with $H(z)$ the Hubble parameter, expressed in terms of the new independent variable --- all in all, the simplest version of this formula yet!

But we already know what $H(z)$ looks like; after all, the dimensionless Friedmann equation (\ref{FriedmannDimensionless}) presents us $H$ explicitly in terms of $x=1/(1+z)$, which is readily rewritten as
\be
H(z) = H_0\cdot\sqrt{ \Omega_M(1+z)^3+\Omega_R(1+z)^4+\Omega_{\Lambda}+\Omega_K(1+z)^2}. 
\ee
Substituting this into our light propagation formula (\ref{LightPropagationz}), we obtain the explicit relationship 
\be
d_e(z) = \frac{c}{H_0}\cdot\int\limits^{z}_0\frac{\Dd z'}{
 \sqrt{ \Omega_M(1+z')^3+\Omega_R(1+z')^4+\Omega_{\Lambda}+\Omega_K(1+z')^2}}
 \label{RedshiftDistance}
\ee
between the redshift $z$ of a light signal reaching us from an object that presently has proper distance $d_e$ from us.

This is the generalization of the approximate redshift-distance relation (\ref{RedshiftDistanceApproximate}) we derived in section
\ref{HubbleDoppler}. It takes into account all the subtleties of light propagation in an expanding universe -- the fact that we are looking into the past as well as the fact that the Hubble parameter is changing over (cosmic) time.

Since proper distance cannot be measured directly, it makes sense to express this in terms of the luminosity distance $d_L$ instead of $d_e$. We know how to get from one to the other using the relation between angular diameter distance $d_{ang}$ and luminosity distance as $d_L = d_{ang}(1+z)^2$ and the relation (\ref{AngularDistanceAndDt0}) between $d_{ang}$ and $d_e\equiv d(t_0)$. In the non-Euclidean case, that relation still contains the present-day value of the scale factor $a_0$; we can, however, replace $a_0$ by a function of $\Omega_K$ using the definition (\ref{OmegaK}) of that quantity. The result is
\be \label{angularDistanceZ}
d_{ang}(z) =\frac{1}{1+z}  \left\{    
\begin{array}{ll}
d_e(z) & \mbox{Euclidean space}\\[0.5em]
D_H/\sqrt{|\Omega_K|}\cdot \sin\left[\sqrt{|\Omega_K|}\cdot d_e(z)/D_H\right] & \mbox{spherical space}\\[0.5em]
D_H/\sqrt{|\Omega_K|}\cdot \sinh\left[\sqrt{|\Omega_K|}\cdot d_e(z)/D_H\right] & \mbox{hyperbolic space},
\end{array}
\right.
\ee
or 
\be
d_{L}(z) =(1+z) \cdot  \left\{    
\begin{array}{ll}
d_e(z) & \mbox{Euclidean space}\\[0.5em]
D_H/\sqrt{|\Omega_K|}\cdot \sin\left[\sqrt{|\Omega_K|}\cdot d_e(z)/D_H\right] & \mbox{spherical space}\\[0.5em]
D_H/\sqrt{|\Omega_K|}\cdot \sinh\left[\sqrt{|\Omega_K|}\cdot d_e(z)/D_H\right] & \mbox{hyperbolic space},
\end{array}
\right.
\ee
where $D_H=c/H_0$ is, once more, the Hubble distance. In the simplest case, that of Euclidean space geometry (which corresponds to $\Omega_K=0$), the relation is
\be
d_L(z) = D_H\cdot\int\limits^{z}_0\frac{\Dd z'}{
 \sqrt{ \Omega_M(1+z')^3+\Omega_R(1+z')^4+\Omega_{\Lambda}}}.
 \label{RedshiftLuminosityDistance}
\ee
For standard candles, per definition, $d_L$ can be determined. The redshift is readily measurable for many galaxies, as well. We have formulated a redshift-distance-relation which links the observable quantities $d_L$ and $z$ with the basic parameters that describe the expansion of our universe, $H_0$ and the $\Omega$s. 

\subsection{Comparing distances}

In the previous section, we have seen three different definitions for the distance of a far-away light source whose light is reaching us at the present time $t_0$;
the earlier section \ref{LightTravelDistance} provided a fourth, namely the light-travel distance. Now that we know how to calculate the cosmic scale factor $a(t)$ as a function of time, and have learned about the best current estimates for the Hubble constant and the $\Omega$ parameters in table \ref{ParameterEstimates}, we can see how those distances compare for objects of different redshifts. 

\begin{figure}[htbp]
\begin{center}
\includegraphics[width=0.8\textwidth]{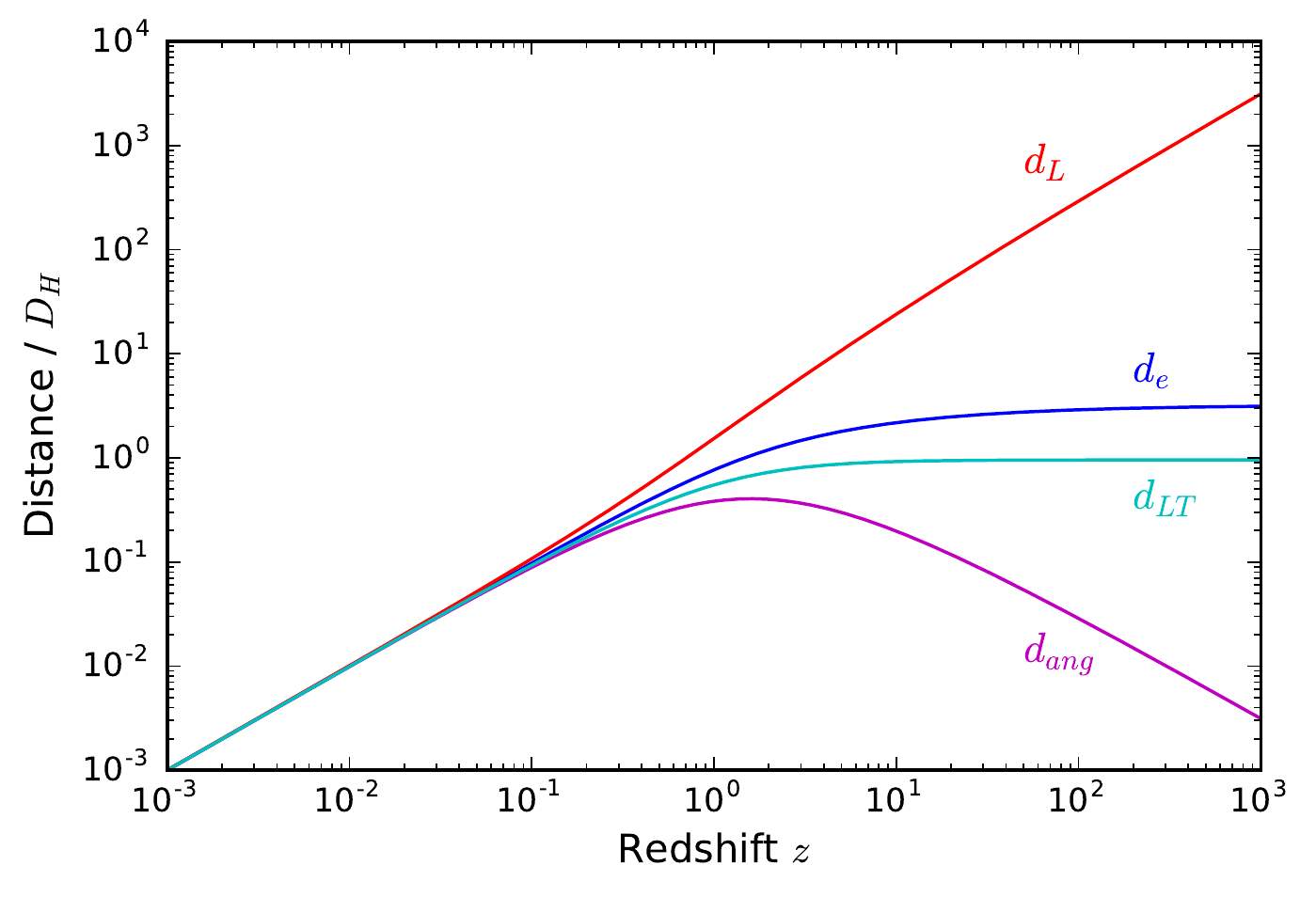}
\caption{Luminosity distance $d_L$, proper distance $d_e$, light-travel distance $d_{LT}$, and angular distance $d_{ang}$, all in terms of the Hubble distance $D_H$, compared for different redshifts $z$, for our own universe as described by the $\Omega$ parameters of table \ref{ParameterEstimates} }
\label{DistanceComparison}
\end{center}
\end{figure}

The distances in question, and their dependence on $z$, are
\begin{center}
\bgroup
\renewcommand{\arraystretch}{1.2}
\begin{tabular}{l|l|l}
$d_e(z)$ & proper distance & relation given by (\ref{RedshiftDistance})\\\hline
$d_{ang}(z)$ & angular distance & relation to $d_e$ given by (\ref{angularDistanceZ})\\\hline
$d_L(z)$ & luminosity distance & $d_L = (1+z)^2\,d_{ang}$\\\hline
$d_{LT}(z)$ & light-travel distance & relation given by (\ref{LightTravelTime})
\end{tabular}
\egroup
\end{center}
For our universe, the relation between these distances in terms of the Hubble distance is shown in figure \ref{DistanceComparison}. 
A handy tool for calculating specific values of these redshift-dependent distances is Ned Wright's Javascript Cosmology Calculator \citep{Wright2006} at
\begin{center}
\href{http://www.astro.ucla.edu/\~wright/CosmoCalc.html}{http://www.astro.ucla.edu/\~{}wright/CosmoCalc.html}
\end{center}
For small distance values, all these different distances coincide --- just as we would expect, given that at small scales, the effects of expansion and the deviation from an ordinary, special-relativistic universe should be small. At larger redshifts, the distances diverge. 

Of particular interest is the angular diameter distance $d_{ang}$, which does not grow monotonously with $z$. Instead, once we go to sufficiently high redshifts beyond about $z=1$, the angular diameter distance $d_{ang}$ shrinks with increasing $z$! There are two competing effects at work here. We see distant objects not as they are, but as they were in the past; in an expanding universe, any object in the Hubble flow was closer to us in the past than it is now. By that reasoning, objects in the past should appear to have a larger angular diameter. On the other hand, there is basic geometry, namely that objects that are farther away from us appear smaller, that is, have a smaller angular diameter. For objects at lower redshift, the latter effect dominates, while for more distant objects, the former effect is dominant, leading to the unusual non-monotonous behaviour seen in figure \ref{DistanceComparison}.

\section{The Redshift-distance diagram revisited}
\label{RedshiftDistanceRevisited}
After a long journey through the concepts and calculations associated with an expanding universe, we can return to our concrete goal of understanding Hubble diagrams, or more precisely: redshift-distance diagrams like our initial figure~\ref{InitialDiagram}, which play a key role in comparing cosmological models with astronomical observations.

Let us collect the ingredients we need to understand that kind of diagram. We will, throughout, assume that we are in a Euclidean universe, $\Omega_K=0$. As described in section \ref{Geometry}, that particular property of the universe is best derived using another type of observations, namely fluctuations in the cosmic microwave background.

Consider standard candles, per definition of known luminosity $L$, whose flux density $I$ is measured by telescope observations. Our aim is to deduce a systematic relationship between those two quantities and the redshift of the light we have received from a standard candle.

Once both $L$ and $I$ are known, we can deduce the object's luminosity distance $d_L$, using the formula
\be
\tag{\ref{LuminosityDistanceInverseSquare}}
I =\frac{L}{4\pi\,d_L^2}.
\ee
From our derivation in section \ref{LuminosityDistanceSection}, we know the relationship between luminosity distance and angular distance, and from the geometry section \ref{Geometry}, we know that in the special case of a Euclidean universe, angular distance is the same as the emitting object's proper distance $d_e$ from us, so that
\be
\tag{\ref{LuminosityDistance}}
d_L \equiv d_e\cdot (1+z_e),
\ee
with $z_e$ the redshift we measure as the distant object's light reaches us. In an expanding universe, $d_e$ and $z_e$ fulfill the redshift-distance relation (\ref{RedshiftDistance}). We neglect the radiation contribution $\Omega_R$ which is only of interest in the very earliest stages of cosmic history and whose absence will not markedly skew our picture; as already mentioned, we assume Euclidean space geometry, so $\Omega_K=0$, and we exploit the fact that for the remaining two 
density parameters, $\Omega_{\Lambda}=1-\Omega_M$ as per the present-day dimensionless Friedmann equation 
(\ref{FriedmannSimplePresent}). 

In consequence, for this special case, the redshift-distance relation takes on the much simpler form
\be
\tag{}
H_0 \cdot d_e = c\cdot\int\limits^{z_e}_0\frac{\Dd z}{
 \sqrt{ \Omega_M(1+z)^3-\Omega_M+1}}.
\ee
The only remaining complication is that astronomers will express the luminosity distance in terms of the distance modulus
\be
\tag{\ref{DistanceModulus}}
\mu(d_L) \equiv 5\log_{10}\left(\frac{d_L}{10\;\mbox{pc}}\right).
\ee
All in all, we expect distance modulus $\mu$ and redshift $z_e$ of a distant standard candle to be related by 
\be
\mu = 5\log_{10}\left[
\frac{c}{H_0}\cdot\frac{1}{\mbox{Mpc}}\cdot (1+z_e)\cdot\int\limits^{z_e}_0\frac{\Dd z}{
 \sqrt{ \Omega_M(1+z)^3-\Omega_M+1}}
\right] + 25.
\ee
In practice, for supernovae, all the quantities inside the logarithm that are not dimensionless are separated, and lumped together with the absolute magnitude; the distance modulus is then fitted to supernova data as
\be
\mu = 5\log_{10}\left[(1+z_e)\cdot\int\limits^{z_e}_0\frac{\Dd z}{
 \sqrt{ \Omega_M(1+z)^3-\Omega_M+1}}
\right] + C,
\label{FittedCurve}
\ee
with two fit parameters: the $\Omega_M$ we are interested in, and an offset $C$. Had we taken the more complicated formulae for more general geometries, we could even attempt to fit both $\Omega_M$ and  $\Omega_{\Lambda}$.

\begin{figure}[htbp]
\begin{center}
\includegraphics[width=0.9\textwidth]{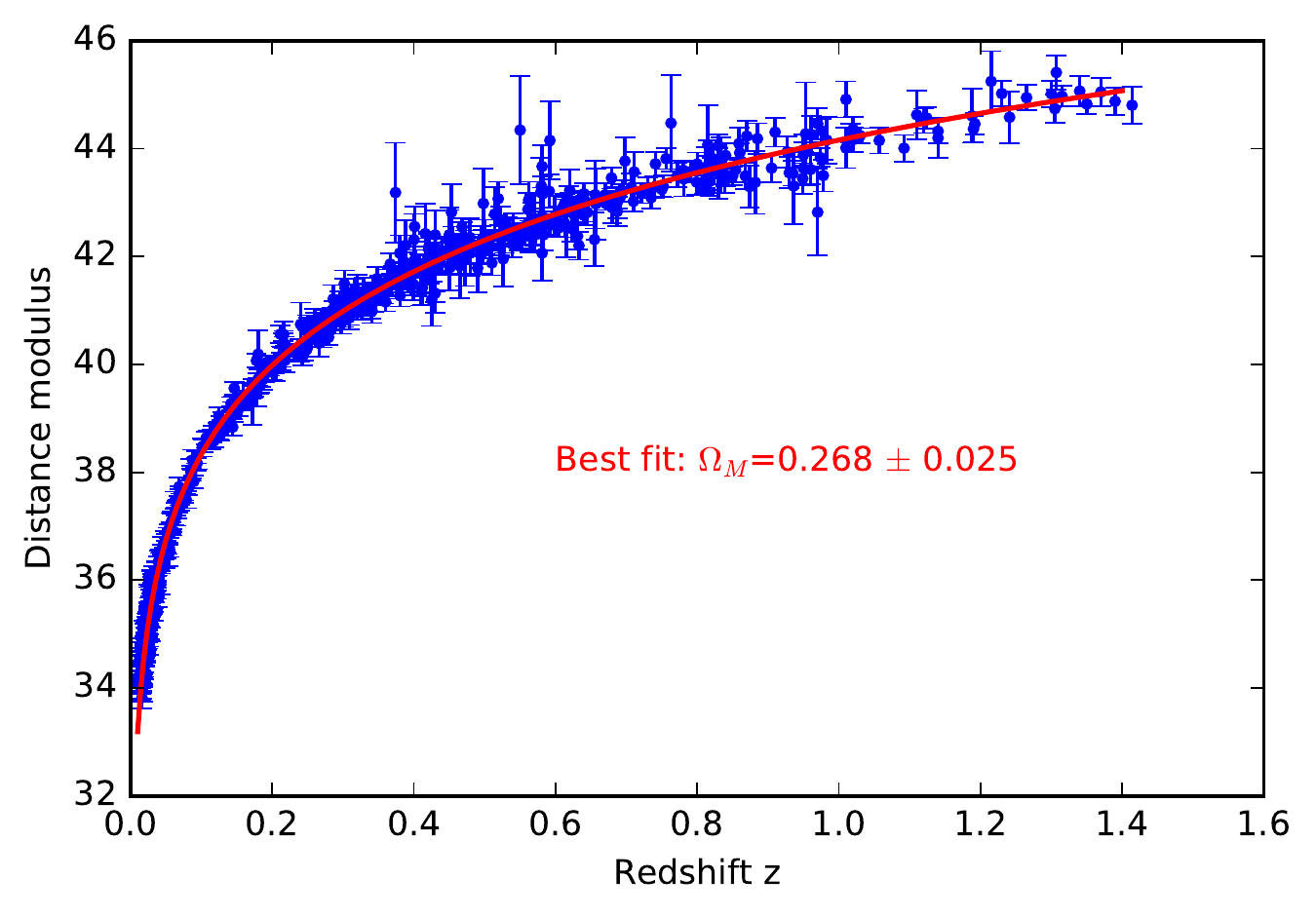}
\caption{Using supernova data to estimate the cosmological parameter $\Omega_M$. Data: Supernova Cosmology Project}
\label{ReconstructedFigure}
\end{center}
\end{figure}

Figure \ref{ReconstructedFigure} is our own version of the initial figure \ref{InitialDiagram}, using the same data from the \Index{Union 2.1 catalogue} of the \Index{Supernova Cosmology Project}, and fitting the function (\ref{FittedCurve}): \index{parameter values, cosmological}
The link for downloading the data and the Python script for fitting, and for producing figure \ref{ReconstructedFigure}, can be found in section \ref{FittingScript}, p.\ \pageref{FittingScript}f. 
The fitted value of $\Omega_M=0.268\pm 0.025$ is compatible with the result of the team that published the data,
$\Omega_M=0.271\pm 0.014$ \citep{SuzukiEtAl2012}. 
\index{redshift-distance relation!generalized|)}
\section{Conclusions}

While modern cosmological models are rooted in the framework of general relativity, at least the simplified, homogeneous and isotropic expanding universe can be understood without recourse to the full formalism of Einstein's theory. We have derived most of the formulae describing light propagation and the dynamics of the universal scale factor by going to small scales, making use of the equivalence principle and the Newtonian limit, taking care to arrive at our results in a way that allows generalisation to the universe as a whole -- either by summing up infinitesimal contributions or because any exact local result for the functional form of the scale factor $a(t)$ automatically applies everywhere the scale factor is involved.

The results have shown us a number of fundamental properties of FLRW universes. In particular, they have allowed us to understand one of the fundamental relations linking the FLRW models with astronomical data, to wit, the redshift-distance relation. 

The scope of this lecture was limited in several respects, and there are several interesting directions in which the results presented here can be extended. We have seen that our expanding universe had a hot, dense Big Bang phase; understanding the basic thermodynamics of the early universe, including the creation of the first light elements and the role of the cosmic background radiation, is an important part of understanding modern cosmology. 

While, in this lecture, we have considered only simplified, perfectly homogeneous and isotropic models in this lecture, the universe around us is inhomogeneous in many interesting ways. An understanding of the origin of the large-scale structure in the universe is yet another building block of modern cosmology. Last but not least, there are aspects of the homogeneous universe we have neglected, for instance the number statistics of distant objects as a function of distance.

Overall, I hope to have demonstrated that it is possible to understand key elements of cosmology using not much more than high-school level mathematics and fairly basic physics, supplemented with some facts from general relativity. Given that science is universal, I believe it is of paramount importance to make the results of cutting-edge research -- in particular in those areas that exert considerable fascination on the general public -- accessible as widely as possible.

\iftoggle{arxiv}{

}{
\printbibliography
}

\printindex

\begin{appendix}
\section{Fitting the supernova data}
\label{FittingScript}
The data used to plot figure \ref{ReconstructedFigure} is from the Supernova Cosmology Project, namely the Union 2.1 catalogue available for download at
\begin{center}
\href{http://supernova.lbl.gov/union/figures/SCPUnion2.1_mu\_vs_z.txt}{http://supernova.lbl.gov/union/figures/SCPUnion2.1\_mu\_vs\_z.txt}
\end{center}

\begin{lstlisting}
#!/usr/bin/env python2
# -*- coding: utf-8 -*-
# created by Markus Poessel, November 2, 2017

import numpy as np
import matplotlib.pyplot as plt
from scipy.integrate import quad
from scipy.optimize import curve_fit

# Union.2.1 catalogue
# File from 
# http://supernova.lbl.gov/union/figures/SCPUnion2.1_mu_vs_z.txt
# An ASCII table with tab-separated columns: Supernova Name, Redshift, 
# Distance Modulus, and Distance Modulus Error. For Union 2.1, there is 
# an additional column for the probability that the supernova was 
# hosted by a low-mass galaxy. 
snz,snmu,dsnmu=np.genfromtxt("SCPUnion2.1_mu_vs_z.txt",
                             usecols=(1,2,3),unpack="True")

# Argument for the integral in the distance modulus function    
def intArgument(zval,OmegaM):
    return 1/np.sqrt(OmegaM*(1+zval)**3-OmegaM+1)

# Computing the distance modulus as a function of redshift z, 
#given a value for OmegaM and an offset CC
def mu(zval,OmegaM,CC):     
    return np.array([5*np.log10((1+zz)*quad(intArgument,0,zz,
                                            args=(OmegaM))[0]) 
                                         + CC for zz in zval])
    
# Fit the redshift-distance curve to the data:
popt, pcov = curve_fit(mu, snz,snmu)
perr=np.sqrt(np.diag(pcov))
    
# Plot both the data and the fitted curve:
zVals = np.linspace(0.01,1.4,1000)
muVals = mu(zVals,popt[0],popt[1])
plt.clf()
plt.errorbar(snz,snmu,dsnmu,fmt="b.",zorder=0)
plt.plot(zVals,muVals,"r",lw=1.5,zorder=1)
plt.text(0.6,38,"Best fit: $\Omega_M$=
 %.3f $\pm$ %.3f" % (popt[0],perr[0]), color="r")
plt.ylabel("Distance modulus")
plt.xlabel("Redshift z")
plt.savefig("sn_fit.pdf",bbox_inches="tight")
\end{lstlisting}

\section{What we need from special relativity}
\label{SpecialRelativity}
\index{special relativity!summarized|(}
In order to make these lecture notes more accessible, I briefly summarize here a number of statements from special relativity that are used in the main text. It should be clear, however, that this brief version can be no more than a crutch --- the best option for interested readers not familiar with special relativity is to change that state of affairs by reading books such as \citenp{Mermin2005}. In the main text, the equivalence principle, cf.\ section \ref{EquivalencePrincipleSection}, plays an important role. In brief, that principle states that for an observer in free fall in a gravitational field described by Einstein's general theory of relativity, the laws of physics are (approximately) the same as those of special relativity. So what are the pertinent laws of special relativity?

\subsection{Inertial reference frames}

For the usual description of moving particles (and more complex objects), you need a frame of reference. \index{reference frame} Most commonly, you describe particle positions by introducing a Cartesian coordinate system, \index{Cartesian coordinates} defined in just the right way so that coordinate distances correspond to physical distances, as measured with a ruler or more modern measuring device. In classical physics, one usually does not think too much about introducing a time coordinate. Time is a parameter that describes how change happens within your system; ideal clocks divide time up into unit intervals, one second the same as the next, one hour the same as any other. (The latter is easily said; linking it to actual measurements is tricky, though --- how can you tell that a clock is ideal?)

In some reference frames, if you let a particle drift, making sure that none of the known forces such as gravity or electromagnetism exert any significant influence, such free particles will nevertheless be seen to accelerate, and in a systematic way: If you track their motions, you will be able to separate acceleration into a component that acts on all particles in the same way, a component that depends on the distance from a certain line in space (the rotation axis), a component that depends on the velocity component orthogonal to that same line, and a component orthogonal both to the line and to the particle's direction toward that line. These are linear, centrifugal, Coriolis, and Euler acceleration, respectively, jointly known as {\em inertial accelerations}. \index{inertial acceleration}

A reference frame in which these inertial accelerations are absent, and free particles move at constant speeds along straight lines, is called an {\em inertial frame}. \index{inertial frame} These are the most common frames used in classical mechanics, even though a reference frame at rest relative to the Earth's surface -- to name a very common example -- is only approximately inertial, seeing that the Earth itself is a rotating reference frame.

The nomenclature regarding inertial frames or, more general, reference frames, varies; the terms ``frame,'' ``reference frame,'' ``frame of reference,'' ``observer'' and ``coordinate system'' are used fairly interchangeably. \index{reference frame} \index{frame of reference} \index{observer!in special relativity} \index{coordinate system!in special relativity} The rationale, presumably, is that each observer will have defined a personal reference frame (that is, a frame in which the observer is at rest) to record their observations, and that descriptions within a given reference frame are made in terms of space and time coordinates. The elementary concept in such a setting is an {\em event}, defined as something identifiable happening at a specific location in space, at a specific point in time. (And yes, this is an idealization --- just like the concept of a space point is an idealization.)

An inertial frame in which an object is at rest is often called that object's {\em rest frame}. \index{rest frame}

\subsection{Relativity principle and constant light speed}
\label{Simultaneity}

Einstein's special theory of relativity posits that there is such a thing as inertial frames, and that the following two principles hold:
\begin{enumerate}
\item {\bf Principle of relativity}: the laws of physics have the same simple form in every inertial frame. \index{principle of relativity} \index{relativity principle}
\item {\bf Constancy of the speed of light}: the speed of light in vacuum has the same value in each inertial frame. In particular, it does not depend on the motion of the light source. \index{speed of light!constancy}
\end{enumerate}
Inertial frames can be in constant motion relative to each other; the first principle states that, when it comes to the laws of physics, no inertial frame is distinguished from any other. This is in contrast to earlier notions of an absolute space, which set a standard of absolute rest. 

The expression ``the same simple form'' refers to physical laws linking specific quantities. Assume, for instance, that the energy $E$ and momentum $p$ of a massless particle are linked as $E=pc$, with $c$ the (vacuum) speed of light, in one inertial frame, that is: with the energy $E$ and momentum $p$ measured in that specific frame. Then the same form holds in any other inertial frame, as well: the energy $E'$ and momentum $p'$ as measured in such another inertial frame are, again, linked, as $E'=p'c$. The energy-momentum-relation has the same mathematical form in all inertial frames.

A direct consequence of the principle of relativity is that, if you set up a particular experiment in one inertial frame, and then set up the same experiment in another inertial frame, the results will be the same. There is no way of distinguishing between inertial frames using local experiments --- that is, experiments that are defined solely within an inertial frame, and do not include any outside influences. (A counter-example would be two different inertial frames in one and the same external magnetic field; unless the magnetic field were to have exactly the same properties in both inertial frames, local magnetic experiments in each frame would not be equivalent.)

We will make use of the constancy of the speed of light when we describe light propagation in an expanding universe in section \ref{LightPropagationSection}.

The constancy of the speed of light is directly linked to notions of {\em simultaneity}. \index{simultaneity} Einstein's definition of simultaneity of two events is as follows. Imagine an observer who is at rest in the chosen reference frame $S$, and who is located half-way between the locations of two events (in other words: if we join the two locations by a straight line, then the distance between our observer and the first event location will be the same as the observer's distance to the second event location, with all distances measured in the inertial frame $S$). Now imagine that directly at each event, a light signal is sent from the event location towards our observer. The two events in question are simultaneous as defined in the inertial frame $S$ if and only if these two light signals arrive at the half-way observer at the same time. It is a key property of special relativity that with this definition, simultaneity is relative: \index{simultaneity!relativity of}\index{relativity!of simultaneity} two events that are simultaneous in one inertial system $S$ are not, in general, simultaneous in another inertial system $S'$ if those two systems are in relative motion.

With an (ideal) master clock, at rest in a specific inertial frame, and specific light signals, we can chart that reference frame's spacetime geometry: We can measure distances by the travel time of a reflected light signal (as in radar distance ranging). And we can find the time coordinate value (vulgo ``the time'') to assign to an event as follows, in a variation of the above definition of simultaneity: if a light signal leaves our master clock at $t=T_1$, reaches the location of the event $\cal E$ at just the time the event is happening, is reflected and arrives back at the master clock at time $t=T_2$, then the event $\cal E$ was happening at time
\be
t ({\cal E})= \frac{T_1+T_2}{2}.
\ee
In this way, one can reconstruct the whole of the geometry of space and time, as defined by special relativity \citep{Mermin2005a,Liebscher2005}. This approach, using radar coordinates and what has become known as the $k$-calculus, was pioneered by Hermann Bondi \citep{Bondi1964}.

\subsection{Transformations}
\label{LorentzTransformation}

Clearly, the constancy of the speed of light is incompatible with classical mechanic's recipe of relating two reference frames in relative motion. In classical mechanics, relative velocities add up directly. If a spaceship is chasing a light signal, and if we, in our reference frame, see the light signal moving at speed $c$ and the spaceship at speed $v$, both in the same direction, we would deduce that, for an observer in the spaceship, the light signal would move at the reduced speed $c-v$. This is contradicted by the principle of the constancy of the speed of light. Clearly, the transition between one inertial reference frame and another is somewhat more complicated in special relativity than in classical mechanics.

We will not go into detail about the {\em \Index{Lorentz transformations}}, that is, about the formulae defining the transition from one inertial frame to another in special relativity. It is worth mentioning, however, that they link space and time in such a way that it makes more sense to talk about the set of all events in space and time, in short: about four-dimensional spacetime. 

There is a helpful analogy that makes use of our usual notion of space. We can introduce an x, y and z axis in order to describe three-dimensional space. If we rotate the resulting coordinate system, space will still be the same, but the x, y and z directions are now different. Thus, it makes no sense to take the division into x, y and z too seriously -- much more important is the underlying three-dimensional structure. In the same way, the Lorentz transformations show how different inertial observers (namely inertial observers in relative motion) define the time direction and space differently -- much more important is the underlying four-dimensional structure of spacetime, not so much its observer-dependant slicings into space and time. And while the main text does not go into any detail at that point, these observer-dependent ways of slicing spacetime into space and time are intimately connected to the different kinds of space geometry discussed for an expanding universe in section \ref{Geometry}.

\subsection{Time dilation}
\label{TimeDilation}
 \index{time dilation!special relativity}
 
One consequence of special relativity that is of interest for the cosmic time coordinate defined in section \ref{CosmicTime} is {\em time dilation}. In short, any inertial observer will judge a moving clock to run more slowly than the clocks at rest in that observer's inertial frame.

\begin{figure}[htbp]
\begin{center}
\begin{tikzpicture}
\def\ss{4}
\def\marm{0.2}
\begin{scope}
\begin{scope}[shift={(-4,+1.2)}]
\fill[magenta!80!white] (-1,-0.5) -- (-1,0.5) -- (1,0.5) -- (1,-0.5) -- cycle;
\fill[magenta!10!white] (-1+\marm,-0.5+\marm) -- (-1+\marm,0.5-\marm) -- (1-\marm,0.5-\marm) -- (1-\marm,-0.5+\marm) -- cycle;
\node at (1-0.2,0) [anchor=east] {\large\tt 26.2};
\node at (0,-0.8) [anchor = center] {C1};
\end{scope}
\begin{scope}[shift={(-4,+2.4)}]
\fill[cyan!80!white] (-1,-0.5) -- (-1,0.5) -- (1,0.5) -- (1,-0.5) -- cycle;
\fill[cyan!10!white] (-1+\marm,-0.5+\marm) -- (-1+\marm,0.5-\marm) -- (1-\marm,0.5-\marm) -- (1-\marm,-0.5+\marm) -- cycle;
\node at (1-0.2,0) [anchor=east] {\large\tt 8.1};
\node at (0,+0.8) [anchor = center] {D};
\draw[->,ultra thick] (1.1,0) -- (2.2,0);
\node at (1.6,0.3) {$v$};
\end{scope}
\begin{scope}[shift={(-0,+1.2)}]
\fill[magenta!80!white] (-1,-0.5) -- (-1,0.5) -- (1,0.5) -- (1,-0.5) -- cycle;
\fill[magenta!10!white] (-1+\marm,-0.5+\marm) -- (-1+\marm,0.5-\marm) -- (1-\marm,0.5-\marm) -- (1-\marm,-0.5+\marm) -- cycle;
\node at (1-0.2,0) [anchor=east] {\large\tt 26.2};
\node at (0,-0.8) [anchor = center] {C2};
\end{scope}
\begin{scope}[shift={(4,+1.2)}]
\fill[magenta!80!white] (-1,-0.5) -- (-1,0.5) -- (1,0.5) -- (1,-0.5) -- cycle;
\fill[magenta!10!white] (-1+\marm,-0.5+\marm) -- (-1+\marm,0.5-\marm) -- (1-\marm,0.5-\marm) -- (1-\marm,-0.5+\marm) -- cycle;
\node at (1-0.2,0) [anchor=east] {\large\tt 26.2};
\node at (0,-0.8) [anchor = center] {C3};
\end{scope}
\end{scope}
\begin{scope}[shift={(0,-\ss)}]
\begin{scope}[shift={(-4,+1.2)}]
\fill[magenta!80!white] (-1,-0.5) -- (-1,0.5) -- (1,0.5) -- (1,-0.5) -- cycle;
\fill[magenta!10!white] (-1+\marm,-0.5+\marm) -- (-1+\marm,0.5-\marm) -- (1-\marm,0.5-\marm) -- (1-\marm,-0.5+\marm) -- cycle;
\node at (1-0.2,0) [anchor=east] {\large\tt 27.2};
\node at (0,-0.8) [anchor = center] {C1};
\end{scope}
\begin{scope}[shift={(4,+2.4)}]
\fill[cyan!80!white] (-1,-0.5) -- (-1,0.5) -- (1,0.5) -- (1,-0.5) -- cycle;
\fill[cyan!10!white] (-1+\marm,-0.5+\marm) -- (-1+\marm,0.5-\marm) -- (1-\marm,0.5-\marm) -- (1-\marm,-0.5+\marm) -- cycle;
\node at (1-0.2,0) [anchor=east] {\large\tt 8.6};
\node at (0,+0.8) [anchor = center] {D};
\draw[->,ultra thick] (1.1,0) -- (2.2,0);
\node at (1.6,0.3) {$v$};
\end{scope}
\begin{scope}[shift={(-0,+1.2)}]
\fill[magenta!80!white] (-1,-0.5) -- (-1,0.5) -- (1,0.5) -- (1,-0.5) -- cycle;
\fill[magenta!10!white] (-1+\marm,-0.5+\marm) -- (-1+\marm,0.5-\marm) -- (1-\marm,0.5-\marm) -- (1-\marm,-0.5+\marm) -- cycle;
\node at (1-0.2,0) [anchor=east] {\large\tt 27.2};
\node at (0,-0.8) [anchor = center] {C2};
\end{scope}
\begin{scope}[shift={(4,+1.2)}]
\fill[magenta!80!white] (-1,-0.5) -- (-1,0.5) -- (1,0.5) -- (1,-0.5) -- cycle;
\fill[magenta!10!white] (-1+\marm,-0.5+\marm) -- (-1+\marm,0.5-\marm) -- (1-\marm,0.5-\marm) -- (1-\marm,-0.5+\marm) -- cycle;
\node at (1-0.2,0) [anchor=east] {\large\tt 27.2};
\node at (0,-0.8) [anchor = center] {C3};
\end{scope}
\end{scope}
\draw (-7,\ss) -- (7,\ss) -- (7,-\ss) -- (-7,-\ss) -- cycle;
\draw (-7,0) -- (7,0);
\node at (-7+0.5,\ss-0.5) {(a)};
\node at (-7+0.5,-0.5) {(b)};
\end{tikzpicture}
\caption{Time dilation for a moving clock. Panel (a) shows matters as they stand at time $t=26.2$ seconds as determined in system $S$; panel (b) shows the moment $t=27.2$ seconds, also as determined in system $S$}
\label{TimeDilationFigure}
\end{center}
\end{figure}
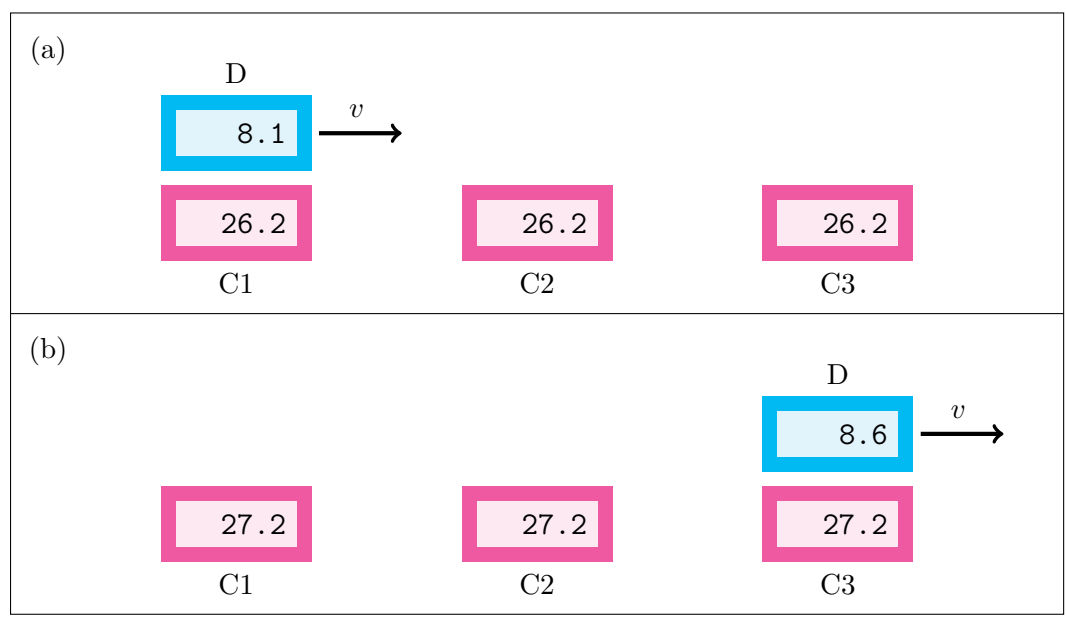

We consider the situation in more detail, as shown in figure \ref{TimeDilationFigure}: Our inertial observer has placed three clocks C1,C2,C3 in his inertial frame $S$, all of which are synchronized, their hands indicating the system's time coordinate $t$. These clocks can be seen, in a row, near the bottom of panel \ref{TimeDilationFigure} (a), and naturally they all show the same time, in this case $t=26.2$ seconds. For simplicity, we assume that those clocks do not follow our somewhat cumbersome everyday division of time into seconds, minutes, hours, days and so on, but merely count seconds.

Now consider an additional clock D, which is moving at some constant speed $v$ relative to the inertial frame $S$. The panel
\ref{TimeDilationFigure} (a) captures the moment where the moving clock D is just passing the first clock C1. Assume that, at this particular moment, the moving clock D is showing the time $8.1$ seconds on its display.

The situation one second later (as determined in the inertial system $S$) is shown in the panel \ref{TimeDilationFigure} (b). Apparently, clock placement happens to be such that one second was all it took for clock D to move from the location of C1 to that of C3, so we see the moment where D passes C3. The fact that, in the system $S$, one second has passed, is dutifully recorded by the three clocks at rest in $S$, which now show $t=27.2$ seconds.

But when we compare D with the clock C3 (easily done since both are at the same location at that particular moment), we see that D is now showing $8.7$ seconds. On clock D, only half a second has passed between the two panels. In this sense, clock D is running more slowly than clocks C1, C2 and C3. In our specific example, D has been running only half as fast as C1, C2,C3. 

For this situation, the relation between a time interval $\Delta t$ as shown by clocks that are synchronized in $S$ and the time interval $\Delta t'$ shown by the moving clock D, which is moving at speed $v$ relative to the inertial system $S$, is given by
\be
\Delta t' = \Delta t\cdot\sqrt{1-(v/c)^2},
\ee
where $c$ is the speed of light (in vacuum). From the fact that, in our example, $\Delta t' = 0.5\cdot\Delta t$, we can deduce that clock D must have been moving at the speed $v= \sqrt{3/4}\, c \approx 0.866\,c. $

A key result of special relativity is that {\em time dilation is mutual}. We can turn the tables: Clock D is at rest in a suitable inertial system, which we can call $S'$. In this system, we can introduce additional clocks D2, D3, which are synchronized to the system's time coordinate $t'$. We can line the three synchronized clocks up in a row so that each of C1, C2, C3 passes by D, D2 and D3. In that situation, a direct comparison of each of the C clocks with the D clocks along the line will give the result that, compared with a row of D clocks, the C clock is going more slowly! 

How can each set of clocks be slower than the other? This is resolved once on realizes that each judgement involves a standard of comparison, namely a notion of simultaneity --- and that simultaneity is not the same in the rest frame of the C clocks and the D clocks. In fact, there is a simple analogy using the geometry of the two-dimensional plane that shows how the apparent contradiction is resolved \citep{Poessel2010a,Poessel2010b}.

In the context of cosmological models, the fact that clocks in relative motion measure time differently is mainly important as a warning, showing us that the cosmic time coordinate introduced in section \ref{CosmicTime} is unusual, and certainly not the usual time coordinate used in, say, classical mechanics. 

\subsection{Doppler effect}
\label{SRDoppler}

The basics of the Doppler effect are well-known from classical mechanics. From the point of view of an inertial observer, the maxima and minima of a light wave will move at the speed of light; for a light source moving away from such an observer, this makes for longer wavelengths than for a  source that is at rest relative to the observer.

For example, follow the emission of one full wavetrain, from one crest to the next, by a source moving away from the observer at constant speed $v$. Assume that it takes the time $T$ for a stationary light source $L$ to emit one such wavetrain towards an observer $O$, from crest to crest. In that time, the first crest we observe has moved from the location of the source $P$ to the location $Q$. Since each crest moves at speed $c$, this corresponds to a distance of $cT$ between one crest and the next, in other words: to a \Index{wavelength} $\lambda_0 = cT$:
\begin{center}
\begin{tikzpicture}
\def\wwt{0.9}
\def\hht{0.35}
\def\mmt{0.7}
\fill[white!80!black] (0,-\hht) -- (0,\hht) -- (\wwt,\hht) -- (\wwt,-\hht) -- cycle;
\draw (0,-\hht) -- (0,\hht) -- (\wwt,\hht) -- (\wwt,-\hht) -- cycle;
\draw[yellow!80!red, ultra thick, domain=0:-3,samples=200] plot (\x, {0.2*sin(360*\x+90)});
\draw[<-,yellow!80!red,ultra thick] (-3.7,0) -- (-3.2,0);
\draw[dotted,thick] (0,0) -- (-3,0);
\draw[dotted,thick] (0,-1.2) -- (0,1);
\draw[dotted,thick] (-1,-1.2) -- (-1,1);
\draw[<->,thick] (0,-0.7) -- (-1,-0.7);
\fill (-7,0) circle (1.5pt);
\node at (-7.5,0) [anchor=west] {\footnotesize $O$};
\node at (-0.5,-1.0) {\scriptsize $cT$};
\node at (0,1.2)  {\footnotesize $P$};
\node at (-1,1.2) {\footnotesize $Q$};
\node at (0.45,0) {\footnotesize $L$};
\path (\mmt+\wwt,0) -- (\mmt+\wwt,-1);
\end{tikzpicture}
\end{center}
Next, imagine the same situation, except that the light source $L$ is moving away from the observer $O$ at a speed $v$. After the time $T$, again, the first crest we consider has moved the distance $cT$ to the left from the emission location $P$. But during the same time, the light source $L$ has moved the distance $vT$ to the right, to the location $R$! All in all, we have the following situation:
\begin{center}
\begin{tikzpicture}
\def\wwt{0.9}
\def\hht{0.35}
\def\mmt{0.7}
\fill[white!80!black] (\mmt,-\hht) -- (\mmt,\hht) -- (\mmt+\wwt,\hht) -- (\mmt+\wwt,-\hht) -- cycle;
\draw (\mmt,-\hht) -- (\mmt,\hht) -- (\mmt+\wwt,\hht) -- (\mmt+\wwt,-\hht) -- cycle;
\draw[yellow!80!red, ultra thick, domain=\mmt:-3,samples=200] plot (\x, {0.2*sin(360/(1+\mmt)*(\x-\mmt)+90)});
\draw[<-,yellow!80!red,ultra thick] (-3.7,0) -- (-3.2,0);
\draw[dotted,thick] (0,0) -- (-3,0);
\draw[dotted,thick] (0,-1.2) -- (0,1);
\draw[dotted,thick] (-1,-1.7) -- (-1,1);
\draw[<->,thick] (0,-0.7) -- (-1,-0.7);
\draw[dotted,thick] (\mmt,-1.7) -- (\mmt,1);
\fill (-7,0) circle (1.5pt);
\node at (-7.5,0) [anchor=west] {\footnotesize $O$};
\node at (-0.5,-1.0) {\scriptsize $cT$};
\node at (0,1.2)  {\footnotesize $P$};
\node at (-1,1.2) {\footnotesize $Q$};
\node at (\mmt,1.2) {\footnotesize $R$};
\node at (\mmt+0.45,0) {\footnotesize $L$};
\draw[<->,thick] (0,-0.7) -- (\mmt,-0.7);
\node at (0.5*\mmt,-1.0) {\scriptsize $vT$};
\draw[<->,thick] (-1,-1.5) -- (\mmt,-1.5);
\node at ({1-0.5*(\mmt+1)-0.3},-1.8)[anchor = center] {\scriptsize $(c+v)T$};
\end{tikzpicture}
\end{center}
In this case, the wavelength evidently is $\lambda = (c+v)T$. Compared with the wavelength of light from a moving source, there has been a wavelength shift
\be
\lambda = \left(1+ \frac{v}{c}\right)\cdot \lambda_0 \equiv (1+z)\cdot \lambda_0,
\ee
where the rightmost expression introduces the wavelength shift $z$, given by  \index{z (wavelength shift)}
\be
z \equiv \frac{\lambda - \lambda_0}{\lambda_0},
\ee
which in our case is 
\be
z = \frac{v}{c}.
\label{ClassicDoppler}
\ee
As we will see below, this is the {\em classical Doppler formula}. \index{Doppler effect!classical}

For $v>0$, that is, for a light source moving away from us, $z>0$, and all wavelengths get transformed into longer wavelengths. This is called a {\em redshift} since for visible light it corresponds to a shift towards the red end of the spectrum.\footnote{The ``for visible light'' is important in this definition. Note that, for instance, infrared light is definitely {\em not} shifted towards red visible light by a redshift; infrared light, too, is shifted towards longer wavelengths, away from the red visible portion of the spectrum. Just use ``redshift'' as a synonym for ``shift towards longer wavelengths,'' and all should be fine.}

For $v<0$, one can make a similar argument, but this time the light source would be moving towards the observer. We would find $z<0$, so all wavelengths get shorter. This is called a {\em blueshift}, since for visible light it corresponds to a shift towards the blue end of the spectrum. 

But there is still one factor we have left out. Our time $T$, corresponding to the period of the light, was measured in our own reference frame. But as recounted in section \ref{TimeDilation}, time dilation means that the period $T'$ determined by an observer on the light source will actually be shorter, namely
\be
T' = T\cdot\sqrt{1-(v/c)^2}.
\label{TimeDilationRevisited}
\ee
The corresponding wavelength of the light, as measured by an observer at rest relative to the light source, will be $\lambda' = cT'$.  In classical physics, where all speeds are much slower than that of light, $v\ll c$, this effect is negligible, and we do end up with the classical Doppler formula (\ref{ClassicDoppler}). But once we take into account (\ref{TimeDilationRevisited}), we find that
\be
\lambda = (c+v)\cdot T = \frac{(c+v)\;T'}{\sqrt{1-(v/c)^2}} = \frac{(1+v/c) \;cT'}{\sqrt{(1+v/c)(1-v/c)}} = \sqrt{\frac{1+v/c}{1-v/c}}\; \lambda'
\ee
corresponding to
\be
z = \frac{\lambda-\lambda'}{\lambda'} =  \sqrt{\frac{1+v/c}{1-v/c}} -1.
\ee
This is the (longitudinal) special-relativistic Doppler effect. \index{Doppler effect!special-relativistic} Note that, as the speed of the source approaches the speed of light, $v\to c$, the wavelength shift $z$ grows beyond all bounds. 

There is also a transverse special-relativistic Doppler effect, although that will not play a role in our cosmological considerations. In case a light source is moving neither towards us nor away from us, but transversally, namely at right angles to our line of sight (equivalently: to the direction of light propagation), there will still be a wavelength shift due to time dilation alone -- by (\ref{TimeDilationRevisited}), we have
\be
\lambda = cT = \frac{cT'}{\sqrt{1-(v/c)^2}} = \frac{\lambda'}{\sqrt{1-(v/c)^2}} \index{Doppler effect!transverse}
\ee
for the wavelength $\lambda'$ as measured at the light source and the wavelength $\lambda$ as measured by an inertial observer relative to whom the light source is moving transversally at the speed $v$. 

\subsection{$E=mc^2$}
\label{RelativisticMassSection}

Perhaps the most famous consequence of special relativity is Einstein's formula $E=mc^2$, the equivalence of mass and energy. It can be derived by applying the formulae of special-relativistic mechanics to e.g. simple collisions \citep{Peters1986,Rohrlich1990}. As a first step, one finds that an increase $\Delta E$ in energy for an object, or system, will necessarily result in an equivalent increase $\Delta m$ in the inertial mass, that is, in the mass factor that determines how strongly an object will react to an external force by accelerating. That {\em relativistic (inertial) mass} \index{mass!relativistic} relates velocity $\vec{v}$ and momentum $\vec{p}$ in an inertial system,
\be
\vec{p} = m\cdot \vec{v},
\label{RelativisticMass}
\ee
and is involved in linking the force $\vec{F}$ acting on an object and the change in momentum experienced by that object as
\be
\vec{F} = \frac{\Dd \vec{p}}{\Dd t}.
\ee
For an object moving at a speed $v$ in an inertial reference frame, the relativistic mass is
\be
m = \frac{m_0}{\sqrt{1-(v/c)^2}},
\ee
where $m_0$ is the {\em rest mass}, \index{mass!rest mass} \index{rest mass} that is, the relativistic mass as measured in an inertial frame in which that object is at rest. \index{relativistic mass}

With this definition of mass, $E=mc^2$ states that mass and energy are equivalent: one cannot have energy without inertial mass, or inertial mass without a corresponding energy. Both physical concepts turn out to measure the same quantity; the only difference is one of units (as is shown by the required conversion factor $c^2$).

There is an alternative way of looking at things, which identifies the concept of mass with the rest mass $m_0$, and omits mention of the relativistic inertial mass as defined by (\ref{RelativisticMass}) as far as possible, if not altogether. In that case, one would rewrite the momentum definition as
\be
\vec{p} =  \frac{m_0}{\sqrt{1-(v/c)^2}}\cdot \vec{v},
\label{RelativisticMass2}
\ee
and keep the $v$-dependent extra factor (which is known as the gamma factor) explicit. In that context, Einstein's famous formula describes not an equivalence, but an additional form of energy: the energy
\be
E_0 = m_0c^2
\ee
associated with an object's rest mass; whenever one is talking about energy conservation, this extra form of energy needs to be taken into account. A new form of energy entails new forms of energy conversion: rest mass energy can be converted into other forms of energy, such as electromagnetic energy, or thermal energy.

These two different ways of talking about mass -- mass as relativistic mass, or mass as rest mass -- have led to intense discussions in the physics education community over the past decades, e.g.\ \citenp{Adler1987,1990PhT....43e..13R,Sandin1991,Okun1989,Okun2009,Hecht2009a,Hecht2009b}. I would argue that, given the state of that discussion, a good working knowledge of special relativity should include an awareness of both views of the matter, and of the controversy itself.

\subsection{Energy, momentum, and pressure}
\label{MomentumPressure}

As mentioned in section \ref{LorentzTransformation}, time and space, as described using inertial frames, are really two aspects of a four-dimensional entity called spacetime. Analogously, as it turns out, energy and momentum are really just two aspects of a four-dimensional vector, called \index{four-momentum} {\em four-momentum}. 

In relativistic theories that respect this four-dimensional structure, it does not make sense to formulate equations, say, that single out only energy, and leave momentum aside. Such an equation will only hold for certain inertial observers, but not for others (in the language of relativity, they are not {\em covariant}); this is at odds with the relativity principle which states that any physical law should have the same form in all inertial frames. So why not choose that particular form to describe physics in the first place?

Whenever we formulate equations that are valid in any inertial reference frame, energy and momentum will always occur together, or in the shape of four-momentum. 

This has consequences when, at a more advanced level, namely in general relativity, we try to find the proper way of describing the mass/energy content of the universe: finding the right physical concept to describe sources of relativistic gravity, in other words: the {\em active mass} in general relativity. \index{mass!active}  How do we describe the mass/energy content of a universe filled with, say, a gas consisting of particles, or of photons, whizzing chaotically in every conceivable direction? The first generalization is straightforward. Instead of merely considering the mass directly associated with matter, we will consider all forms of energy as sources of gravity. That is why, in section \ref{EinsteinsEquations} on the generalization of Newton's equation of gravity, our density $\rho$ will include all kinds of energy contributions (equivalently: will be the density of the relativistic mass introduced in section \ref{RelativisticMassSection}).

But even with that addition, we are not yet fully relativistic. After all, energy and momentum cannot be separated from each other in an observer-independent way; if energy is a source of gravity, momentum must be a source term as well. For a gas made of particles, we will get not only a contribution from all these particles' rest masses, kinetic energy and so on, but also from the particle momenta. When the momenta only describe the chaotic motion of a gas that is, on average, at rest, then these momentum contributions turn out to be directly proportional to the pressure $p$ exerted by the gas.  \index{pressure!contribution to four-momentum} That is why, in section \ref{EinsteinsEquations}, there is a contribution $3p/c^2$: a pressure contribution, converted into the proper units of a mass density via the factor $c^2$, with one contribution from each of the three space directions.\footnote{Interestingly enough, while this pressure term is important in cosmology, there is a 
complication in the situation where we are closest to our everyday notion of what pressure is, that is, when we are looking at a gas inside a container with suitable walls. There, the stress experienced by the walls exactly cancels the contribution from the $3p/c^2$ term of the gas \citep{MisnerPutnam1959,2006AmJPh..74..607E}. }

A simple derivation of the pressure contribution to the momentum density can be found in \citenp{Jagannathan2009}; for a more complete understanding, one needs to understand both four-vectors and the tensor description of the properties of a fluid or similar medium; a good introduction can be found in \citenp{Schutz1985}.
\index{special relativity!summarized|)}

\end{appendix}

\section*{Acknowledgements}
\addcontentsline{toc}{section}{Acknowledgements}

The summer school ``Astronomy from Four Perspectives'' from August 26 to September 2 in Heidelberg, at which this lecture was held, was made possible through generous funding from the W. E. Heraeus Foundation.

I would like to thank Thomas Müller for his thorough and, in parts, repeated reading of this manuscript and his many corrections and suggestions; I would also like to thank Martin Wetz, Björn Malte Schäfer and Stefan Brems for helpful comments and useful suggestions concerning previous versions of this manuscript.

\end{document}